\documentclass[12pt]{article}
\pdfoutput=1
\usepackage{mheck}

\institution{HARVARD}{\   Jefferson Physical Laboratory, Harvard University, Cambridge, MA 02138, USA}
\institution{UvA}{\   Institute for Theoretical Physics, University of Amsterdam,  Amsterdam, The Netherlands}

\title{On Exceptional Instanton Strings}

\authors{Michele Del Zotto \worksat{\HARVARD}\footnote{e-mail: {\tt delzotto@physics.harvard.edu}} and Guglielmo Lockhart \worksat{\HARVARD,\UvA}\footnote{e-mail: {\tt lockhart@uva.nl}}}

\abstract{According to a recent classification of 6d $(1,0)$ theories within F-theory there are only six ``pure'' 6d gauge theories which have a UV superconformal fixed point. The corresponding gauge groups are $SU(3),SO(8),F_4,E_6,E_7$, and $E_8$. These exceptional models have BPS strings which are also instantons for the corresponding gauge groups. For $G$ simply-laced, we determine the 2d $\cn=(0,4)$ worldsheet theories of such BPS instanton strings by a simple geometric engineering argument. These are given by a twisted $S^2$ compactification of the 4d $\cn=2$ theories of type $H_2, D_4, E_6, E_7$ and $E_8$ (and their higher rank generalizations), where the 6d instanton number is mapped to the rank of the corresponding 4d SCFT. This determines their anomaly polynomials and, via topological strings, establishes an interesting relation among the corresponding $T^2 \times S^2$ partition functions and the Hilbert series for moduli spaces of $G$ instantons. Such relations allow to bootstrap the corresponding elliptic genera by modularity. As an example of such procedure, the elliptic genera for a single instanton string are determined. The same method also fixes the elliptic genus for case of one $ F_4 $ instanton. These results unveil a rather surprising relation with the Schur index of the corresponding 4d $\cn=2$ models.}

\begin{document}

\maketitle

\tableofcontents

\section{Introduction}
Recently, many new results have been obtained in the context of 6d $(1,0)$ theories \cite{Heckman:2013pva,Gaiotto:2014lca,Ohmori:2014pca,DelZotto:2014hpa,Heckman:2014qba,Ohmori:2014kda,DelZotto:2014fia,Heckman:2015bfa,Apruzzi:2015wna,DelZotto:2015isa,Ohmori:2015pua,DelZotto:2015rca,Heckman:2015ola,Louis:2015mka,Cordova:2015fha,Heckman:2015axa,Ohmori:2015pia,Ohmori:2015tka,Bhardwaj:2015oru,Cremonesi:2015bld,Heckman:2016ssk,Cordova:2016xhm,Font:2016odl,Morrison:2016djb,Benvenuti:2016dcs}; nonetheless, many of their properties remain rather mysterious. A distinctive feature of these theories is that among their excitations they have self-dual BPS strings preserving 2d $(0,4)$ supersymmetry on their worldsheet (see e.g. \cite{Seiberg:1996vs}). The 2d $(0,4)$ theories on the worldsheets of the BPS strings give an interesting perspective on the physics of the 6d $(1,0)$ models \cite{Haghighat:2013gba,Haghighat:2013tka,Hohenegger:2013ala,Haghighat:2014pva,Hosomichi:2014rqa,Kim:2014dza,Cai:2014vka,Haghighat:2014vxa,Kim:2015gha,Gadde:2015tra,Kim:2015fxa,Iqbal:2015fvd,Hohenegger:2015btj,Hohenegger:2016eqy,Haghighat:2016jjf,Kim:2016foj,Shimizu:2016lbw}. Often, such 2d worldsheet theories can be determined using brane engineerings in IIA or IIB superstrings \cite{Brunner:1997gk,Brunner:1997gf,Hanany:1997gh,Hollowood:2003cv}; however, these perturbative brane engineerings are less helpful in the case of 6d (1,0) systems with exceptional gauge groups, a fact which is related to the absence of an ADHM construction for exceptional instanton moduli spaces \cite{Atiyah:1978ri,Douglas:1995bn,Douglas:1996uz,Witten:1994tz}.\footnote{ For a review, see \cite{Dorey:2002ik}.} On the other hand, it is well-known that systems with exceptional gauge symmetries are ubiquitous in the landscape of 6d $(1,0)$ models realized within F-theory\cite{Aspinwall:1997ye}, which rely upon the gauge symmetries of non-perturbative seven-brane stacks \cite{Vafa:1996xn,Morrison:1996na,Morrison:1996pp,Bershadsky:1996nh}. The main aim of this paper is to begin filling this gap, shedding some light on the 2d $(0,4)$ sigma models with target space the exceptional instanton moduli spaces.

The rank of a 6d SCFT is defined to be the dimension of its tensor branch, i.e. the number of independent abelian tensor fields. Each tensor field is paired up with a BPS string which sources it. As our aim is to characterize the exceptional instanton strings, we prefer to avoid the complications arising from bound states of strings of different types, and we choose to work with rank one theories.  The list of 6d $(1,0)$ rank one theories realized within F-theory can be found in section 6.1 of \cite{Heckman:2015bfa}. It is rather interesting to remark that there are only six ``pure'' gauge theories of rank one which can be completed to SCFTs. The corresponding gauge groups are $SU(3),SO(8),F_4,E_6,E_7$ and $E_8$, while the Dirac pairing of the corresponding strings is $n=3,4,5,6,8,12$.

One of the most intriguing features of the 6d $(1,0)$ theories which arise in F-theory is that some gauge groups are ``non-Higgsable'' \cite{Bershadsky:1997sb,Morrison:2012np}, which is the case for the exceptional models above. These models arise, for instance, in the context of the Heterotic $E_8 \times E_8$ superstring compactified on K3 with instanton numbers $(12-n,12+n)$ for the two $E_8$ factors. Whenever $n\neq 0$, the Heterotic string has a strong coupling singularity \cite{Seiberg:1996vs,Duff:1996rs,Duff:1996cf}, which for $3\leq n \leq 12$ supports a 6d (1,0) SCFT of rank one with non-Higgsable gauge symmetries\cite{Morrison:1996na,Witten:1996qb,Morrison:1996pp}. For $n=7,9,10,11$, the non-Higgsable models include some extra degrees of freedom.

It is interesting to remark that the rank one models with $n=3,4,6,8,12$ are realized in F-theory as orbifold singularities of the form $X_n\equiv (\C^2 \times \mathbb{T}^2)/\mathbb{Z}_n$ \cite{Morrison:1996na,Witten:1996qb}: such models are precisely the rank one 6d SCFTs with pure simply-laced gauge group and no additional matter. In what follows we are going to argue that the 2d $(0,4)$ worldsheet theories describing a bound state of $k$ BPS instantonic strings for such theories arise from well--known 4D $\cn=2$ theories compactified on $\PP^1$ with Kapustin's $\beta$-twist \cite{Kapustin:2006hi}: for $n=3,4,6,8,12$ we obtain (respectively) the $\beta$-twisted rank $k$ version of the 4d $\cn=2$ theories $H_2, D_4, E_6, E_7, E_8$ with flavor symmetry $SU(3)$, $SO(8)$, $E_6$, $E_7$, $E_8$ respectively, plus a decoupled free hypermultiplet. In what follows we are going to denote these 4d $\cn=2$ theories simply by $\widetilde{H}_{G}^{(k)}$.

\medskip

Let us denote by $\mathbb{E}_k(X)$ the elliptic genus of the 2d $(0,4)$ worldsheet theories for a bound state of $k$ strings of the 6d SCFT engineered by F-theory on the local elliptic threefold $X$.\footnote{ In general $k$ is a vector of integers labeling various possible bound states of different types of BPS strings. For rank one theories, however, it is a single integer, which coincides with the instanton number for the models we are considering.} The topological string partition function $Z_\text{top}(X)$ of the elliptic threefold has an expansion in terms of the $\mathbb{E}_k(X)$ \cite{Haghighat:2014vxa} which takes the schematic form
\begin{equation}
Z_\text{top}(X) = Z_0(X) \, \left(1 +\sum_{k\geq1} \mathbb{E}_k(X) \, Q^{k}\right).\label{eq:topell}
\end{equation}
Let $\widetilde{X}_n $ be a crepant resolution of $X_n$ within the moduli space of M-theory on $X_n$. From our simple geometric engineering argument it follows, in particular, that
\begin{equation}\label{themainclaim}
\mathbb{E}_k \left( \widetilde{X}_n \right)\Bigg|_{n=3,4,6,8,12} = Z_{(S^2 \times T^2)_{\beta}}\left(\widetilde{H}_{G}^{(k)}\right)\Bigg|_{G=SU(3),SO(8),E_{6,7,8}},
\end{equation}
where the RHS denotes the partition function of the 4d $\cn=2$ theory $\widetilde{H}_{G}^{(k)}$ on the background $S^2 \times T^2$, with Kapustin's $\beta$-twist on $S^2$ \cite{Kapustin:2006hi,Closset:2013vra,Closset:2013sxa,Benini:2015noa,Putrov:2015jpa,Gadde:2015wta}. This gives a rather intriguing relation among the $\beta$-twisted $S^2 \times T^2$ partition function for the 4d $\cn=2$ theories $\widetilde{H}_G^{(k)}$ and the topological strings on $\widetilde{X}_n $. One of the main consequences of this relation is that the Hilbert series \cite{Benvenuti:2006qr} for the moduli spaces of instantons, also known as the Hall-Littlewood limit \cite{Gadde:2011uv} of the superconformal index \cite{Kinney:2005ej} for the $\widetilde{H}^{(k)}_G$ theories \cite{Keller:2012da,Hanany:2012dm}, arise in the limit $q\to 0$ of the $Z_{(S^2 \times T^2)_{\beta}}$ partition function, where $q=e^{2 \pi i \,\tau}$, and the complex structure modulus of the $T^2$ is $\tau$.\footnote{ This fact was remarked in \cite{Gadde:2015xta,Putrov:2015jpa} for the $\widetilde{H}^{(1)}_{E_6}$ and the $\widetilde{H}^{(1)}_{SO(8)}$ theories respectively by a direct computation. Our geometric engineering argument predicts that must be the case for all the $\widetilde{H}^{(k)}_{G}$ theories.} This is because the topological string partition function is equivalent to a 5d BPS count \cite{Gopakumar:1998ii,Gopakumar:1998jq,Dedushenko:2014nya} that, in the limit where the elliptic fiber grows to infinite size, reduces to a 5d Nekrasov partition function \cite{Nekrasov:2003rj,Iqbal:2003ds}, which, for pure gauge theories, coincides with the Hilbert series of the instanton moduli spaces (see Section 2.1 of \cite{Keller:2011ek} for a simple derivation of this fact). This interesting property, combined with the key remark that the elliptic genera are Jacobi forms of fixed index and weight zero,\footnote{ Jacobi forms of given type are elements of bi-graded rings, whose grading is governed by two integers, the weight and the index \cite{MR781735,MR1163219}. These rings are, in particular, finitely generated. For fixed weight and index therefore, each Jacobi form is determined by a finite expansion in the generators.} can be used to ``bootstrap'' the elliptic genus by modularity. The index is determined by the anomaly of the elliptic genus under a modular transformation $ S: \tau \to -1/\tau $; this modular anomaly is captured by the `t Hooft anomalies for the 2d theories, which one can read off from their 4-form anomaly polynomials. The latter have been computed recently for all strings of 6d $(1,0)$ theories by means of anomaly inflow \cite{Kim:2016foj,Shimizu:2016lbw}. Our geometric engineering argument gives an alternative derivation for $G$ simply-laced. Using the knowledge of the anomaly polynomial coefficients for the 2d theories and their $q\to 0$ limits, one can formulate an Ansatz in the appropriate ring of weak Jacobi forms which allows to bootstrap the elliptic genera for the 2d $(0,4)$ models of interest --- including the case $ G = F_4 $.\footnote{The 2d $(0,4)$ worldvolume theory of the BPS instanton strings for the 6d $(1,0)$ pure $G=F_4$ gauge theory can be determined by a generalization of the methods of \cite{Martucci:2014ema}, by inserting two appropriate surface defects for the $H^{(k)}_{E_6}$ theories on ${\mathbb P}^1$. A detailed study of this model (and other models obtained by similar techniques) goes beyond the scope of the present work and will be discussed elsewhere \cite{DZL2}. Nevertheless, in this paper we will compute the elliptic genus for one $F_4$ string by using modular bootstrap and basic properties of this 2d CFT.} In this paper we determine the modular anomaly for all $ G $ and for any number $ k $ of strings; for the case $ k = 1 $, we uniquely determine the elliptic genera for all $ G $ by modularity, which is one of the main results of this paper. 

\medskip

The modular bootstrap approach outlined above is inspired by recent progress in topological string theory, where modularity, in combination with other geometric considerations, provides a very powerful approach for solving topological string theory on elliptic Calabi-Yau threefolds.\footnote{ In fact, at the level of genus-zero invariants, a similar approach was used to study the topological string partition function for the local half-K3 surface already in \cite{Klemm:1996hh}.} In that context, the modular anomaly of the elliptic genera translates to the holomorphic anomaly equation of topological string theory. By using modularity and the holomorphic anomaly equation and making an Ansatz for the topological string partition analogous to Equation \eqref{eq:topell}, the authors of \cite{Huang:2015sta,Huang:2015ada} were able to solve topological string theory on various compact elliptic Calabi-Yau threefolds to all genus, for very large numbers of curve classes in the base of the elliptically-fibered Calabi-Yau and arbitrary degree in the fiber class, for geometries where the elliptic fibers are allowed to develop degenerations of Kodaira type $I_1$. From the topological string theory perspective, our approach for computing elliptic genera of 6d SCFTs with gauge group corresponds to a generalization of the techniques developed in \cite{Huang:2015sta,Huang:2015ada} to a particular class of non-compact Calabi-Yau threefolds with more singular degenerations of the elliptic fiber. An interesting question is to further extend this approach to generic elliptic Calabi-Yau threefolds, which one may take to be either compact or non-compact (in which case the refined topological string partition function can be computed), corresponding respectively to 6d $ (1,0) $ theories  with or without gravity, with a variety of allowed spectra of tensor, vector, and hypermultiplets; this wider class of theories is currently under study and will be discussed elsewhere \cite{KlockB}.

\medskip

Remarkably, we find also a connection between the explicit expressions for the $ (T^2\times S^2)_\beta $ partition functions and the Schur indices of the $ H^{(1)}_G $ theories. For $ G = SU(3) $, the Schur index can be obtained as a specific limit of $ Z_{(S^2 \times T^2)_{\beta}} $; for other choices of $ G $ the relation is more involved, but nonetheless we find that both the Schur index and $ Z_{(S^2 \times T^2)_{\beta}} $ can be computed out of an auxiliary function, $ L_G(v,q) $. Naively it would be tempting to identify this function with the Macdonald limit of the index, especially because 1) it reduces to the Hall-Littlewood index in the limit $ q\to 0$ and 2) in an appropriate limit it specializes to the Schur index. However, it is easy to check that this is not the case. We find that the function $ L_G(v,q) $ is a power series in $ v,q $ whose coefficients are sums of dimensions of representations of the global symmetry group $ G $ with positive multiplicities. It would be very interesting to relate these results to BPS spectroscopy along the lines of \cite{Cecotti:2010fi,Cordova:2015nma,Cecotti:2015lab,Cordova:2016uwk}.

\medskip

We leave open the problem of determining the 2d SCFTs corresponding to $n=5,7$. This is related to the fact that the corresponding geometries involve pointwise singularities of higher order \cite{Morrison:2012np}, which generate non-trivial monodromies for $\tau_E$. This entails in particular that these theories are not simple $\beta$-twists of the type considered above. Another line of investigation which we leave open is the computation of the elliptic genera for our models from the 2d TQFT of \cite{Putrov:2015jpa}.

\medskip

This paper is organized as follows: in Section \ref{6drev} we briefly review some salient features of the F-theory backgrounds that engineer the 6d SCFTs we study in this paper; Section \ref{2d4d6d} contains a review of the main properties of the 4d $\mathcal{N}=2$ theories of type $H_G^{(k)}$ and the geometric engineering argument identifying the twisted compactification leading to the 2d $(0,4)$ worldsheet theories; in Section \ref{sec:worldsheet} we discuss general properties of the 2d SCFTs which follow from the engineering: the central charges, the anomaly polynomial, and the elliptic genera; in Section \ref{sec:topstrings} we review the topological string argument sketched above; in Section \ref{sec:modular} we derive our Ansatz from the modularity properties of the elliptic genera; finally, in Section \ref{sec:Sch} we remark on an intriguing relation among the elliptic genera derived in Section \ref{sec:modular} and the Schur index of the corresponding $\cn=2$ theories.

\section{Minimal 6d (1,0) SCFTs}\label{6drev}
\subsection{F-theory engineering of 6d SCFTs in a nutshell}
In this section we quickly review the geometric setup of \cite{Heckman:2013pva}, which provides the geometric engineering of 6d (1,0) SCFTs from F-theory, including the minimal ones which are the focus of this paper. For our purposes, an F-theory background can be viewed either as M-theory on an elliptically fibered Calabi-Yau $X$ with section: %the section can be dropped
\begin{equation}\begin{matrix} E &\hookrightarrow& X\\&&\downarrow \\ & & B
\end{matrix}\quad\end{equation}
 in the limit where the elliptic fiber $E$ has shrunk to zero size or, dually, as a compactification of Type IIB string theory on a K\"ahler internal manifold $B$ which is stable and supersymmetric thanks to non-trivial axio-dilaton monodromies sourced by seven-branes \cite{Vafa:1996xn}. In particular, the IIB seven-branes are dual to shrunken singular elliptic fibers in the M-theory realization and the complex structure parameter of the elliptic curve $\tau_E$ is dual to the axio-dilaton field in IIB. In order to engineer a 6d system, one takes $B$ to have complex dimension 2. As the system is decoupled from gravity, its volume has to be infinite, and hence $X$ must be a local Calabi-Yau threefold.\footnote{ The infinite-volume limit has to be taken with care, see the discussion in \cite{Cordova:2009fg,Bhardwaj:2015oru}.} Consider a local Weierstrass model for the elliptic fibration of $X$, 
\be
y^2  = x^3 + f x + g
\ee
where $f$ and $g$ are sections of $\mathcal{O}(-4 K_B)$ and $\mathcal{O}(-6K_B)$ respectively. The discriminant of the fibration is $\Delta \equiv 4 f^3 + 27 g^2 \in \mathcal{O}(-12K_B)$, and $\Delta = 0$ is the locus where the fiber degenerates, which is dual to the position of the IIB seven-branes. To engineer a minimal 6d SCFT one needs a geometry which has no intrinsic scale and an isolated special point $p \in B$ such that at least one of the following holds
\begin{itemize}
\item[$a.)$] The order of vanishing of $(f,g,\Delta)\geq (4,6,12)$ at $p \in B$;
\item[$b.)$] The K\"ahler base of the Calabi-Yau 3-fold is an orbifold of type $\C^2/\Gamma_{HMV}$ where $\Gamma_{HMV}$ is a discrete subgroup of $U(2)$ of HMV type \cite{Heckman:2013pva} and the point $p$ is fixed by the orbifold group action.
\end{itemize}
Examples where the point $p$ is smooth in the base $B$ but $a.)$ is satisfied are provided by the theories on the worldvolumes of a stack of $N$ Heterotic $E_8$ instantonic 5-branes \cite{Aspinwall:1997ye} which corresponds in F-theory to a point $p$ with a singular fiber with order of vanishing  of $(f,g,\Delta)\equiv(4N,6N,12N)$. Examples where the fiber at $p$ is smooth but $b.)$ is satisfied are the $(2,0)$ theories engineered in IIB as orbifolds by discrete subgroups of $SU(2)$. For most $(1,0)$ SCFTs realized in F-theory both $a.)$ and $b.)$ occur \cite{Heckman:2013pva,Heckman:2015bfa}. The Calabi-Yau condition on $X$ imposes rather strong constraints on the allowed discrete subgroups $\Gamma_{HMV} \subset U(2)$ in $b.)$ --- see \cite{Heckman:2013pva}. In particular, to each allowed $\Gamma_{HMV}$ corresponds a minimal model of non-Higgsable type \cite{Heckman:2013pva}. The models so obtained are minimal in the sense that they sit at the end of a chain of gauge-group Higgsings and the corresponding gauge symmetries cannot be Higgsed further \cite{Morrison:2012np}. If the SCFT has a non-Abelian flavor symmetry, this is engineered by a flavor divisor through $p$, \textit{i.e.} a non-compact divisor belonging to the discriminant $\Delta$ which contains $p$ along which the order of vanishing of $(f,g,\Delta)$ in the Weierstrass model are strictly less than $(4,6,12)$ \cite{Aspinwall:1997ye,DelZotto:2014hpa}. Abelian flavor symmetries are more subtle, being related to the Mordell-Weyl group of the elliptic fibration \cite{Morrison:2012ei}.\footnote{ In some cases it is possible to determine the abelian factors of the flavor groups by means of Higgs branch RG flows, see \cite{Heckman:2016ssk}.}

Resolving the singularity in the base by blow-ups, removing all points where the order of vanishing of $(f,g,\Delta)$ in the Weierstrass model is $\geq (4,6,12)$ while keeping the elliptic fiber shrunk to zero size, corresponds to flowing along the tensor branch of the 6d model, which is parametrized by the vevs of the tensor multiplet scalars dual to the K\"ahler classes of the divisors of the resolution.  On the tensor branch the 6d theories develop a sector of BPS strings, which are engineered by D3-branes wrapping the divisors resolving the singularity at the point $p$ in the base. For the geometries corresponding to SCFT tensor branches, the resolution divisors have always the topology of $\mathbb{P}^1$s \cite{Heckman:2013pva}.\footnote{ For the geometries corresponding to tensor branches of LSTs this does not always occur \cite{Bhardwaj:2015oru}.} The K\"ahler volume of each such divisor is proportional to the tension of the corresponding BPS string. In particular, such strings become tensionless at the singularity. Whenever one such divisor $C$ is also an irreducible component of the discriminant of the elliptic fibration, this signals that in the IIB picture we have a wrapped seven-brane along it. The seven-brane topology is $\mathbb{R}^{1,5} \times C \subset \mathbb{R}^{1,5} \times \widetilde{B}$, where $\widetilde{B}$ is the resolved base corresponding to the 6d tensor branch. Along the flat $\mathbb{R}^{1,5}$ directions the strings on the seven-brane give rise to a gauge SYM sector with gauge coupling $1/g^2 \sim \text{vol } C$. The precise form of the gauge group is encoded in the corresponding singularity for the elliptic fiber along $C$ --- see e.g. table 4 of \cite{Grassi:2011hq} for a coincise review. If this is the case the wrapped D3-branes have the dual r\^ole of instantons for the 6d gauge group induced by the wrapped seven-brane.

\subsection{Minimal 6d (1,0) SCFTs from F-theory orbifolds}
In order to avoid complications with threshold bound states among BPS strings of different types, we focus on 6d theories of rank one.  Consider a resolution of the singularity at $p\in B$. As the model is of rank one, the corresponding resolution is based on a single compact divisor of the base $B$ with negative self-intersection. Let us call such curve $\Sigma$. It is easy to see that $\Sigma$ must have the topology of a $\mathbb{P}^1$ (see the appendix B of \cite{Heckman:2013pva} for a derivation). The negative of the self-intersection number of $\Sigma$ gives the Dirac pairing of the BPS string obtained by wrapping a D3-brane on $\Sigma$, which distinguishes between different ``flavors'' of BPS strings. Naively, one would expect that all possible self Dirac pairings are allowed, but this is not the case \cite{Morrison:2012np}. First of all, whenever the irreducible divisor $\Sigma$ in the resolution of $p\in B$ has self--intersection $\leq -3$ the Calabi-Yau condition on $X$ forces the elliptic fiber to degenerate along $\Sigma$. Moreover, this also puts a bound $ \Sigma\cdot \Sigma \geq - 12$: a more negative self--intersection number would lead to fibers which are too singular, so that $c_1(X)$ cannot vanish. In the IIB picture, this has the interpretation that the backreaction on the geometry arising from too many wrapped seven-branes destabilizes the background \cite{Aspinwall:1996mn}. For $ -12 \leq \Sigma \cdot \Sigma \leq -3$, $\Sigma$ is necessarily an irreducible component of the discriminant of the elliptic fibration, hence in the engineering it corresponds to a non-Higgsable coupled tensor-gauge system and the wrapped D3-branes gives rise to BPS instanton strings. The field content of the six-dimensional theories obtained via geometric engineering is such that the 6d gauge anomalies are automatically canceled via the Green-Schwarz mechanism \cite{Green:1984bx,Sadov:1996zm,Grassi:2000we}. For $ \Sigma\cdot \Sigma = -9, -10,-11$, the corresponding models needs respectively 3,2,1 further blow-ups to flow on the tensor branch, so these models map respectively to rank 4,3,2 SCFTs.

\medskip

In all these cases, shrinking $\Sigma$ to a point gives rise to a Hirzebruch-Jung singularity in the K\"ahler base. Recall that an $HJ_{p,q}$ singularity is the K\"ahler orbifold of $\C^2$ corresponding to the action
\be
HJ_{p,q}\,\,\colon \qquad (z_1,z_2) \to (\omega z_1, \omega^q z_2) \qquad \omega^p = 1.
\ee
The rank one theories correspond to bases with Hirzebruch-Jung orbifold singularity of types $(p,q)=(n,1)$ with $n=1,2,3,4,5,6,7,8,12$ \cite{Reid,Heckman:2013pva}: these singularities  can indeed be resolved with a single blow up in the base, leading to a single divisor of self-intersection $-n$. The resolved base is
\begin{equation}\label{thenormalbundah}
\widetilde{B} = \text{Tot}\left(\mathcal{O}(-n) \to \mathbb{P}^1\right) \qquad 1\leq n \leq 12,
\end{equation} 
where the K\"ahler class of the base $\mathbb{P}^1$ corresponds to the vev of the tensor multiplet scalar parametrizing the 6d tensor branch. In Table \ref{HJns} we list the minimal non-Higgsable gauge groups corresponding to such singularities \cite{Morrison:2012np}.

\begin{table}
\begin{center}
\begin{tabular}{|c|ccccccccc|}
\hline
$HJ_{n,1}$& $HJ_{1,1}$ &$HJ_{2,1}$&$HJ_{3,1}$&$HJ_{4,1}$&$HJ_{5,1}$&$HJ_{6,1}$&$HJ_{7,1}$&$HJ_{8,1}$&$HJ_{12,1}$\phantom{$\Big|$}\\ 
\hline
fiber & $I_0$ & $I_0$  & $IV$ & $I_0^*$ & $IV^*_{ns}$ & $IV^*$ & $III^*$ & $III^*$ & $II^*$\phantom{$\Big|$}\\ 
$\mathfrak{g}_{min}$ & none & none &$\mathfrak{su}_3$&$\mathfrak{so}_8$&$\mathfrak{f}_4$&$\mathfrak{e}_6$&$\mathfrak{e}_7\oplus \tfrac{1}{2}\mathbf{56}$&$\mathfrak{e}_7$&$\mathfrak{e}_8$\phantom{$\Big|$}\\
\hline
\end{tabular}
\end{center}\caption{ Minimal gauge groups for the 6d theories of rank 1. For $n=1$ one obtains the E-string theory, the theory describing a single heterotic $E_8$ instanton that has shrunk to zero size. As $H_{2,1}$ is a Du Val singularity of type $A_1$, the surface is a local CY 2-fold and one obtains the $A_1$ (2,0) SCFT.  The model corresponding to $HJ_{7,1}$ contains some charged matter in the $\tfrac{1}{2}\mathbf{56}$ representation of $\mathfrak{e}_7$.}\label{HJns}
\end{table}

In most of this paper we focus on the models corresponding to $n=3,4,6,8,12$ which can be realized as orbifolds in F-theory of the form \cite{Witten:1996qb,Morrison:1996pp,Heckman:2013pva}
\be\label{orbo1}
X_n\equiv (T^2 \times \C^2)/\mathbb{Z}_n, \qquad n=3,4,6,8,12.
\ee 
Denoting by $\lambda$ the $T^2$ coordinate and $(z_1,z_2)$ the $\mathbb{C}^2$ coordinates the orbifold action is 
\be\label{orbo2}
(\lambda,z_1,z_2) \to (\omega^{-2}\, \lambda ,\omega \, z_1 ,\omega \, z_2) \qquad \omega^n = 1.
\ee
The models with $ n = 3,6,8,12 $ deserve special attention as they correspond, respectively, to the gauge groups $SU(3)$ and $E_{6,7,8}$ in 6d: the naive ADHM quiver for $SU(3)$ gives rise to an anomalous 2d $(0,4)$ system \cite{Kim:2016foj}, while it is well-known that there is no ADHM construction for the instanton worldsheet theories of the $E_{6,7,8}$ theories.

\section{Instanton strings and $\widetilde{H_G}^{(k)}$ theories}\label{2d4d6d}

\subsection{A lightning review of $\widetilde{H}_G^{(k)}$ models}

\begin{figure}
\begin{center}
\begin{tabular}{c|cccccccccc}
\, & \multicolumn{4}{c}{$\substack{\displaystyle{\,\,\mathbb{C}^2_\parallel}\\\vspace{-.05in}\\\overbrace{\hspace{.9in}}}$} & \multicolumn{4}{c}{$\substack{\displaystyle{\,\,\mathbb{C}^2}\\\vspace{-.05in}\\\overbrace{\hspace{.9in}}}$} &  \multicolumn{2}{c}{$\substack{\displaystyle{\,\,\mathbb{C}_\perp}\\\vspace{-.05in}\\\overbrace{\hspace{.2in}}}$} \\
 & 0 &1 & 2 & 3 & 4 & 5 & 6 & 7 & 8 & 9\\
\hline
seven-brane & X & X&  X & X & X & X & X & X & - & - \\
D3 & - & - & - & - & X & X & X & X & - & - \\
\end{tabular}
\end{center}
\caption{IIB brane engineering of the $\widetilde{H}^{(k)}_G$ models.}\label{37branes}
\end{figure}

The 4d $\cn=2$  theories  of type $H_G^{(k)}$ can be constructed in a variety of ways (see e.g. \cite{Argyres:1995xn,Banks:1996nj,Minahan:1996fg,Ganor:1996xd,Minahan:1996cj,Ganor:1996pc,Douglas:1996js,Fayyazuddin:1998fb,Aharony:1998xz,Benini:2009gi}). In F-theory these models (and their higher rank generalization) arise as the worldvolume theories of a stack of D3-branes probing a stack of exotic seven-branes. In M-theory such exotic seven-branes correspond to local elliptic K3s, with shrunk fibers of Kodaira type respectively $IV, I_0^*,IV^*,III^*,$ and $II^* $. The corresponding seven-branes have gauge symmetries respectively of types $G = SU(3),SO(8),E_{6,7,8}$.

Let us consider for the moment the Type IIB picture (see Figure \ref{37branes}). The low energy worldvolume theory on the seven-brane is an 8d SYM gauge theory. The instantons of such eight-dimensional gauge theories are identified with D3 branes which are parallel to the seven-branes.

Consider the case of a single D3 brane probe. The transverse geometry to the stack of seven-branes is identified with the Coulomb branch of the probe theory \cite{Sen:1996vd,Banks:1996nj}, which has a nontrivial deficit angle encoding the axio-dilaton monodromy induced by the seven-branes. The Higgs branch of the probe D3 brane theory corresponds to dissolving the D3 brane into a gauge flux on the seven-brane. With a single D3-brane probe one obtains rank-1 SCFTs with flavor symmetries corresponding to the gauge algebras on the seven-branes worldvolumes and Higgs branch which equals the reduced moduli space of one $G$ instanton.

Traditionally, these models have been denoted as $H_2,D_4,E_6,E_7$ and $E_8$, but we prefer to denote them as $H_G^{(1)}$, since all these models arise from $T^2$ compactifications of the theory of one Heterotic $E_8$ instanton with Wilson lines for the flavor symmetry \cite{Ganor:1996xd,Ganor:1996pc}.\footnote{ There are two additional types of exotic seven-branes corresponding to the Kodaira fibers of type $II$ and $III$, which give rise to the models $H_{\varnothing}^{(k)}$ and $H_{SU(2)}^{(k)}$. These branes however cannot be consistently compactified on a $ \mathbb{P}^1 $ unless they intersect other seven-branes. For this reason they do not play a role in the construction of the 6d minimal models we are considering in this paper --- cf. Footnote \ref{fn:others}.\label{fn:sbranah}}

Corresponding to $ k > 1 $ instantons on the seven-branes are stacks of $k$ parallel D3 branes, whose worldvolume support rank $k$ generalizations of the rank one 4d $\cn=2$ models above which we denote $H_G^{(k)}$. We summarize some of their properties in Table \ref{tb:D3Table}. The $k$-dimensional Coulomb branches of the $H_G^{(k)}$ models are symmetric products of the Coulomb branches of the $H_G^{(1)}$ theories, while the Higgs branches of the $H_G^{(k)}$  theories are given by the reduced moduli spaces of $k$ $G$-instantons \cite{Douglas:1996js,Fayyazuddin:1998fb,Aharony:1998xz}. In particular, the Coulomb branch operators of the $H_G^{(k)}$ theories have dimensions $\{ j \Delta_G\}_{j=1,2,\dots,k}$, where $\Delta_G$ is the dimension of the Coulomb branch operator of the rank one model $H_G^{(1)}$ (cfr. Table \ref{tb:D3Table}).

\medskip

To be more precise, for any $ k\geq 1 $ the D3 worldvolume theory also includes a decoupled free hypermultiplet associated to the center of mass motion of the instantons in $\mathbb{C}^2_{\parallel}$. Let   $\widetilde{H}_G^{(k)}$ denote the 4d $\cn=2$ SCFT corresponding to the direct sum of the $H_G^{(k)}$ SCFT with the SCFT of a decoupled free hyper. The Higgs branch of the $\widetilde{H}_G^{(k)}$ theory is the moduli space of $k$ $G$-instantons, which is going to play an important role in what follows.

\begin{table}[t!]
\begin{center}
\begin{tabular}{|c|cc||ccccc|}
\hline
G $\phantom{\Big|}$& $-$ & $SU(2)$ & $SU(3)$ & $SO(8)$ & $E_6$ & $E_7$ & $E_8$\\
\hline
Kodaira fiber $\phantom{\Big|}$& $II$ & $III$ & $IV$ & $I_0^*$& $IV^*$& $III^*$ & $II^*$\\
$\Delta_G$ $\phantom{\Big|}$ &$6/5$ & $4/3$ & $3/2$ & $2 $& $3 $& $4 $& $6$ \\
$n_h- n_v$ $\phantom{\Big|}$ &$6k/5-1$&$2k-1$&$3k-1$&$6k-1$&$12k-1$&$18k-1$&$30k-1$ \\
\hline
\end{tabular}
\end{center}
\caption{Properties of $H_G^{(k)}$ theories. The type of Kodaira fiber associated to the $H_G^{(k)}$ theory is listed, as well as the scaling dimension $ \Delta_G$ of the lowest dimensional Coulomb branch operator and the difference between the effective numbers of hyper and vector multiplets.}
\label{tb:D3Table}
\end{table}
\medskip
The global symmetries of the $\widetilde{H}_G^{(k)}$ theories can be read off from Figure \ref{37branes}. The strings stretched between the stack of D3 branes and the seven-branes give rise to a $G$-type flavor symmetry which couples the $\widetilde{H}_G^{(k)}$ theory to the seven-brane gauge theory. The motion of the stack of D3 branes in the ${\mathbb C}^2_{\parallel}$ directions endows the system with an $SU(2)_L \times SU(2)_R$ global symmetry, while ${\mathbb C}_{\perp}$ gives a $U(1)_r$ symmetry. The group $SU(2)_R \times U(1)_r$ is identified with the R-symmetry of the 4d $\cn=2$ superalgebra, while $SU(2)_L$ is an additional flavor symmetry of the system. For $k=1$ only the center of mass free hypermultiplet transforms under $SU(2)_L$, and the flavor symmetry of the $H^{(1)}_G$ factor is just $G$. For $k>1$ the flavor symmetry of the $H^{(k)}_G$ models is $SU(2)_L \times G$.

\subsection{The $\beta$-twisted $\widetilde{H}_G^{(k)}$ models and 6d instanton strings}
\label{sec:btwisted}

\begin{figure}
\begin{center}
\begin{tabular}{c|cccccccccc}
IIB & \multicolumn{6}{c}{$\substack{\displaystyle{\,\,\mathbb{R}^{1,5}} \\ \vspace{-0.05in}  \\\overbrace{\hspace{1.2in}} \\  \vspace{-0.01in}}$}& \multicolumn{4}{c}{$\substack{\displaystyle{\,\,\widetilde{B}} \\ \vspace{-0.05in} \\\overbrace{\hspace{.9in}} \\  \vspace{-0.01in}}$} \\
background & \multicolumn{4}{c}{$\substack{\displaystyle{\,\,\mathbb{C}^2_\parallel}\\\vspace{-.04in}\\\overbrace{\hspace{.9in}}}$} & \multicolumn{2}{c}{$\substack{\displaystyle{\,\,\mathbb{R}^{1,1}}\\\vspace{-.04in}\\\overbrace{\hspace{.4in}}}$} & \multicolumn{2}{c}{$\substack{\displaystyle{\,\,\mathbb{P}^1}\\\vspace{-.04in}\\\overbrace{\hspace{.4in}}}$} &  \multicolumn{2}{c}{$\substack{\displaystyle{\,\,\mathcal{O}(-n)}\\\vspace{-.04in}\\\overbrace{\hspace{.4in}}}$} \\
 & 0 &1 & 2 & 3 & 4 & 5 & 6 & 7 & 8 & 9\\
\hline
seven-brane & X & X&  X & X & X & X & X & X & - & - \\
D3 & - & - & - & - & X & X & X & X & - & - \\
\end{tabular}
\end{center}
\caption{IIB description of the tensor branch of the 6d (1,0) theory.}\label{wrapped37branes}
\end{figure}

Compactification of the seven-brane worldvolume theory on $\PP^1$ gives rise to a six dimensional $(1,0)$ SYM sector, with $1/g_{YM}^2 \sim \text{vol } \PP^1$. Furthermore, from the reduction of Type IIB fields on the $ \mathbb{P}^1 $ one obtains a tensor multiplet with scalar vev $ \langle\phi \rangle \sim \text{vol } \mathbb{P}^1 $, coupled to the SYM sector \emph{\`a la} Green-Schwarz \cite{Green:1984bx,Sadov:1996zm}, automatically cancelling the anomalies. To this tensor multiplet are coupled strings of tension $ t \sim \langle \phi \rangle $ which arise by wrapping the D3 branes on the $\mathbb{P}^1$. From such engineering it is clear that the worldsheet theories of the 6d instantonic strings for the minimal models with $n=3,4,6,8,12$ are just given by an appropriate twisted compactification on $ \mathbb{P}^1 $ of the $\widetilde{H}_G^{(k)}$ theories (see Figure \ref{wrapped37branes}). There are several possible twists for an $\cn=2$ theory on a $\mathbb{P}^1$; the twist which is relevant for us can be determined by the structure of the ambient geometry. Consider the case of a single D3 brane probe. The normal direction to the 7 branes is identified with the Coulomb branch of the probe D3 brane \cite{Sen:1996vd,Banks:1996nj}. In wrapping the $\mathbb{P}^1$, the normal direction to the seven-brane becomes the fiber of a nontrivial line bundle over it of the form in equation \eqref{thenormalbundah}, and therefore the Coulomb branch of the probe D3 brane supporting the $\widetilde{H}^{(1)}_G$ theory also becomes non-trivially fibered over the $ \mathbb{P}^1 $. This suggest to choose a twist for which
\begin{equation}\label{eq:DeltaG}
n =  -R_{G} = - 2 \Delta_G.
\end{equation}
Moreover, the D3 branes engineer instantons for the same gauge groups in 8d and 6d: dissolving the D3s into flux must give rise to identical Higgs branches, \emph{i.e.} the instanton moduli space for the corresponding gauge group. This signals that the $SU(2)_R$ symmetry is left untouched by the twist. These two facts together with the requirement of 2d $(0,4)$ symmetry, fix the twist to be just an embedding of the $U(1)_r$ R-symmetry group of the 4d $\cn=2$ SCFTs in the holonomy of $\mathbb{P}^1$. Supersymmetric twistings of 4d $\cn=2$ theories on four manifolds which are products of Riemann surfaces are well known \cite{Bershadsky:1995vm,Kapustin:2006hi}: The twisting above is precisely a Kapustin $\beta$-twist on the four manifold $\mathbb{R}^{1,1} \times \mathbb{P}^1$ \cite{Kapustin:2006hi}. Let us proceed by briefly reviewing such construction.

Recall that an $\cn=2$ SCFT has a global R-symmetry $U(1)_r \times SU(2)_R$. Consider a four-manifold of the form $\Sigma \times C$, with $\Sigma$ a two dimensional flat Lorentzian or Euclidean manifold and $C$ a Riemann surface with holonomy group $U(1)_C$. To preserve some supersymmetry on $\Sigma$, one needs to identify $U(1)_C$ with a $U(1)$ subgroup of the R-symmetry. There are two canonical choices: the $\alpha$-twist identifies $U(1)_C$ with a Cartan subgroup of $SU(2)_R$, the $\beta$-twist identifies it with $U(1)_r$. Fixing complex structures on $\Sigma$ and $C$, left handed spinors are sections of
\begin{equation}
S_- = K_\Sigma^{-1/2}\otimes K_C^{1/2} + K_\Sigma^{1/2}\otimes K_C^{-1/2}
\end{equation}
while right handed spinors are sections of
\begin{equation}
S_+ = K_\Sigma^{-1/2}\otimes K_C^{-1/2} + K_\Sigma^{1/2}\otimes K_C^{1/2}.
\end{equation}
The 8 supercharges of the 4d $\cn=2$ superalgebra transform as an $SU(2)_R$ doublet of left-handed spinors with $U(1)_r$ charge $+1$ and an $SU(2)_R$ doublet of right handed spinors with $U(1)_r$ charge $-1$. By the $\beta$-twist, these become sections of
\begin{equation}
\begin{aligned}
&S_- \otimes K_C^{1/2} = K_\Sigma^{-1/2}\otimes K_C + K_\Sigma^{1/2}\otimes \mathcal{O}_C \qquad\quad U(1)_r \text{ charge } +1,\\
&S_+\otimes K_C^{-1/2} =K_\Sigma^{-1/2}\otimes K_C^{-1} + K_\Sigma^{1/2}\otimes \mathcal{O}_C \qquad U(1)_r \text{ charge } -1.
\end{aligned}
\end{equation}
Of the 8 supercharges, only 4 transform as scalars along $C$. All four supercharges have the same chirality on $\Sigma$, leading to 2d $(0,4)$ supersymmetry. In the language of \cite{Festuccia:2011ws,Dumitrescu:2012ha,Dumitrescu:2012at,Closset:2013vra}, the $\beta$-twist can be viewed as a curved rigid supersymmetry background preserving four supercharges. In particular, we are interested in backgrounds of the form $\mathbb{R}^{1,1} \times S^2$ or $T^2 \times S^2$ for theories with a $U(1)$ R-symmetry \cite{Closset:2013sxa,Nishioka:2014zpa,Honda:2015yha,Benini:2015noa,Gadde:2015wta}. One starts with a background for the new minimal $\cn=1$ supergravity that has a non-trivial unit background $U(1)$ R-symmetry flux on the $S^2$ \cite{Closset:2013vra,Closset:2013sxa,Nishioka:2014zpa,Benini:2015noa}, and identifies the R-symmetry background gauge field of the supergravity with the $U(1)_r$ symmetry of the $\cn=2$ theory.\footnote{ In Section 4 of \cite{Gadde:2015wta} the $\beta$-twist is referred to as the \emph{Higgs reduction}. See also appendix F of \cite{Putrov:2015jpa} for more details.} In presence of this R-symmetry monopole one obtains consistent geometries only if the $U(1)_r$ charges are quantized over the integers \cite{Closset:2013vra,Closset:2013sxa}.\footnote{ Notice that the theories $H^{(1)}_{\varnothing}$ and $H^{(1)}_{SU(2)}$ have Coulomb branch operators with R-charges respectively $12/5$ and $8/3$, hence if $\beta$-twisted these would not lead to consistent geometries and in order to compactify them on spheres a different background is necessary --- cf. Footnote \ref{fn:sbranah}.\label{fn:others}} Another interesting comment is that the two-dimensional theory does not have a Coulomb branch. This is consistent with the fact that under the $ \beta $-twist the degrees of freedom that correspond to moving the D3 brane within $ \widetilde B $ are projected out.

\medskip

Notice that this very same reasoning applies straightforwardly to higher instantonic charge $k$, \emph{mutatis mutandis}. In the case of a D3 brane stack, the vevs of the Coulomb branch operators for the $H_G^{(k)}$ theories, being symmetric products of the transverse direction to the 7 branes, also become fibers of nontrivial bundles over $\mathbb{P}^1$ of the form $\bigoplus_{j=1}^k{\mathcal O}(- 2 j \Delta_G)$. Moreover, the Higgs branches of the theory on a stack of $k$ wrapped D3 branes are still given by dissolving instanton into flux, and therefore coincide with $k$-instanton moduli spaces for the corresponding gauge groups. Following the same argument as for the $ k=1 $ case, this forces the theories on the worldsheet of the wrapped D3 branes to be $\beta$-twisted $\widetilde{H}^{(k)}_G$ theories on $\mathbb{R}^{1,1}\times \mathbb{P}^1$. More precisely, the $\beta$-twist of the $\widetilde{H}_G^{(k)}$ models on $\mathbb{R}^{1,1}\times \mathbb{P}^1$ gives rise to the 2d $(0,4)$ theories which flow in the IR to the worldsheet theories for the 6d BPS instantons of charge $k$. In what follows we denote the latter 2d $(0,4)$ IR SCFTs by $\widetilde h_G^{(k)}$, and we also denote by $ h^{(k)}_{G} $ the same theories with the the decoupled center of mass (0,4) hypermultiplet removed.

\medskip

By construction, in the limit in which the volume of the $ \mathbb{P}^1 $ goes to zero, a $ \beta $-twisted 4d $ \mathcal{N}=2 $ theory gives a $(0,4)$ sigma model into its Higgs branch \cite{Kapustin:2006hi,Putrov:2015jpa,Gadde:2015wta}. Of course, the Higgs branches of the $\widetilde{H}^{(k)}_G$ models are precisely the hyperk\"ahler moduli spaces of $k$ $G$ instantons $\mathcal{M}_{G,k}$. The condition for obtaining a gauge anomaly free $(0,4)$ SCFT are equivalent to the condition for having a non-anomalous $U(1)_r$ symmetry for the 4d $\cn=2$ theory we began with \cite{Kapustin:2006hi}.

\section{Some generalities about the 2d $(0,4)$ $\widetilde h_G^{(k)}$ SCFTs }\label{sec:worldsheet}

Typically, the models obtained by the procedure outlined in Section \ref{sec:btwisted} are not 2d (0,4) SCFTs. As the BPS instanton strings arise at low energies on the tensor branch of the 6d theory, the D3 branes are wrapping a $\mathbb{P}^1$ of finite size. Sending the volume of the $\mathbb{P}^1$ to zero (and hence sending the 6d gauge coupling to infinity) corresponds to reaching the 6d superconformal point; this simultaneously captures an RG flow of the worldsheet theories of the strings to an IR fixed point. A crucial consequence of this fact is that whole equivalence classes of 2d theories which flow to the same IR fixed point can correspond to the same BPS worldsheet theory, which in a certain way mimics what happens in the context of the supersymmetric quantum mechanics description of BPS states in 4d $\cn=2$ theories \cite{Cecotti:2011rv,Alim:2011kw}. In particular, whenever the $\widetilde  h^{(k)}_G$ models have different dual descriptions we can use that to our advantage. Recently, progress in this direction has been achieved on two fronts: on one hand it was shown that 4d $\cn=2$ S-dualities \cite{Gaiotto:2009we} induce 2d $(0,4)$ Seiberg-like dualities \cite{Putrov:2015jpa}, and on the other hand it was shown that there are 4d $\cn=1$ Lagrangian theories which flow to 4d $\cn=2$ fixed points, with supersymmetry enhancements at the fixed point\cite{Gadde:2015xta,Maruyoshi:2016tqk,Maruyoshi:2016aim}. 
Using these novel 2d dualities, we can reconstruct some protected properties of the IR 2d $(0,4)$ SCFTs of type $\widetilde h^{(k)}_G$  from their geometric engineering discussed above.\footnote{Understanding the geometric counterparts of such flows is an extremely interesting question, but is also outside the scope of the present paper. We plan to return to this issue in the future.}

The global symmetry of a 2d $(0,4)$ theory of type $\widetilde h^{(k)}_G$ is $SU(2)_L\times SU(2)_R \times SU(2)_r \times G$, where $SU(2)_L\times SU(2)_R$ combine to the $SO(4)$ isometry of a transverse $\mathbb{C}^2_{\parallel}$ to the 2d worldsheet,  $SU(2)_r$ is the superconformal R-symmetry for the small $\cn=4$ SCA of the supersymmetric chiral sector, and $G$ is a global symmetry \cite{Witten:1994tz}. From our engineering, we see clearly the contribution of  $SU(2)_L\times SU(2)_R \times G$ (see Figure \ref{wrapped37branes}), however we do not see directly the $SU(2)_r$ symmetry which has a geometrical origin and emerges when we shrink the $\mathbb{P}^1$ to zero size (i.e. at the 6d conformal point).

\subsection{Central Charges $(c_L,c_R)$}

The $\beta$-twisted compactification provides a relation between the central charges of the 2d theory $(c_L,c_R)$ and the 4d conformal anomalies $(a,c)$ \cite{Putrov:2015jpa}. In particular, for the models we consider in this paper, one has \cite{Putrov:2015jpa}:
\begin{equation}
(c_L , c_R) = (4 , 6 ) \times 24 (c - a)
\end{equation}
The superconformal central charges $(a, c)$ have been determined for all $H_G^{(k)}$ theories \cite{Aharony:2007dj,Shapere:2008zf}:
\begin{equation}\label{ACC!}
\begin{aligned}
&a =  {1 \over 4} k^2 \Delta_G + {1\over 2} k (\delta_G - 1) - {1 \over 24}\\
&c =  {1 \over 4} k^2 \Delta_G + {3\over 4} k (\delta_G - 1) - {1 \over 12}
\end{aligned}
\end{equation}
which gives
\begin{equation}
24 (c - a) \big|_{H_G^{(k)}}= k h_G - 1,
\end{equation}
where $ h_G $ is the Coxeter number of the group $G=SU(3),SO(8),E_{6,7,8}$. Including the contribution of a center of mass hypermultiplet, for which $c=1/12$, $a=1/24$ and $24 (c - a) = 1$, one obtains
\begin{equation}\label{cLcR}
(c_L, c_R) = (4 , 6 ) \, k \, h_G = (4,6) \, \dim_{\mathbb{H}} \cm_{G,k},
\end{equation}
where $\cm_{G,k}$ is the moduli space of $k$ instantons for the group $G$, or equivalently the Higgs branch of the theory $\widetilde{H}^{(k)}_G$.

\subsection{Anomaly polynomial}
The anomaly polynomials for the 2d $(0,4)$ theories on the worldsheets of the BPS instanton strings of 6d $(1,0)$ theories have been computed elegantly by an anomaly inflow argument \cite{Kim:2016foj,Shimizu:2016lbw}. For the $\widetilde h^{(k)}_G$ theories one obtains, in particular:
\begin{equation}
\begin{aligned}
A_{2d} = & \, {k^2 n - k \left(n - 2\right) \over 2} \, c_2(F_{SU(2)_L}) - {k^2 n + k \left(n - 2\right)\over 2} \, c_2(F_{SU(2)_R}) + {k n \over 4} \text{tr} \, F^2_G\\ & \, \, + k h^\vee_G \left({1 \over 12} \, p_1(T\Sigma) + c_2(F_{SU(2)_r})\right).
\label{eq:a2d}
\end{aligned}
\end{equation}
Alternatively, the central charges $c_L$ and $c_R$ of the 2d theory we computed in the previous section determine the contribution of the gravitational anomaly as follows:
\begin{equation}
- {c_L - c_R \over 24} \, p_1(T\Sigma) = {k h^\vee_G \over 12} \, p_1(T\Sigma) = {k(n-2) \over 4} \, p_1(T\Sigma)
\end{equation}
and moreover one also determines the coefficient of $c_2(F_{SU(2)_r})$ from a $(0,4)$ Ward identity \cite{Putrov:2015jpa}. The remaining parts of the 2d anomaly polynomial also match against the known `t Hooft anomalies of the 4d $\widetilde{H}^{k}_G$ theories \cite{Aharony:2007dj,Shapere:2008zf}. In particular, the 4d `t Hooft anomaly coefficients for the $SU(2)_L\times G$ global symmetries $k_L$ and $k_G$ are
\begin{equation}\label{eq:ks}
\begin{aligned}
&k_G = 2 k \Delta_G = k n\\
&k_L = k^2 \Delta_G - k (\Delta_G - 1) = {k^2 n \over 2} - k \left({n\over 2} - 1\right).\\
\end{aligned}
\end{equation}
These correspond respectively to the global anomaly terms for the flavor symmetries $SU(2)_L$ and $G$ in Equation \eqref{eq:a2d}. Similarly, the `t Hooft anomaly for the $SU(2)_R$ symmetry of the 4d $\cn=2$ theory is given by $n_v \equiv 8 a - 4 c $. For the models at hand 
\begin{equation}
n_v = 8 a - 4 c = k^2 \Delta_G + k (\Delta_G - 1) = {k^2 n \over 2} + k \left({n\over 2} - 1\right),
\end{equation}
which matches the $SU(2)_R$ term of Equation \eqref{eq:a2d}. This follows because the $SU(2)_L \times SU(2)_R \times G$ contributions to the anomaly polynomial can be determined directly from the 4d anomaly polynomial by integrating it on the $\mathbb{P}^1$, following the same ideology of e.g.\cite{Benini:2009mz,Alday:2009qq}, which gives an alternative derivation for $A_{2d}$.

\subsection{Elliptic genus}

Another interesting BPS property of the 2d $(0,4)$ IR SCFTs of $k$ instantonic strings which can be reconstructed from our engineering argument is their (flavored) elliptic genus. Following \cite{Closset:2013vra}, we realize the $T^2 \times S^2$ background as a quotient of $\C \times \mathbb{P}^1$ with metric
\begin{equation}
ds^2 = d w d{\bar w} + {4 r^2 \over (1 + z {\bar z})^2} d z d{\bar z},
\end{equation}
where $w$ and $z$ are coordinates on $\C$ and on $\mathbb{P}^1$ respectively, while $r$ is the $\mathbb{P}^1$ radius. We identify
\begin{equation}
(w,z) \sim (w + 1, z e^{i \alpha}) \sim (w+\tau,z e^{i \beta}),
\end{equation}
where $\tau$ is complex and $\alpha$ and $\beta$ are real angles with periodicity $2 \pi$. The identification of $w$ gives rise to a torus $T^2$ with complex structure $\tau$, while the identification on $z$ indicates how the $\mathbb{P}^1$ rotates as we go around the two cycles of the torus. As we mentioned in the previous section, we have a unit monopole R-symmetry flux though $\mathbb{P}^1$, which implies the quantization of the $U(1)_r$ charges. The complex structure moduli for this background are related to $(\tau,\alpha,\beta)$ and have been determined by \cite{Closset:2013vra}: these are $\tau$, the complex structure of $T^2$, and $\sigma \equiv \alpha \tau - \beta$ (with fixed $\tau$). The partition function on such background depends locally holomorphically on $\tau,\sigma$ \cite{Closset:2013vra}. On top of this, the partition function can depend on fugacities and fluxes for the other global symmetries of the theory: indeed, one can easily add Abelian background gauge fields for the Cartan of the global symmetry group of the model. The gauge field must be flat on $T^2$ \cite{Closset:2013vra}. The corresponding holomorphic line bundles are labeled by their first Chern class $c_1 \in \mathbb{Z}$ ($\equiv$ flux through $\mathbb{P}^1$) and a single holomorphic modulus, whose real and imaginary parts correspond to Wilson lines wrapping the cycles of the torus. Only the $ U(1)_r $ R-symmetry has a flux through the $\mathbb{P}^1$; on the other hand, we can turn on fugacities for the other global symmetries along the cycles of the $T^2$. 

The $T^2 \times S^2$ backgrounds discussed above are 1/2 BPS and defined for any 4d supersymmetric theory with at least four supercharges and a $U(1)$ R-symmetry. In general such partition functions on $T^2 \times S^2$ localize over (infinite) sums over distinct elliptic genera \cite{Benini:2015noa}, labeled by gauge flux sectors on the two sphere.\footnote{ More precisely by triples given by flat connections on $T^2$ commuting with a given gauge flux through $S^2$ \cite{Benini:2015noa}.} Under favorable circumstances, however, such infinite sums can truncate to finite sums \cite{Gadde:2015wta}. In particular, for backgrounds without global symmetry fluxes (other than the $U(1)$ R-symmetry monopole) and with a choice of $U(1)$ R-symmetry such that all the elementary fields have non-negative R-charges, this sum turns out to consist of a single term \cite{Gadde:2015wta}, which one can identify with a RR elliptic genus for a 2d $(0,2)$ theory, of the kind defined in \cite{Benini:2013nda,Benini:2013xpa}. This is precisely the case for the Kapustin $\beta$-twist of the $\widetilde{H}^{(k)}_G$ theories discussed above, where the amount of supersymmetry is doubled and the $(T^2 \times S^2)_\beta$ partition function of the $\widetilde H^{(k)}_G$ theory localizes to an elliptic genus for the $\widetilde h^{(k)}_G$ model. Schematically
\begin{equation}
\mathbb{E}_{\widetilde h_{G}^{(k)}} = Z_{(T^2 \times S^2)_\beta}\left(\widetilde{H}_G^{(k)}\right),
\end{equation}
where $\mathbb{E}_{\widetilde h_{G}^{(k)}}$ is the (flavored) RR elliptic genus \cite{Benini:2013nda,Benini:2013xpa} of the $\widetilde h^{(k)}_G$ theory. The leading order term in the $ q- $expansion of $\mathbb{E}_{\widetilde h_{G}^{(k)}}$ is proportional to $ q^{-c_L/24} $, and therefore, in cases where elliptic genera are effectively computable, one can read off the left central charge $c_L$ of the CFT directly from them and verify \eqref{cLcR}.

Furthermore, the $ \beta $ twist behaves particularly nicely with respect to the 2d (0,4) dualities \cite{Putrov:2015jpa}, and this gives rise to a strategy for computing the elliptic genera of the $\widetilde h^{(k)}_G$ $(0,4)$ models, even when they do not have a Lagrangian formulation. In order to fix the precise map among the elliptic genus fugacities and the $\beta$-twisted $T^2 \times S^2$ partition function of the $\widetilde{H}^{(k)}_G$ theories, it is helpful to consider the Lagrangian case corresponding to $G=SO(8)$. In particular, this case gives an interesting consistency check for our geometric engineering argument as the corresponding 2d BPS worldsheet theories have already been determined from a different perspective in \cite{Haghighat:2014vxa} .

\subsubsection{Strings of the $SO(8)$ 6d $(1,0)$ minimal SCFT revisited}

\begin{figure}
\begin{center}
\includegraphics[width=0.5\textwidth]{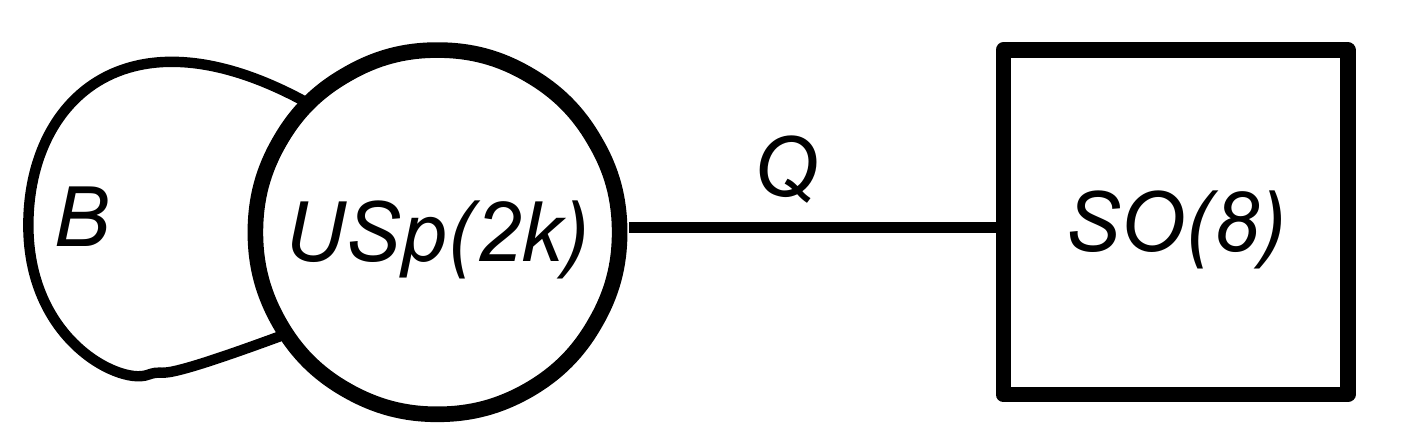}
\caption{2d quiver corresponding to the $H_{SO(8)}^{(k)}$ theory.}\label{fig:D4}
\end{center}
\end{figure}

It is well known that the $H^{(k)}_{SO(8)}$ theories, which correspond to D3 branes probing the seven-brane associated to a $I^*_0$ singularity, are Lagrangian SCFTs \cite{Banks:1996nj,Douglas:1996js}. In particular, $H^{(1)}_{SO(8)}$ is just $SU(2)$ SYM with four hypermultiplets in the fundamental representation, while the $H^{(k)}_{SO(8)}$ theories for $k>1$ are given by an $USp(2k)$ gauge theory with four hypermultiplets in the fundamental representation and one hypermultiplet in the antisymmetric. In the $\beta$-twisted reduction on $\PP^1$ of any Lagrangian theory, each vector (resp. hyper) multiplet in 4d leads to a (0,4) vector (hyper) multiplet in 2d \cite{Kapustin:2006hi}. From this it follows at once that the 2d (0,4) quiver gauge theory describing the strings consists of an $ USp(2k) $ vector multiplet $ \Upsilon $, a hypermultiplet $ B $ transforming in the anti-symmetric representation of $ USp(2k) $, and a $ USp(2k)\times SO(8) $ bifundamental hypermultiplet $ Q $, as summarized by the quiver in Figure \ref{fig:D4}, which indeed coincides with the one obtained in \cite{Haghighat:2014vxa} from a different brane engineering in Type IIB string theory. This serves as a first consistency check for our claim. The elliptic genus of the theory can be computed from the results of \cite{Closset:2013sxa,Nishioka:2014zpa,Putrov:2015jpa,Benini:2015noa}. For $k=1$ we obtain the Jeffrey-Kirwan residue of the following 1-form one-loop determinant that matches exactly with Equation (3.21) of \cite{Haghighat:2014vxa}:
\begin{equation}
d \zeta (2 \pi i ) \left({\eta(\tau)^2 \over \theta_1(v \,x^{\pm};\tau)}\right) \times \left(\prod_{i=1}^4{  \eta(\tau)^4\over \theta_1(v \, \mu_i^{\pm} \, z^{\pm};\tau)})\right) \times \left({ \theta_1(z^{\pm 2};\tau) \theta_1(v^2 \, z^{\pm 2};\tau)\theta_1(v^2;\tau) \over \eta(\tau)^3}\right),
\end{equation}
where the first term in parentheses comes from the decoupled center of mass hyper of the 4d $\cn=2$ system, the second corresponds to the four hypermultiplets in the fundamental which, having $U(1)_r$ charge zero, contribute as $(0,4)$ hypers, while the third term corresponds to the contribution of the $SU(2)$ vector multiplet. The parameters $ x = e^{2\pi \epsilon_{-}} $ and $ v = e^{2\pi i\epsilon_+} $ are exponentiated fugacities for $ SU(2)_L $ and $ SU(2)_R $ respectively; $ \zeta $ is the exponential of the holonomy of the gauge field; finally, $ \mu_{1},\dots \mu_{4} $ are exponentiated fugacities for the $ SO(8) $ flavor symmetry.

\subsubsection{The $E_6$ case: $k=1$}\label{sec:e6rev}
According to our geometric engineering argument, the elliptic genus for one BPS istantonic string of the minimal $(1,0)$ SCFT with  $E_6$ gauge symmetry coincides with the $\beta$-twisted partition function of the $\widetilde{H}^{(1)}_{E_6}$ 4d $\cn=2$ SCFT, which is the well-known rank one $E_6$ Minahan-Nemeschansky theory $ H^{(1)}_{E_6} $, plus a decoupled free hypermultiplet. Luckily, the $\beta$-twisted partition function for the $E_6$ MN theory has been computed recently, with two different insightful methods \cite{Gadde:2015xta,Putrov:2015jpa}. In one approach, the $E_6$ MN theory is realized as a fixed point with enhanced supersymmetry of a Lagrangian 4d $\cn=1$ theory. Upon compactification on $S^2$ the 4d $\cn=1$ model gives rise to a 2d $(0,2)$ theory which flows in the IR to a fixed point with enhanced $(0,4)$ supersymmetry. The elliptic genus in this case has been computed in  \cite{Gadde:2015xta} by localization from the $(0,2)$ matter content. The second approach involves the 2d $(0,4)$ avatar of Gaiotto $\cn=2$ dualities developed in \cite{Putrov:2015jpa}: the elliptic genera of the theories compactified on $ T^2\times S^2 $ are captured by correlators of a TQFT on the Gaiotto curve of the 4d parent theory $ \widetilde H^{(k)}_G $.\footnote{ This provides in principle a way to compute elliptic genera of all the $ \widetilde{H}^{(k)}_G $ theories. However, the tools required to compute generic TQFT correlators are not yet available. We leave this to future work.} In particular, the elliptic genus of the $ H^{(1)}_{E_6} $ theory has been computed in \cite{Putrov:2015jpa} by exploiting the duality of this theory with the $ SU(3),\, N_f=6 $ theory \cite{Argyres:2007cn}.

\begin{figure}[t!]
\begin{center}\includegraphics[width=.6\textwidth]{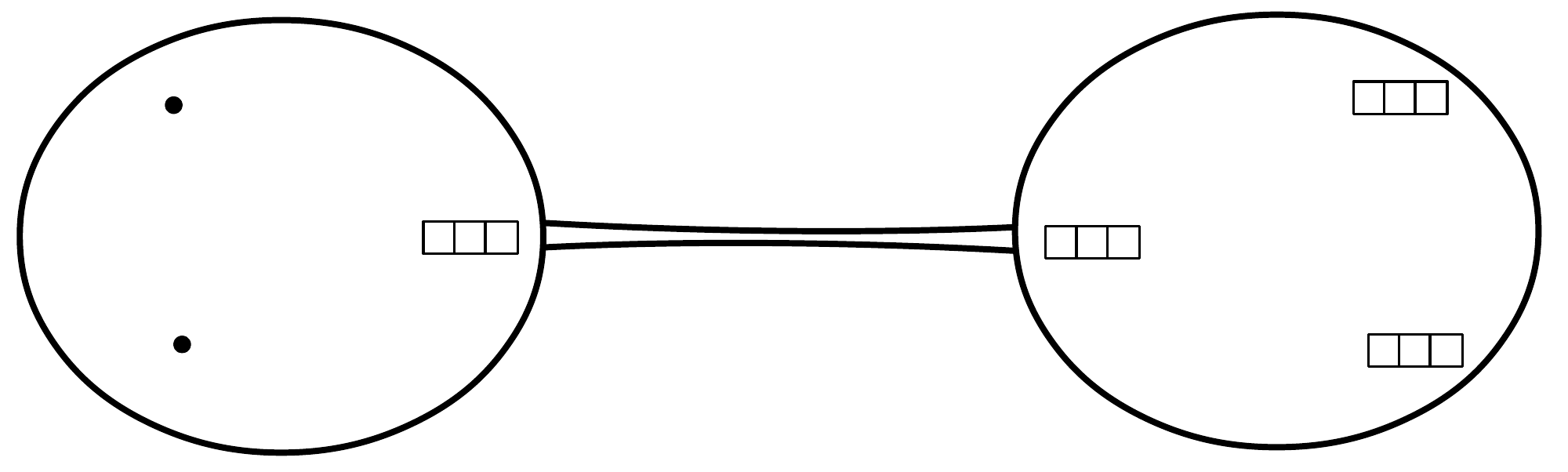}\end{center}
\caption{Gaiotto $ T_3 $ theory from degeneration of the $ SU(3), N_f = 6 $ curve.}
\label{fig:E6}
\end{figure}

Let us briefly review the computation of the $ H_6^{1} $ elliptic genus performed in \cite{Putrov:2015jpa}. The elliptic genus of the $ SU(3),N_f = 6 $ theory can be obtained starting with the $ H_{E_6}^{(1)} $ elliptic genus. This theory has a manifest $ SU(3)^3\subset E_6 $ global symmetry group; one can weakly gauge an $ SU(2) $ subgroup of a $SU(3)$ factor and couple  a hypermultiplet to this gauge group, as in Figure \ref{fig:E6}. This implies the following relation at the level of elliptic genera:
\begin{equation} \mathbb{E}_{SU(3),N_f = 6}(\mathbf{a},\mathbf{b},x,y) = \frac{1}{2}\int \frac{d\zeta}{2\pi i \zeta}\frac{\eta^2 \theta(\zeta^{\pm 2})\theta(v^2)\theta(v^2\zeta^{\pm 2})}{\theta(v s^\pm \zeta^\pm)}\mathbb{E}_{h_{E_6}^{(1)}}(\mathbf{a},\mathbf{b},\mathbf{c}).\end{equation}
Here, the integration is performed by picking up the Jeffrey-Kirwan residues of the integrand, and $ \mathbf{a},\mathbf{b},\mathbf{c} $ are $ SU(3)_{\mathbf{a}}\times SU(3)_{\mathbf{b}} \times SU(3)_{\mathbf{c}} $ fugacities. Moreover $ \zeta  $ is the fugacity for the gauged SU(2) subgroup of $ SU(3)_{\mathbf{c}} $; the hypermultiplet is also charged under an additional $SU(2)$ whose fugacity is denoted by $ s $. The $ x,y $ fugacities associated to the $ U(1)\times U(1) $ global symmetry of the $ SU(3),N_f=6 $ theory are determined in terms of $ \mathbf{c}$ and $ s $ as follows:
\begin{equation} (c_1,c_2,c_3)= (r\zeta,r\zeta^{-1},r^{-2});\qquad x = s^{1/3}/r;\qquad y = s^{-1/3}/r.\end{equation}
Following \cite{Putrov:2015jpa}, this formula can be inverted to give the $ h_{E_6}^{(1)} $  elliptic genus in terms of the known elliptic genus for $ SU(3), N_f = 6 $, according to the following formula:
\begin{equation} \mathbb{E}_{h_{E_6}^{(1)}}(\mathbf{a},\mathbf{b},\mathbf{c}) = \frac{1}{2\theta(v^2\zeta^{\pm 2})}\int \frac{ds}{2\pi i s}\frac{\theta(s^{\pm2})\theta(v^{-2})}{\theta(v s^\pm \zeta^\pm)}\mathbb{E}_{SU(3),N_f = 6}(\mathbf{a},\mathbf{b},x,y).\end{equation}
This results in a sum of a somewhat large number of terms, but it can be shown that it can be expressed in terms of $ E_6 $ characters as the following expansion:
\begin{align}\mathbb{E}_{h_{E_6}^{(k)}} &=  v^{11}\,q^{-11/6} \bigg((1+\chi^{E_6}_{78} v^2+\chi^{E_6}_{2430}v^4+\dots)\nonumber\\
&+ q\,((1+\chi^{E_6}_{78})+(1+2\chi^{E_6}_{78}+\chi^{E_6}_{2430}+\chi^{E_6}_{2925})v^2+\dots) +\dots\bigg).\end{align}

\section{Topological strings and elliptic genera}\label{sec:topstrings}

\begin{figure}
\begin{center}
\begin{tabular}{c|cccccccccc}
IIB & \multicolumn{6}{c}{$\substack{\displaystyle{\,\,(\mathbb{T}^2 \times \mathbb{R}^4)_{\varepsilon_1,\varepsilon_2}} \\ \vspace{-0.05in}  \\\overbrace{\hspace{1.2in}} \\  \vspace{-0.01in}}$}& \multicolumn{4}{c}{$\substack{\displaystyle{\,\,\widetilde{B}} \\ \vspace{-0.05in} \\\overbrace{\hspace{.9in}} \\  \vspace{-0.01in}}$} \\
background & \multicolumn{4}{c}{$\substack{\displaystyle{\,\,\mathbb{R}^4_{\varepsilon_1,\varepsilon_2}}\\\vspace{-.04in}\\\overbrace{\hspace{.9in}}}$} & \multicolumn{2}{c}{$\substack{\displaystyle{\,\,T^2}\\\vspace{-.04in}\\\overbrace{\hspace{.4in}}}$} & \multicolumn{2}{c}{$\substack{\displaystyle{\,\,\mathbb{P}^1}\\\vspace{-.04in}\\\overbrace{\hspace{.4in}}}$} &  \multicolumn{2}{c}{$\substack{\displaystyle{\,\,\mathcal{O}(-n)}\\\vspace{-.04in}\\\overbrace{\hspace{.4in}}}$} \\
 & 0 &1 & 2 & 3 & 4 & 5 & 6 & 7 & 8 & 9\\
\hline
seven-brane & X & X&  X & X & X & X & X & X & - & - \\
D3 & - & - & - & - & X & X & X & X & - & - \\
\end{tabular}
\end{center}
\caption{Schematic IIB description of the 6d (1,0) $\Omega$ background.}\label{wrapped37branes}
\end{figure}

\subsection{6d BPS strings and topological strings}

Combining the geometric engineering picture in F-theory with the duality between F-theory and M-theory \cite{Vafa:1996xn} and the Gopakumar-Vafa formula \cite{Gopakumar:1998ii,Gopakumar:1998jq,Dedushenko:2014nya} gives a canonical relation among the spectrum BPS states of 6d $(1,0)$ theories and the closed topological string partition function \cite{Haghighat:2013gba,Haghighat:2013tka,Haghighat:2014vxa }. In the present context, this relation reads
\begin{equation}\label{eq:topstring}
Z_\text{top}\left(\widetilde{X}_n\right)\Bigg|_{n=3,4,6,8,12} = Z_{G,0} \, \left(1 +\sum_{k\geq1} \mathbb{E}_{\widetilde h_G^{(k)}} \, Q^{k}\right)\Bigg|_{G=SU(3),SO(8),E_{6,7,8}}.
\end{equation}
where $\widetilde{X}_n$ is a resolution of $X_n$, the orbifold singularity in Equations \eqref{orbo1}--\eqref{orbo2}, $Q$ is a fugacity proportional to $e^{-t}$ where $t$ is the K\"ahler class of the base $\mathbb{P}^1$, and $Z_{G,0}$ is a factor that encodes the spectrum of BPS particles arising from KK reduction of the 6d hyper-, tensor, and vector multiplets, and crucially is independent of $t$.\footnote{ Some details about the geometry of $\widetilde{X}_n$ can be found in \cite{Haghighat:2014vxa} and references therein.} The topological string free energy admits a genus expansion
\begin{equation}
\text{log } Z_{top}(X) = - {1\over \varepsilon_1 \varepsilon_2} \sum_{g,n\geq 0}
(-\varepsilon_1 \varepsilon_2)^g (\varepsilon_1+\varepsilon_2)^m F_{g,m}(X),
\end{equation}
and in \cite{Haghighat:2014vxa} B-model techniques were used to compute $F_{0,0}(\widetilde{X}_n)$ for $3 \leq n \leq 12$. In particular, for rank one models $F_{0,0}$ has the following expansion
\begin{equation}
F_{0,0}(\widetilde{X}_n) = \sum_{k\geq 0} e^{- k \, t} F^{(k)}_{0,0}(\tau,m_i),
\end{equation}
where $t$ is the K\"ahler class of the base $\mathbb{P}^1$ in $\widetilde{X}_n$, $\tau$ is the K\"ahler class of the elliptic fiber of $\widetilde{X}_n$, and $m_i$ correspond to the K\"ahler classes resolving the singular elliptic fibers of $\widetilde{X}_n$. This gives nontrivial relations among the genus zero invariants $ F^{(k)}_{0,0} $ and elliptic genera of BPS strings \cite{Haghighat:2014vxa}. An especially simple one is the following:
\begin{equation}\label{eq:FvsE}
F^{(1)}_{0,0}(\widetilde{X}_n)\Bigg|_{n=3,4,6,8,12} = \lim_{\varepsilon_1,\varepsilon_2 \to 0} \varepsilon_1 \varepsilon_2 \, \mathbb{E}_{\widetilde h_{G}^{(1)}} \Bigg|_{G=SU(3),SO(8),E_{6,7,8}}.
\end{equation}
This has been checked for $G=SO(8)$ in \cite{Haghighat:2014vxa}; we have checked that analogous results hold for $ G = E_6 $ at one string. See also \cite{Kim:2016foj} for $ G = SU(3) $.

\subsection{Elliptic genera and Hilbert series}
\label{sec:ellhilb}
The elliptic genera of the 2d (0,4) SCFTs that were obtained above display some interesting properties which have a natural explanation in light of geometry. For instance, it was first observed in \cite{Gadde:2015wta} for one $ E_6 $ string and in \cite{Putrov:2015jpa} for $ SO(8) $ strings that the leading order term in the elliptic genus of the $ h^{(1)}_{E_6} $ and $ h^{(1)}_{SO(8)} $ theories coincides with the Hall-Littlewood index of the 4d $H^{(1)}_{E_6} $ or $ H^{(1)}_{SO(8)} $ theories respectively, or alternatively with the Hilbert series of the reduced moduli space of one $ E_6$ or $ SO(8) $ instanton. The connection with topological string theory can be used to derive and generalize such relation between the elliptic genera and the Hall-Littlewood index from geometry.

From the perspective of the 6d $(1,0)$ theories, the computations outlined in Section \ref{sec:topstrings} are suggestive of a localization computation in 6d on an $\Omega$--background of the form $(\mathbb{T}^2 \times \mathbb{R}^4)_{\varepsilon_1,\varepsilon_2}$, where $\varepsilon_1$ and $\varepsilon_2$ are identified with the Cartan generators of the $SO(4)$ isometries of $\mathbb{R}^4$. Similar to what happens for 4d $\cn=2$  and 5d $\cn=1$ theories, the partition function of the 6d $(1,0)$ theory on the $\Omega$--background localizes on generalized elliptic  equivariant characters of the instanton moduli spaces, which are computed precisely by the elliptic genera of the BPS instanton strings \cite{Nekrasov:2003rj,Iqbal:2003ds}. This $\Omega$--background lifts to an F-theory background of the form 
\begin{equation}
F / (X\times S_{6d}^1 \times S_{5d}^1 \times \mathbb{R}^4)_{\varepsilon_1,\varepsilon_2} \quad\longleftrightarrow\quad M / (X\times S_{5d}^1 \times \mathbb{R}^4)_{\varepsilon_1,\varepsilon_2} \label{FFF}\end{equation} 
where the F/M theory duality exchanges the radius $R_{6d}$ of $S_{6d}^1$ on the F-theory side with the volume $\text{Im }\tau \sim 1/R_{6d}$ of the elliptic fiber of $X$.\newline

In the limit $\text{Im }\tau \to \infty$, all the KK modes in the reduction from 6d to 5d decouple and one is left with a genuine 5d $\cn=1$ theory. In the case of the $\widetilde{X}_n$ models with $n=3,4,6,8,12$, the geometry of the 2-cycles is given by an affine Dynkin graph of type $\widehat{A}_2,\widehat{D}_4,\widehat{E}_{6,7,8}$ respectively, and one can take the limit $\text{Im }\tau \to \infty$ in such a way that only one 2-cycle with Coxeter-Dynkin label 1 in the affine diagram is sent to infinite size, while the others are kept of finite size. Proceeding this way, one obtains an M-theory geometry corresponding to pure 5d $\cn=1$ gauge theory, with gauge groups respectively $SU(3),SO(8),E_{6,7,8}$ (as well as a $ U(1) $ vector multiplet coming from compactification of the tensor multiplet, which decouples since $ g_{U(1)}\simeq R_{6d}^{1/2} \to 0 $).\footnote{ Notice that there is a leftover contribution from the Green-Schwarz term in six dimensions, which gives rise to a 5d Cern-Simons coupling of the form $A_{U(1)} \wedge \text{Tr} (F_G \wedge F_G)$. This term, however, upon reduction on $S^1_{5d}$ can be absorbed by a shift in the $\theta$ angle for the gauge group $G$ in the resulting $4d$ effective theory.} In particular, in this limit the topological string partition function reduces to the 5d $\cn=1$ Nekrasov partition function for a pure SYM theory with gauge group $G$, which we denote by $ Z_{\text{inst}}^{5d}(G) $, times a perturbative contribution coming from the $ G $ vector multiplet and the decoupled free abelian vector multiplet which is independent of the base K\"ahler parameter $ t $:
\begin{equation}\label{eq:Nekrasov}
Z_{top}(\widetilde{X}_n) \Big|_{n=3,4,6,8,12} \xrightarrow{\text{Im }\tau \to \infty} Z^{5d}_{\text{pert}}(G\times U(1))\cdot Z^{5d}_{\text{inst}}(G)\Big|_{G=SU(3),SO(8),E_{6,7,8}}.
\end{equation}
The instantonic piece $Z^{5d}_{\text{inst}}$ can be written as \cite{Keller:2011ek,Nakajima:2003pg,Nekrasov:2004vw}:
\begin{equation}\label{eq:hilb}
Z^{5d}_{\text{inst}}(G) = 1+\sum_{k\geq1} \widetilde{Q}^k \,\mathcal{H}(\mathcal{M}_{G,k}),
\end{equation}
where $\mathcal{H}(\mathcal{M}_{G,k})$ is the Hilbert series of the moduli space of $k$ $G$-instantons.
Combining Equations \eqref{eq:topstring},\eqref{eq:Nekrasov} and \eqref{eq:hilb}, we obtain
\begin{equation}
\lim_{\text{Im }\tau \to \infty}  \mathbb{E}_{\widetilde h_G^{(k)}} Q^k =  \mathcal{H}(\mathcal{M}_{G,k}) \, \widetilde{Q}^k.\label{eq:limite}
\end{equation}

This explains and generalizes the results of \cite{Putrov:2015jpa} for the $ SU(2),\, N_f = 4 $ theory (corresponding to one SO(8) instanton) and \cite{Putrov:2015jpa,Gadde:2015xta} for the $ h^{(1)}_{E_6} $ theory (corresponding to one $ E_6 $ instanton). With the results already available in the literature, we can check that this relation extends to other cases as well. For instance, from the expression for the elliptic genus of two $ SO(8) $ instantons in Section 3.1 of \cite{Haghighat:2014vxa}, taking the $ q\to 0 $ limit and setting $ x = 1 $ for simplicity, one finds:
\begin{align}& \lim_{q\to 0} q^2\, \mathbb{E}_{\widetilde h^{(2)}_{SO(8)}}(\epsilon_+,\tau) = v^{25}\bigg[\frac{1}{(1-v^2)^{24}(1+v^2)^{12}(1\!+v^2+\!v^4)^{11}}\bigg(1+v^2+20v^4+65 v^6 + 254 v^8 \nonumber\\
&+ 841 v^{10} + 2435 v^{12} + 6116 v^{14} + 14290 v^{16} + 
 29700 v^{18} + 55947 v^{20} \!+\! 96519 v^{22} \!+\! 152749 v^{24}\nonumber\\& + 220408 v^{26} + 
 293226 v^{28} + 359742 v^{30} + 406014 v^{32} + 421960 v^{34} + 
 406014 v^{36} + 359742 v^{38} \nonumber\\ &+ 293226 v^{40} + 220408 v^{42} + 
 152749 v^{44} + 96519 v^{46} + 55947 v^{48} + 29700 v^{50} + 14290 v^{52} \nonumber\\ &+ 
 6116 v^{54} + 2435 v^{56} + 841 v^{58} + 254 v^{60} + 65 v^{62} + 
 20 v^{64} + v^{66} + v^{68}\bigg)\bigg]\cdot \frac{1}{(1-v)^2},\end{align}
 where the term in square brackets agrees with the Hilbert series of the reduced moduli space of two $ SO(8) $ instantons (see Equation (5.20) of \cite{Hanany:2012dm}, where their parameter $ t $ is to be identified with $ v^2 $), and $ \frac{1}{(1-v)^2} $ is the contribution of the center of mass hypermultiplet in the limit $ x\to 1 $.

Likewise, we find that in the same limit the elliptic genera for one and two SU(3) instantons, computed using the results of \cite{Kim:2016foj}, are given respectively by:
\begin{align}& \lim_{q\to 0} q^{1/2}\, \mathbb{E}_{\widetilde h_{SU(3)}^{(1)}}(\epsilon_+,\tau) = v^{3}\bigg[\frac{1+4v^2+v^4}{(1-v^2)^4}\bigg]\cdot \frac{1}{(1-v)^2}\end{align}
and
\begin{align} \lim_{q\to 0} q\, \mathbb{E}_{\widetilde h_{SU(3)}^{(2)}}(\epsilon_+,\tau) = v^{13}&\bigg[\frac{1}{(1-v^2)^{12}(1+v^2)^6(1+v^2+v^4)^5}\bigg(1 + v^2 + 6 v^4 + 17 v^6 + 31 v^8 \nonumber\\&+ 52 v^{10} + 92 v^{12} + 110 v^{14} + 
 112 v^{16} + 110 v^{18} + 92 v^{20} + 52 v^{22} \nonumber\\&+ 31 v^{24} + 17 v^{26} + 
 6 v^{28} + v^{30} + v^{32}\bigg)\bigg]\cdot \frac{1}{(1-v)^2};\end{align}
 the expressions in square brackets agree respectively with the Hilbert series of the reduced moduli space of one and two $ SU(3) $ instantons (which can be read off from Equations (3.12) of \cite{Benvenuti:2010pq} and (3.21) of \cite{Hanany:2012dm}).

 An alternative field theoretical derivation of this relation would go as follows. The elliptic genus corresponds to the partition function of the 4d $\cn=2$ theory on $T^2 \times S^2$. Let us write $T^2 = S^1_{R_1} \times S^1_{R_2}$. Taking $\text{Im } \tau$ to infinity is equivalent to sending $R_1 \to 0$, thus reducing to a partition function on $S^1 \times S^2$ for the corresponding 3d $\cn=4$ theory. For an $S^1$ reduction, the Higgs branch does not receive corrections \cite{Seiberg:1996nz}. From the results above, it is tempting to conjecture that the corresponding partition function for the 3d $\cn=4$ theory computes the Higgs limit of the superconformal index, which is the Hilbert series of the Higgs branch \cite{Razamat:2014pta}.

\section{Modular bootstrap of the elliptic genera}\label{sec:modular}

\subsection{From anomaly four-form to modular transformation}

In this section we explain the relation between the anomaly four-form polynomial of the 2d SCFTs and the modular transformation of their flavored elliptic genus, a result which will be useful in Section \ref{sec:hilb} for determining the elliptic genera of instanton strings.

In our computation of the elliptic genus we keep track of the dependence on the fugacities of the global symmetry group $ F $, which for the theories at hand we can write schematically as the product of various non-Abelian factors:
\begin{equation} F = \prod_{a} F_a.\end{equation}
We denote by $ \{\vec{z}_a\} $ the fugacities associated to the Cartan of the non-Abelian factors.

Under a modular transformation $ \tau\to -1/\tau $, the elliptic genus transforms as a weight-zero Jacobi form of several elliptic variables:
\begin{equation} \mathbb{E}(\vec{z}_a/\tau,-1/\tau) = e^{\frac{2\pi i}{\tau}f(\vec{z}_a)}\mathbb{E}(\vec{z}_a,\tau).\end{equation}
We refer to the phase $f(\vec{z}_a)$ as the modular anomaly of the elliptic genus. This is a quadratic form of the various fugacities:
\begin{equation} f(\vec{z}_a) = \frac{1}{2}\sum_a k_a \, (\vec{z}_a|\vec{z}_a)_a,\end{equation}
where the $ k_a $ have the physical interpretations as coefficients in the OPE of the currents associated to the various global symmetries, as in \cite{Benini:2012cz,Benini:2013cda}, while $(x|y)_a\equiv \tfrac{1}{2 h^{\vee}_G} \sum_{\alpha \in R} \langle \alpha^\vee, x\rangle\langle \alpha^\vee, y\rangle$ is the Weyl-invariant symmetric bilinear form on the root lattice of the group $F_a$ normalized such that the short roots have length 2 \cite{MR1890629}.

The modular anomaly can be read off directly from the anomaly four-form $ \mathcal{A}_4 $, which includes terms of the form \cite{Bobev:2015kza}:
\begin{equation}\sum_{a}k_a ch_2(\mathcal{F}_a).\end{equation}

We find that for the $ \widetilde{h}^{(k)}_{G} $ theory the modular anomaly for the elliptic genus can be determined from Equation \eqref{eq:a2d}, by the following replacements:
\begin{equation}
c_2(F_{SU(2)_R}) \to - \varepsilon_+^2 \qquad c_2(F_{SU(2)_r}) \to - \varepsilon_+^2 \qquad c_2(F_{SU(2)_L})  \to - \varepsilon_-^2
\end{equation}
\begin{equation}
{1\over 2} \text{tr} \, F_G^2 \to - {1 \over 2 h^\vee_G} \sum_{\alpha\in\Delta_G} (m_\alpha)^2 \qquad p_1 \to 0,
\end{equation}
where $\vec{m}\equiv (m_1,...,m_r)$ are the fugacities associated to the global symmetry group $G$, and, for a root $\alpha = n_1 \alpha_1 + \dots + n_r \alpha_r \in \Delta$, $ m_\alpha \equiv \sum_{i} n_i m_i $. To make contact with the literature about Jacobi forms, it is useful to switch to the root lattice, which amounts to the change of variables $m_i = (C_G)_{ij} y_j$, where $C_G$ is the Cartan matrix of $G$. Then,
\begin{equation} 
{1\over 2 h^\vee_G} \sum_{\alpha\in\Delta_G} (m_\alpha)^2 \equiv  (\, \vec{y} \, | \,\vec{y} \,)_G,
\end{equation}
where $h^\vee_G$ is the dual Coxeter number. It follows that the modular anomaly can be expressed as
\begin{equation} f_{\widetilde h^{(k)}_G}(m_\alpha,\epsilon_+,\epsilon_-) = - k \left(\frac{h_G^\vee}{6}+1\right) (\,\vec{y}\,|\,\vec{y} \,)_G - k\frac{h_G^\vee}{6}(5\epsilon_+^2-\epsilon_-^2)+k^2\left(\frac{h^\vee_G}{6}+1\right)(\epsilon_+^2-\epsilon_-^2).\label{eq:ellindex}\end{equation} 

\subsection{Constraining one-string elliptic genera with modularity}

\label{sec:hilb}
In this section we determine the elliptic genera of all the theories $ \widetilde h_G^{(1)} $ corresponding to one instanton string for $ G = SU(3),SO(8),F_4,E_6,E_7,E_8 $. In order to achieve this we rely heavily on the modular properties of the elliptic genera, as well as their relation to the Hilbert series of one-instanton moduli spaces. We begin this section with a general discussion of our approach, which applies for any number $ k $ of strings, and then employ these techniques to determine the elliptic genera of all the rank 1 theories. This approach closely parallels the one undertaken in \cite{Huang:2015sta,Huang:2015ada,KlockB} in the context of topological string theory on compact elliptic Calabi-Yau threefolds.

\medskip

The Hilbert series of the moduli space of $ k $ $G$-instantons is a ratio of two factors,
\begin{equation} \mathcal{H}(\mathcal{M}_{G,k}) = \frac{N_{G,k}(v,x,m_\alpha)}{D_{G,k}(v,x,m_\alpha)},\end{equation}
where the denominator is a product of factors associated to the generators of the moduli space of $ k $ $ G $-instantons \cite{Cremonesi:2014xha}; the set of such generators is provided explicitly in Section 8.5 of \cite{Cremonesi:2014xha}, and from that one obtains the following expression:
\begin{equation} D_{G,k}(v,x,m_\alpha) = \prod_{i=1}^k\bigg(\prod_{\substack{j=-i\\j-i\text{ even}}}^i(1-v^ix^j)\bigg)\bigg(\prod_{\substack{j=-i+1\\j-i\text{ odd}}}^{i-1}\prod_{\alpha \in \widetilde\Delta_G}(1-v^{i+1}x^{j}e^{2\pi i m_\alpha})\bigg),\label{eq:hdenom}\end{equation}
where $ \widetilde \Delta_G $ includes the positive and negative roots of $ G $, as well as its Cartan vectors, and we denote by $ x = e^{2\pi i \epsilon_-},v = e^{2\pi i \epsilon_+} $ the exponentials of the $ SU(2)_L\times SU(2)_R $ fugacities.

However, the way it is written Equation \eqref{eq:hdenom} contains too many factors. To see this, recall that the topological string partition function in the limit $q\to 0$ takes the form
\begin{equation} Z_{top}(\widetilde X_G)=Z_{0}\left(1 +\sum_{k\geq 1}\widetilde{Q}^k\mathcal{H}(\mathcal{M}_{G,k})\right).\end{equation}
Note that a term of the form 
\begin{equation}(1- v^i x^j)= 1- e^{ i \pi ((i+j)\epsilon_1 + (i-j)\epsilon_2)}\end{equation}
in the denominator of the Hilbert series, Equation \eqref{eq:hdenom}, would lead to a singularity in the topological string free energy $ \mathcal{F}_{top} = \log(Z_{top}) $ at
\begin{equation} (i+j)\epsilon_1 + (i-j)\epsilon_2 = 0.\end{equation}
However, from the genus expansion of the topological string free energy,
\begin{equation}
\mathcal{F}_{top}(X) = \sum_{g,n\geq 0}
(-\epsilon_1 \epsilon_2)^{g-1} (\epsilon_1+\epsilon_2)^n F_{g,m}(X),
\end{equation}
one sees that only poles at $ \epsilon_1 = 0 $ or $ \epsilon_2 = 0 $ are allowed to occur. This implies that all the terms in the denominator of \eqref{eq:hdenom} for which $ i\neq \pm j $ must cancel against analogous factors in the numerator.\footnote{Indeed, such cancelations occur for all the examples we have checked. It would be interesting to find a satisfactory gauge-theoretic explanation for this fact.}${}^{,}$\footnote{An analogous argument has been used by M.-X. Huang, S. Katz, and A. Klemm to formulate an Ansatz for the topological string theory partition function for compact elliptic Calabi-Yau threefolds; we are grateful to them for sharing a copy of the draft of their upcoming paper \cite{KlemmToApp}, and refer to the slides of A. Klemm's talk \href{http://theory.caltech.edu/~ss299/FtheoryAt20/Talks/Klemm.pdf}{`\emph{BPS states on elliptic Calabi-Yau, Jacobi-forms and 6d theories}'} at the ``F-theory at 20'' conference, Caltech, February 2016 for a sketch of their argument.}

 This leads to a somewhat leaner expression for the denominator:
\begin{equation} D'_{G,k}(v,x,m_\alpha) = \prod_{i=1}^k\prod_{s=\pm 1}\bigg((1-(v x^s)^{i})\, \prod_{\substack{j=-i+1\\j-i\text{ odd}}}^{i-1}\prod_{\alpha \in \Delta_+}(1-v^{i+1}x^{j}e^{2\pi i s m_\alpha})\bigg),\label{eq:hleaner}\end{equation}
where the label $ \alpha $ now runs only over the \emph{positive} roots of $ G $.

As discussed in Section \ref{sec:ellhilb}, the elliptic genus is expected to reduce to the Hilbert series in the 5d limit $ q\to 0 $, as in Equation \eqref{eq:limite}. Using ideas similar to the ones developed in \cite{Huang:2015ada,Haghighat:2015ega,Haghighat:2014pva}, we now formulate an Ansatz for the elliptic genus which matches the form of the Hilbert series in the 5d limit. We begin by noting that it is natural to interpret each factor of the form
\begin{equation} (1-e^{2\pi i z})\end{equation}
 in \eqref{eq:hleaner} as the contribution of a zero mode of a bosonic field on the BPS string, and to also include in the elliptic genus the contributions of its excitations. In other words, in order to pass to the elliptic genus one would like to replace any such factor by a factor of $(1-e^{2\pi i z})\prod_{j=1}^\infty(1-q^je^{2\pi i z})(1-q^j x^{-2\pi i z})$, where $ q = e^{2\pi i \tau} $. It is in fact convenient to express the denominator in a modular covariant fashion, so we instead make the following replacement as in \cite{Huang:2015sta,Huang:2015ada,Haghighat:2015ega}:
\begin{equation} (1- e^{2\pi i z}) \mapsto \varphi_{-1,1/2}(z,\tau)\equiv \frac{\theta_1(z,\tau)}{\eta(\tau)^3} = i e^{-\pi i z}(1-e^{2\pi i z})\prod_{j=1}^\infty\frac{(1-q^je^{2\pi i z})(1-q^j x^{-2\pi i z})}{(1-q^j)^2},\end{equation}
which is a weak Jacobi form of modular weight $ -1 $ and index $ 1/2 $. Furthermore, in order to account for the leading order behavior of the elliptic genus
\begin{equation} \mathbb{E}_{\widetilde h_{G}^{(k)}} \simeq q^{-c_L/24}(\dots)=q^{-\frac{kh^\vee_G}{6}}(\dots) \end{equation}
we also include a factor of
\begin{equation} \eta(\tau)^{4 k h^\vee_G} = \left(q^{1/24}\prod_{k=1}^\infty(1-q^k)\right)^{4 k h^\vee_G}\end{equation}
in our expression for the denominator.

We are therefore led to the following Ansatz for the elliptic genus:

\begin{align} &\mathbb{E}_{\widetilde h_{G}^{(k)}}(\epsilon_1,\epsilon_2,m_\alpha,\tau) =\nonumber\\\nonumber\\
\vspace{.2in}
&\frac{\mathcal{N}_{G,k}(\epsilon_1,\epsilon_2,m_\alpha,\tau)}{\eta(\tau)^{4kh^\vee_G}\displaystyle{\prod_{s=\pm1}\prod_{i=1}^k}\left(\varphi_{-1,1/2}(i(\epsilon_++s\,\epsilon_-),\tau)\displaystyle{\prod_{\substack{j=-i+1\\j-i \text{ odd}}}^{i-1}}\prod_{\alpha\in \Delta_+}\varphi_{-1,1/2}( s((i+1)\epsilon_++j \epsilon_-))+m_{\alpha},\tau)\right)}.\label{eq:ansatzG}\end{align}
The numerator should be a weak Jacobi form of several elliptic variables and integer Fourier coefficients. Considerations based on modularity and topological string theory significantly constrain its form. First of all, the requirement that the elliptic genus be a Jacobi form of weight zero implies that the numerator is a holomorphic Jacobi form of weight
\begin{equation} 2 k h_G^\vee -2k -\frac{k(k+1)}{2}(\text{dim}(G)-\text{rk}(G)),\label{eq:weight}\end{equation}
which for $ G $ simply laced reduces to
\begin{equation} 2k (h_G^\vee-1)-\frac{k(k+1)}{2}h_G^\vee \text{rk}(G).\end{equation}
Furthermore, the modular anomaly of the denominator can be easily read off from the modular transformation of the Jacobi theta function,
\begin{equation} \theta_1(z/\tau,-1/\tau) = \sqrt{-i\tau}e^{\frac{2\pi i}{\tau}\frac{z^2}{2}}\, \theta_1(z,\tau),\end{equation}
and is given by:
\begin{align} f^D_{\widetilde h^{(k)}_G}(m_\alpha,\epsilon_+,\epsilon_-) =& \frac{1}{2}k(k+1)h^\vee_G (\,\vec{y}\,|\,\vec{y}\,)_G+\frac{k(k+1)}{6}\big[(2k+1)(\epsilon_+^2+\epsilon_-^2)\nonumber\\
&+(\text{dim}(G)-\text{rk}(G))(2+k)((3k+5)\epsilon_+^2+(k-1)\epsilon_-^2)\big].\label{eq:modden}\end{align}

\medskip

The modular anomaly of the elliptic genus \eqref{eq:ellindex} is simply the difference between the the modular anomaly of the numerator ($f^{N}$) and that of the denominator \eqref{eq:modden}. Therefore
\begin{align} f^N_{\widetilde h^{(k)}_G}(m_\alpha,\epsilon_+,\epsilon_-) &=  f_{\widetilde h^{(k)}_G}(m_\alpha,\epsilon_+,\epsilon_-)+ f^D_{\widetilde h^{(k)}_G}(m_\alpha,\epsilon_+,\epsilon_-)\nonumber\\
&=\mathcal{C}_{flavor}(G,k) \frac{(\vec{y}\,|\,\vec{y})_G}{2}+\mathcal{C}_{\epsilon_+}(G,k)\epsilon_+^2+\mathcal{C}_{\epsilon_-}(G,k)\epsilon_-^2,
\label{eq:modan}\end{align}
where
\begin{align} \mathcal{C}_{flavor}(G,k) &= {1\over 3} \, k \left(h^\vee_G (2 + 3k) - 6\right),\\
 \mathcal{C}_{\epsilon_+}(G,k) &=\frac{k}{24}\Big(2(2+5\text{dim}(G)-10h_G^\vee-5\text{rk}(G))+k(21(\text{dim}(G)-\text{rk}(G))+4(9+h_G^\vee))\nonumber\\&\,\,\,\,+2k^2(4+7(\text{dim}(G)-\text{rk}(G)))+3k^3(\text{dim}(G)-\text{rk}(G)\Big),\\
 \mathcal{C}_{\epsilon_-}(G,k) &= \frac{k}{24}\Big(2(2-\text{dim}(G)+2h_G^\vee+\text{rk}(G))+k(\text{rk}(G)-\text{dim}(G)-4(3+h_G^\vee))\nonumber\\&\,\,\,\,+2k^2(4+\text{dim}(G)-\text{rk}(G))+k^3(\text{dim}(G)-\text{rk}(G)\Big),\end{align}
capture the anomaly with respect to $ G, SU(2)_R, $ and $ SU(2)_L $ respectively. One sees by inspection that for all the choices of $ G $ that arise for (1,0) 6d SCFTs with no matter and for any number $ k $ of strings the coefficient $\mathcal{C}_G$ is an integer, while $\mathcal{C}_{\epsilon_\pm}$ are either integers or half-integers. These coefficients also play the r\^ole of weights for the corresponding Jacobi form.

\medskip

We remark that in the case of one $ G $-instanton the modular anomaly of the numerator simplifies to
\begin{equation}{1\over 6} \left( 5 h^{\vee}_G - 6 \right) \, ( \vec{y} \,| \,\vec{y} )_G +\frac{1}{2}(2\epsilon_+)^2\left(1+\text{dim}(G)-\text{rk}(G)-\frac{h^\vee_G}{3}\right).\end{equation}
In other words, the dependence on the $ SU(2)_L $ fugacity $ \epsilon_- $ drops out. This is indeed consistent with the fact that for a single instanton the $ SU(2)_L $ flavor symmetry only acts on the decoupled hypermultiplet (whose contribution to the elliptic genus is confined to denominator terms). Furthermore, it turns out that the elliptic genus can be expressed in terms of Jacobi forms with elliptic variable $ 2\epsilon_+ $ and index
\begin{equation} \frac{1}{2}\left(1+\text{dim}(G)-\text{rk}(G)-\frac{h^\vee_G}{3}\right),\end{equation}
which always belongs to $\mathbb{Z}/2$. This is a useful fact, since the dimension of the space of Jacobi forms grows rapidly with the index. On a related note, one indeed observes that the Hilbert series of one-instanton moduli spaces only depends on the square of the variable $ v = e^{2\pi i\epsilon_+} $.

\medskip

To make further progress, we express the numerator in terms of the appropriate basis of Jacobi forms, which should capture the invariance of the elliptic genus under the Weyl group of the global symmetry $ G\times SU(2)_L\times SU(2)_R $. The natural set of Jacobi forms to use are therefore the Weyl-invariant Jacobi forms for $ G $, $ SU(2)_L $ and $ SU(2)_R $ whose theory has been developed in \cite{MR0466134,}. We refer to those papers for the precise definition of this class of functions, but we remark here that under a modular transformation $ \tau \to -1/\tau $, a Weyl[G]-invariant Jacobi form $ \Phi(z,\tau) $ of weight $ \ell $ and index $ m $ transforms as follows:
\begin{equation}\Phi(z/\tau,-1/\tau) = \tau^\ell e^{\frac{2\pi i m}{\tau}\frac{(z|z)}{2}}\Phi(z,\tau).\end{equation}
Comparing with Equation \eqref{eq:modan}, one sees that the numerator has integral index with respect to $ G $, and half-integral with respect to $ SU(2)_{L,R} $. Using a slight generalization of corollary 3 to Theorem 8 of Chapter III of \cite{MR781735}, we write the numerator schematically as a finite sum
\begin{equation} \sum_{i} a_i\, g_i(\epsilon_+,\epsilon_-,m_\alpha,\tau),\end{equation}
where each $ g_i $ is a product of powers of Weyl[$SU(2)_L$]-, Weyl[$ SU(2)_R $]-, and Weyl[$G$]- invariant Jacobi forms and Eisenstein series $ E_4(\tau) $ and $ E_6(\tau) $, such that $ g_i $ has the correct modular weight and indices.

\medskip

For $ SU(2) $, the algebra of Weyl-invariant Jacobi forms of integer index is generated by the two well-known functions  \cite{MR781735}
\begin{align} \varphi_{0,1}(z,\tau) &= 4\sum_{k=2}^4\frac{\theta_k(z,\tau)^2}{\theta_k(0,\tau)^2}\\
 \varphi_{-2,1}(z,\tau) &= \frac{\theta_1(z,\tau)^2}{\eta(\tau)^{6}},\end{align}
 where the labels $ k,m $ in $ \varphi_{k,m} $ denote respectively the weight and the index of the Jacobi form.
 
In what follows will also need to make use of half-integral Jacobi forms for SU(2). By a lemma of Gritsenko \cite{Gritsenko:1999fk}, any Jacobi form with integral Fourier coefficients, of even weight $ 2k $ and half-integral index $ m+1/2 $, can be written as
 \begin{equation} \varphi_{0,3/2} = \frac{\theta_1(2z,\tau)}{\theta_1(z,\tau)}\end{equation}
 times a Jacobi form of weight $ 2k $ and integral index $ m -1 $. Likewise, a Jacobi form with integral Fourier coefficients, of odd weight $ 2k+1 $ and half-integral index $ m+1/2 $, can be written as
 \begin{equation} \varphi_{-1,1/2} = \frac{\theta_1(z,\tau)}{\eta(\tau)^3}\end{equation}
 times a Jacobi form of weight $ 2k+2 $ and integral index $ m $.

\medskip 

 For $ G $ simple, the Weyl[$G$]-invariant Jacobi forms of integer index form a polynomial algebra over the ring of modular forms; a set of $ \text{rk}(G)+1 $ generators for this algebra has been constructed in all cases except $ G = E_8 $ \cite{MR1163219}. For any $ G $, of these generators, one has modular weight zero, while the others all have negative weight  (we refer to \cite{MR1163219} for details). In practice, we find that keeping the dependence on the fugacities $ m_{\alpha} $ for $ G $ significantly complicates the task of determining the elliptic genus by modularity arguments alone. This is due to the fact that Jacobi forms of high index arise in the numerators. In this paper, therefore, for simplicity we set $ m_\alpha \to 0 $, and leave the dependence of the elliptic genus on the $ G $ fugacities for future work. In this limit, all the negative weight Weyl[$ G $]-invariant Jacobi forms vanish, while the weight zero Jacobi form reduces to a constant.

The approach outlined here is general and leads to an Ansatz for all of the $ \tilde{h}^k_{G} $ theories. In the remainder of this section, we demonstrate the efficiency of this method in the case $ k = 1 $.  In all cases, we match our Ansatz in the $ q\to 0 $ limit against the Hilbert series, which has the known expansion \cite{Benvenuti:2010pq} 
\begin{equation} \mathcal{H}(\mathcal{M}_{G,1}) = \frac{1}{(1-v x)(1-v x^{-1})}\sum_{\ell\geq 0} \text{dim}(\ell\cdot \text{Adj}_{G})v^{2\ell},\label{eq:hilb2}\end{equation}
where $ \ell\cdot \text{Adj}_G $ denotes the representation whose highest weight is $ \ell $ times the highest root in the adjoint representation of $G$.  As we explain in the rest of this section, will will also need to further impose the vanishing of certain coefficients in the Fourier expansion of the elliptic genus at higher orders in $ q $. In all cases, we will find that such constraints are sufficient to fix all the unknown coefficients in our Ansatz for the numerator.\footnote{It is natural to ask whether also for $ k > 1  $ one can completely determine the elliptic genus by imposing a sufficient number of constraints on the Ansatz; this question will be addressed elsewhere \cite{KlockB}.}

\subsection{Elliptic genus of one $SU(3)$ string}
\label{sec:1su3}
In this section we apply the techniques discussed above to uniquely determine the elliptic genus for one SU(3) instanton. We make the following Ansatz:
\begin{align} \mathbb{E}_{\widetilde h_{SU(3)}^{(1)}}(\epsilon_+,\epsilon_-) = \frac{\mathcal{N}_{SU(3),1}(2\epsilon_+,\tau)}{\eta(\tau)^{12}\varphi_{-1,1/2}(\epsilon_1)\varphi_{-1,1/2}(\epsilon_2)\prod_{\alpha\in\Delta_+^{SU(3)}}\varphi_{-1,1/2}(2\epsilon_+,\tau)^2}.\label{eq:SU3gen}\end{align} The modularity constraints discussed in Section \ref{sec:hilb} imply that the numerator is a Jacobi form of modular weight $ -2 $ and index $ 3 $ with respect to $ 2\epsilon_+ $. This fixes its form up to two unknown coefficients $ a_1,a_2 $:

\begin{align} \mathcal{N}_{SU(3),1}(2\epsilon_+,,\tau) &\,=\, a_1\phi_{-2,1}(2\epsilon_+,\tau)\phi_{0,1}(2\epsilon_+,\tau)^2+a_2 \phi_{-2,1}(2\epsilon_+,\tau)^3E_4(\tau).\end{align}
We next impose the equality \footnote{We find it always necessary to multiply the Hilbert series by a factor of $ v^{h^\vee_G} $, which makes it symmetric under $ v\to v^{-1} $. We view such factor as being part of the relative normalization between $ Q $ and $ \widetilde{Q} $ in Equations \eqref{eq:topstring} and \eqref{eq:hilb}.}
\begin{equation} \lim_{q\to0} q^{\frac{4h^\vee_{SU(3)}}{24}}\mathbb{E}_{\widetilde h_{SU(3)}^{(1)}}(\epsilon_+,\epsilon_-) = v^{h^\vee_{SU(3)}}\,\mathcal{H} (\mathcal{M}_{SU(3),1})\label{eq:match}\end{equation}
between the elliptic genus and the Hilbert series of one SU(3) instanton, where the latter quantity is given by
\begin{equation} \frac{1}{(1-v x)(1-v x^{-1})}\sum_{\ell\geq 0} \text{dim}(\ell\cdot \text{Adj}_{SU(3)})v^{2\ell},\label{eq:hilb3}\end{equation}
where $ \text{dim}(\ell\cdot \text{Adj}_{G}) = \ell^3 $. Imposing Equation \eqref{eq:match} uniquely fixes the coefficients of the numerator, and one finds:
\begin{equation} a_1 = -\frac{1}{24},\qquad a_2 = \frac{1}{24}.\end{equation}
This completely determines the elliptic genus of one SU(3) instanton string. We have checked that our result is in agreement with the genus-zero topological string data given in \cite{Haghighat:2014vxa}. Furthermore, we can remove from the elliptic genus the contribution of the center of mass hypermultiplet,
\begin{equation} \mathbb{E}_{c.m.} = \frac{\eta(\tau)^2}{\theta_1(\epsilon_1,\tau)\theta_1(\epsilon_2,\tau)}, \end{equation}
to obtain the elliptic genus of the $ h^{(1)}_{E_6} $ theory; we have verified up to $ \mathcal{O}(q^{7/2}) $ that this matches with the expression which was recently obtained in \cite{Kim:2016foj} by gauge-theoretic techniques. We observe here that
\begin{equation}  v^{1-h^\vee_{SU(3)}}q^{\frac{4(h^\vee_{SU(3)}-1)}{24}}\mathbb{E}_{ h_{SU(3)}^{(1)}}(\epsilon_+,\epsilon_-) \bigg\vert_{q v^0}=9 = \text{dim}(SU(3)) + 1,\label{eq:dimG}\end{equation}
and we will see later on that a similar statement holds for the other choices of $ G $. In order to efficiently display the numerical coefficients appearing in the elliptic genus, we find it convenient for all $G$ to define a rescaled elliptic genus,
\begin{equation}\mathcal{E}_{G}(p,\tilde p) = q^{\frac{h^\vee_G-1}{6}}v^{1-\frac{h^\vee_G}{3}}\mathbb{E}_{h_{G}^{(1)}}(2\epsilon_+,\tau), \end{equation}
which we can expand in terms of variables $ p=v^2 $ and $ \tilde p = q/v^2 $ as:
\begin{equation} \mathcal{E}_G(p,\tilde p) = \sum_{k,l\geq 0} b^G_{k,l}\, p^k{\tilde p}^l.\end{equation}
The coefficients $b_{k,l}^{SU(3)}$ in the series expansion of the elliptic genus of one SU(3) instanton are displayed in Table \ref{tb:SU3} of Appendix \ref{sec:appdata}.

\subsection{Elliptic genus of one SO(8) string}
Next, we use modularity to fix the elliptic genus of one SO(8) string (with fugacities $ m_\alpha $ set to zero for simplicity). From our discussion in Section \ref{sec:hilb} it follows that the numerator has modular weight $ -14 $ and index $ 23/2 $ with respect to $ 2\epsilon_+ $. We can therefore write the numerator as
\begin{align} \mathcal{N}_{SO(8),1}(2\epsilon_+,0,\tau) &= \phi_{0,3/2}(2\epsilon_+,\tau)\phi_{-2,1}^7(2\epsilon_+,\tau)\bigg(a_1\phi_{0,1}(2\epsilon_+,\tau)^3\nonumber\\
&+a_2E_4(\tau)\phi_{-2,1}(2\epsilon_+,\tau)^2\phi_{0,1}(2\epsilon_+,\tau)+a_6E_6(\tau)\phi_{0,1}(2\epsilon_+,\tau)^3\bigg),\end{align}
which depends on just three undetermined coefficients. Imposing
\begin{equation} \lim_{q\to0} q^{\frac{4h^\vee_{SO(8)}}{24}}\mathbb{E}_{SO(8),1} = v^{h^\vee_{SO(8)}}\,\mathcal{H} (\mathcal{M}_{SO(8),1}),\label{eq:matchso}\end{equation}
where \cite{Benvenuti:2010pq}
\begin{equation}\mathcal{H} (\mathcal{M}_{SO(8),1})=\frac{\left(v^2+1\right) \left(v^8+17 v^6+48 v^4+17 v^2+1\right)}{\left(1-v^2\right)^{10}}\cdot \frac{1}{(1-v)^2},\end{equation}
is sufficient to fix all the undetermined coefficients. We find:
\begin{equation} a_1 = \frac{7}{144};\qquad a_2 = -\frac{1}{16}; \qquad a_3 = \frac{1}{72}.\end{equation}
We have verified that under this choice of coefficient the elliptic genus agrees with the known expression in \cite{Haghighat:2014vxa} to high powers in $ q $.
Analogously to the $ SU(3) $ case, we observe that:
\begin{equation}  v^{-h^\vee_{SO(8)}}q^{\frac{4(h^\vee_{SO(8)}-1)}{24}}\mathbb{E}_{h_{SO(8)}^{(1)}}(m_\alpha,\epsilon_+,\epsilon_-) \bigg\vert_{q v^0}=29 = \text{dim}(SO(8)) + 1.\label{eq:dimG2}\end{equation}

\subsection{Elliptic genera of exceptional instanton strings}
In this section we determine the elliptic genera of the one instanton theories $ \widetilde h^{(1)}_G $ for $ G = F_4, E_6,E_7, $ and $E_8 $.

\subsubsection{$\mathbf{G = F_4}$} We begin by looking at $ G = F_4 $. Although in this case we do not have a geometric engineering construction for this theory, we can still use the anomaly polynomial \eqref{eq:ellindex}, which based on the derivation of \cite{Kim:2016foj,Shimizu:2016lbw} is also valid for the $ F_4 $ strings, to fix the form of the elliptic genus. The numerator of our Ansatz has modular weight $ -32 $ and index $ 23 $ with respect to $ 2\epsilon_+ $. This leads to an expression which is determined up to 9 coefficients. In this case, comparing with the Hilbert series,
\begin{equation} \lim_{q\to0} q^{\frac{4h^\vee_{F_4}}{24}}\mathbb{E}_{\widetilde h_{F_4}^{(1)}} = v^{h^\vee_{F_4}}\,\mathcal{H} (\mathcal{M}_{F_4,1}),\label{eq:matchf}\end{equation}
 only fixes 8 out of the 9 coefficients.

 In order to fix the remaining coefficient $ a $, we now factor out the center of mass hypermultiplet contribution $ \mathbb{E}_{c.m.} $ and look at the subleading order in the $ q $ expansion of the elliptic genus $ \mathbb{E}_{h_{F_4}^{(1)}} $. This is given by $ v^{8}q^{-4/3+1} $ times
\begin{equation}\frac{1}{54}\left(\frac{7}{v^6}a+\frac{98}{v^4}a+\frac{357}{v^2}a+(2862+868a)+\mathcal{O}(v^2)\right).\label{eq:f4o1}\end{equation}
Notice that if we set $a = 0$ all the terms with negative powers of $v$ drop out. Furthermore, one finds that the $v^0$ coefficient is 
\begin{equation}53 = \text{dim}(F_4)+ 1, \end{equation}
analogously to the cases $G=SU(3)$ and $SO(8)$. We find that the numerator of our Ansatz for the theory of one $ F_4 $ instanton is given by:
\begin{align}\mathcal{N}_{F_4,1}=\frac{1}{746496}\phi_{-2,1}^{16} \bigg(&\phi_{-2,1}^6 \phi_{0,1} \left(56 E_6^2-81 E_4^3\right)+45 E_4^2 E_6 \phi_{-2,1}^7+486 E_4^2 \phi_{-2,1}^4 \phi_{0,1}^3\nonumber\\
-&366 E_4 E_6 \phi_{-2,1}^5 \phi_{0,1}^2-453 E_4 \phi_{-2,1}^2 \phi_{0,1}^5+209 E_6 \phi_{-2,1}^3 \phi_{0,1}^4+104 \phi_{0,1}^7\bigg),\hspace{0.3in}\end{align}
where we have omitted the elliptic and modular arguments $ (2\epsilon_+,\tau) $ of the Jacobi and modular forms for brevity.

In Appendix \ref{sec:appdata} we also provide the series expansion coefficients of the elliptic genus. One can verify that the coefficients can be written in terms of sums of dimensions of small numbers of representations of $F_4$ with positive coefficients. We view these facts as a strong indication that the choice $a=0$ gives the elliptic genus of one $F_4$ instanton string.

\subsubsection{$\mathbf{G = E_6}$}The theory of one $ E_6 $ instanton string is the $ T^2\times S^2 $ compactification of Gaiotto's $ T_3 $ theory, whose elliptic genus has been computed via (0,4) dualities \cite{Putrov:2015jpa} (see the overview in Section \ref{sec:e6rev}). We now recover the same result (with fugacities $ m_\alpha $ turned off) by resorting to modularity. The numerator of our Ansatz has modular weight $ -50 $ and index $ 69/2 $ with respect to $ 2\epsilon_+ $. There is a 10-dimensional space of Jacobi forms of such weight and index, and matching against the Hilbert series fixes 8 of the coefficients. We fix the remaining additional coefficient by requiring the $ v^{-2k}q^1 $ terms (with $k>0$) of the expression for $ \mathbb{E}_{h_{E_6}^{(1)}} $ to vanish. This completely fixes the elliptic genus (and in fact gives an over-determined set of constraints on the coefficients), and we again observe that the $q^1v^0$ term is:
\begin{equation} 1 +\text{dim}(E_6) = 79,\end{equation}
analogously to the $G = SU(3), SO(8), $ and  $F_4 $ cases.\newline

\noindent The explicit expression for the numerator of the elliptic genus is:
\begin{align}\mathcal{N}_{E_6,1}=\frac{1}{23887872}\phi_{-2,1}^{25} \phi_{0,3/2} &\bigg(9 \phi_{-2,1}^8 \left(23 E_4^4-64 E_4 E_6^2\right)+4 \phi_{-2,1}^6 \phi_{0,1}^2 \left(512 E_6^2-1845 E_4^3\right)\nonumber\\
&\!\!\!\!+4656 E_4^2 E_6 \phi_{-2,1}^7 \phi_{0,1}+23010 E_4^2 \phi_{-2,1}^4 \phi_{0,1}^4-14880 E_4 E_6 \phi_{-2,1}^5 \phi_{0,1}^3\nonumber\\
&-18564 E_4 \phi_{-2,1}^2 \phi_{0,1}^6+7280 E_6 \phi_{-2,1}^3 \phi_{0,1}^5+4199 \phi_{0,1}^8\bigg).\end{align}
In Appendix \ref{sec:appdata} we also provide the series expansion coefficients of the elliptic genus. One can verify that the coefficients can be written in terms of sums of dimensions of small numbers of representations of $E_6$ with positive coefficients.\newline

We can compare the expression we find with the elliptic genus of the $ T_3 $ theory; we have checked up to $ \mathcal{O}(q^4) $ that the two expressions match, which serves as a check of both our Ansatz and of our geometric engineering argument.
 In all the cases discussed so far we notice that all the coefficients $ b^G_{k,l} $ with $ k\leq h^\vee_G/3-1 $ and $ l \leq h^\vee_G/3-1 $ vanish.\footnote{We refer the reader to Appendix \ref{sec:appdata} for our notation. It would be desirable to find a physical argument for why these coefficients should vanish.} We will assume this to also hold true for $ G= E_7 $ and $ G=E_8 $, which will be crucial for uniquely fixing the elliptic genera.

\subsubsection{$\mathbf{G = E_7}$} In this case, the modular weight of the numerator is -92, and the index with respect to $ 2\epsilon_+ $ is $ \frac{121}{2} $. This fixes the Ansatz for the numerator up to 21 undetermined coefficients. Comparing the leading order terms in the $ q $-expansion with the Hilbert series of the moduli space of one $ E_7 $ instanton fixes 13 coefficients, leaving 8 undetermined. We fix these by imposing the vanishing of coefficients $ b^{E_7}_{k,l} $ for $ k \leq 5,l\leq 5 $. This is an overdetermined set of constraints that leads to a unique solution, which is given by:
\begin{align}&\mathcal{N}_{E_7,1}=\frac{1}{2972033482752} \phi_{-2,1}^{46}  \phi_{0,3/2}\bigg(12 (6399 E_4^5 E_6 - 
      10528 E_4^2 E_6^3) \phi_{-2,1}^{13} \nonumber\\
      &+ (1472256 E_4^3 E_6^2\!-\!151875 E_4^6 \!-\! 
      60416 E_6^4) \phi_{-2,1}^{12} \phi_{0,1} 
      - 180 E_4 E_6 (26739 E_4^3 - 8704 E_6^2) \phi_{-2,1}^{11} \phi_{0,1}^2 \nonumber\\
      &+ 
   18 E_4^2 (258993 E_4^3 - 627040 E_6^2) \phi_{-2,1}^{10} \phi_{0,1}^3 + 
   280 E_6 (106623 E_4^3 - 5680 E_6^2) \phi_{-2,1}^9 \phi_{0,1}^4 \nonumber\\
   &- 
   567 E_4 (45667 E_4^3 - 29056 E_6^2) \phi_{-2,1}^8 \phi_{0,1}^5 - 51471000 E_4^2 E_6 \phi_{-2,1}^7 \phi_{0,1}^6 \nonumber\\
   &+ 
   228 (217503 E_4^3 - 25648 E_6^2) \phi_{-2,1}^6 \phi_{0,1}^7 + 31668516 E_4 E_6 \phi_{-2,1}^5 \phi_{0,1}^8 \nonumber\\
   &- 
   40739325 E_4^2 \phi_{-2,1}^4 \phi_{0,1}^9 - 6249100 E_6 \phi_{-2,1}^3 \phi_{0,1}^{10} + 14827410 E_4 \phi_{-2,1}^2 \phi_{0,1}^{11} - 
   1964315 \phi_{0,1}^{13}\bigg).\end{align}
We display the series expansion coefficients of the elliptic genus thus obtained in Appendix \ref{sec:appdata}. Again, we verify that the $q^1v^0$ coefficient is given by
\begin{equation} \text{dim}(E_7)+1=134.\end{equation}

\subsubsection{$\mathbf{G = E_8}$} Finally, we proceed in the same manner for the case of one $ E_8 $ instanton. The modular weight of the numerator is -182 and the index with respect to $ 2\epsilon_+ $ is $ \frac{231}{2} $. This determines the Ansatz up to 56 coefficients, of which 23 are fixed by matching with the Hilbert series. As for the $ E_7 $ case, we impose the vanishing of $ b^{E_8}_{k,l} $ for $ k\leq 9 $ and $ l \leq 9 $. This again gives an overdetermined set of constraints on the series coefficients $ b^{E_8}_{k,l} $ which uniquely determines the form of our Ansatz. We provide the expression for the numerator, which is rather unwieldy, in Appendix \ref{sec:appdata}, along with the series expansion coefficients of the elliptic genus. We also verify in this example that the $q^1v^0 $ coefficient is given by $\text{dim}(E_8)+1=249$.

\section{Relation with the Schur index of $ H^{(1)}_G $}\label{sec:Sch}

In this section we comment on a surprising relation between the elliptic genus of one instanton string $ \mathbb{E}_{h_{G}^{(1)}} $, which is the $\beta$-twisted partition function on $ T^2\times S^2 $, and the Schur index of the $ H_{G}^{(1)} $ theory, which is a partition function on $ S^1\times S^3 $.\\

\subsection{The case $ G = SU(3) $}
\label{sec:caseSU3}
We begin by discussing the $G = SU(3) $ theory, which is the same as the $(A_1,D_4)$ Argyres-Douglas theory. More precisely, we remove the contribution of a free hypermultiplet  
\begin{equation} \eta(q)^4\phi_{-1,1/2}(\epsilon_1) \phi_{-1,1/2}(\epsilon_2) \end{equation}

from the denominator of Equation \eqref{eq:SU3gen}, so we consider
\begin{equation} \mathbb{E}_{h^{(1)}_{SU(3)}}(\epsilon_+,\tau) =\frac{\mathcal{N}_{G,1}(2\epsilon_+,\tau)}{\eta(\tau)^{8}\prod_{\alpha\in\Delta_+^{SU(3)}}\varphi_{-1,1/2}(2\epsilon_+,\tau)^2},\end{equation}
 and make the specialization
 \begin{equation} \epsilon_+ = \tau/4.\end{equation}
 In this limit, one has:
\begin{equation} \phi_{-2,1}(2\epsilon_+,\tau) \mapsto q^{-1/4}\frac{\theta_4(0,\tau)^2}{\eta(\tau)^6}, \qquad \phi_{0,1}(2\epsilon_+,\tau) \mapsto 4q^{-1/4}\left(\frac{\theta_2(0,\tau)^2}{\theta_3(0,\tau)^2}+\frac{\theta_3(0,\tau)^2}{\theta_2(0,\tau)^2}\right),\end{equation}
and 
\begin{equation} \phi_{-1,1/2}(2\epsilon_++z,\tau)\mapsto \frac{i}{e^{\pi i z}q^{1/8}}\frac{\theta_4(z,\tau)}{\eta(\tau)^3}.\end{equation}
We then expand the elliptic genus as a $ q $-series and find the following result:
\begin{align}\mathbb{E}_{h^{(1)}_{SU(3)}}(\tau/4,\tau) = 2 q^{1/6}&\bigg(1+8q^{1/2}+36 q+128 q^{3/2}+394 q^2+1088q^{5/2}+2776 q^3+6556 q^{7/2}\nonumber\\&+15155 q^4 + 33056 q^{9/2} + 69508 q^5 + 141568 q^{11/2} + 
 280382 q^6 \nonumber \\&+\! 541696 q^{13/2}\! +\! 1023512 q^7\! +\! 1895424 q^{15/2}\!+\! 
 3446617 q^8\!+\!\mathcal{O}(q^{17/2})\bigg),\end{align}
or in other words, up to $ \mathcal{O}(q^{17/2}) $,
\begin{align} \mathbb{E}_{h_{SU(3)}^{(1)}}(\tau/4,\tau) = 2 q^{1/6} \mathcal{I}_{H^{(1)}_{SU(3)}}(q^{1/2}),\label{eq:reln}\end{align}
where $ \mathcal{I}_{H^{(1)}_{SU(3)}}(q) $ is the Schur limit of the superconformal index of the $ H^{(1)}_{SU(3)} $ theory (in the limit $ m_{\alpha} \to 0 $). This theory coincides with the $ (A_1,D_4) $ Argyres-Douglas theory, and the explicit expression for its Schur index has been obtained in \cite{Cordova:2015nma,Buican:2015ina}.\newline

\begin{table}[h!]
\begin{center}
\begin{tabular}{c| lllllll }\,\,\,\,\,\,\,\,l\,\,\,\,\,\,\,k&0&1&2&3&4&5&6\\
\hline
0&0& 1& 8& 27& 64& 125& 216\\%& 343& 512& 729& 1000& 1331& 1728& 2197& 
%  2744& 3375\\
1&1& 0& 9& 64& 216& 512& 1000\\%& 1728& 2744& 4096& 5832& 8000& 10648& 
%  13824& 17576& 21952\\
2&8& 9& 0& 53& 360& 1188& 2816\\%& 5500& 9504& 15092& 22528& 32076& 
%  44000& 58564& 76032& 96668\\
3&27& 64& 53& 0& 245& 1600& 5211\\%& 12288& 24000& 41472& 65856& 98304& 
%  139968& 192000& 255552& 331776\\
4&64& 216& 360& 245& 0& 971& 6168\\%& 19818& 46528& 90750& 156816& 
%  249018& 371712& 529254& 726000& 966306\\
5&125& 512& 1188& 1600& 971& 0& 3435\\%& 21312& 67716& 158208& 308125& 
%  532224& 845152& 1261568& 1796256& 2464000\\
6&216& 1000& 2816& 5211& 6168& 3435& 0\\%& 11139& 67800& 213219& 495872& 
%  964000& 1664280& 2642472& 3944448& 5616216\\
%7&343& 1728& 5500& 12288& 19818& 21312& 11139& 0& 33667& 201536& 
%  628074& 1454080& 2821500& 4867776& 7727447& 11534336\\
%8&512& 2744& 9504& 24000& 46528& 67716& 67800& 33667& 0& 96004& 
%  566496& 1750815& 4036096& 7816125& 13474296& 21384335\\
%9&729& 4096& 15092& 41472& 90750& 158208& 213219& 201536& 96004& 0& 
%  260564& 1518016& 4656339& 10690048& 20660750& 35586432\\
%10&1000& 5832& 22528& 65856& 156816& 308125& 495872& 628074& 566496& 
%  260564& 0& 677704& 3903616& 11890854& 27192832& 52450625\\
%11&1331& 8000& 32076& 98304& 249018& 532224& 964000& 1454080& 1750815& 
%  1518016& 677704& 0& 1698120& 9681344& 29302047& 66760704
\end{tabular}
\caption{Expansion coefficients $ b^{SU(3)}_{k, l} $ for one $SU(3)$ instanton.}\label{tb:SU3xx}
\end{center}
\end{table}

The factor of $ 2 $ can be accounted for by looking at the data in Table \ref{tb:SU3}. We reproduce a small region of that table in Table \ref{tb:SU3xx}. The spectrum of states that contribute to the elliptic genus consists of two identical sets, whose degeneracies are captured by the coefficients $ b^{SU(3)}_{k,l} $: those for $ k > l $ and those for $ k <l $. Let us denote by 
 \begin{equation} L_{SU(3)}(v,q) = \sum_{l\geq 0}\sum_{k>l}b^{SU(3)}_{k,l} v^{2k}(q/v^2)^{2l}\end{equation}
the half of the elliptic genus expansion associated to the upper right half of the table. In other words,
\begin{equation} L_{SU(3)}(v,q) = (1+8v^2+27 v^4+\dots) + q(9+64 v^2+\dots)+\dots\end{equation}
 
  Our prescription for matching with the Schur index requires setting $ v = q^{1/4} $. In this limit, the coefficients of the $ q $ expansion are obtained by summing along the anti-diagonals in Table \ref{tb:SU3}, and it is clear that each of the two sets of states contributes an identical term
  \begin{equation} L_{SU(3)}(q^{1/4},q) = \left(1+ 8 \,q^{1/2} + (27+9)\,q +(64+64)\,q^{3/2}+\dots \right)= \mathcal{I}_{H^{(1)}_{SU(3)}}(q^{1/2}).\end{equation}
to the elliptic genus.
\subsection{Generalization to other $ G $}
It is natural to ask whether a similar relation between elliptic genus and Schur index continues to hold for other $ G $ (at least for the simply laced cases, which have a clear four-dimensional origin as discussed in Section \ref{2d4d6d}). A first hint that the relation might not just be a coincidence comes by comparing the elliptic genus of a free 2d $ (0,4) $ hypermultiplet, which is given by
\begin{equation} \mathbb{E}_{h.m.}(\epsilon_+,\epsilon_-,\tau)=-\frac{\eta(\tau)^2}{\theta_1(\epsilon_++\epsilon_-,\tau)\theta_1(\epsilon_+-\epsilon_-,\tau)},\end{equation}
to the Schur index of a 4d hypermultiplet,
\begin{equation} \mathcal{I}_{h.m.}(\epsilon_-,\tau) = \exp\left(\sum_{n=1}^\infty\frac{1}{n}\frac{q^{n/2}}{1-q^n}(z^n+z^{-n})\right),\end{equation}
where $ z = e^{2\pi i \epsilon_-} $. Setting $ \epsilon_+ = \tau/4 $, one indeed finds that
\begin{equation}\mathbb{E}_{h.m.}(\tau/4,\epsilon_-,\tau) = \mathcal{I}_{h.m.}(\epsilon_-,\tau/2).\end{equation}
Furthermore, the same relation also holds between the elliptic genus of a free 2d $ (0,4) $ vector multiplet,
\begin{equation} \mathbb{E}_{v.m.}(\epsilon_+,\tau)=\eta(\tau)^2\frac{\theta_1(2\epsilon_+,\tau)}{\eta(\tau)},\end{equation}
and the Schur index of a 4d vector multiplet:
\begin{equation} \mathcal{I}_{v.m.}(\tau) = \eta(\tau)^2.\end{equation}
In other words, we find:
\begin{equation} \mathbb{E}_{v.m.}(\tau/4,\tau)= \mathcal{I}_{v.m.}(\tau/2).\end{equation}

\begin{table}[t]
 \begin{center}
\begin{tabular}{c| lllllll }\,\,\,\,\,\,\,\,l\,\,\,\,\,\,\,k&0&1&2&3&4&5&6\\
\hline
0&0& 0& \color{red}1& \color{red}28& \color{red}300& \color{red}1925& \color{red}8918\\%& 32928& 102816& 282150& 698775&   1591876\\%& 3383380\\%& 6782139\\
1& 0& 0& 0& \color{red}29& \color{red}707& \color{red}6999& \color{red}42889\\%& 193102& 699762& 2156994& 5864958&   14426643\\%& 32695949\\%& 69214145\\
2 &-1& 0& 0& \color{blue}2& \color{red}464& \color{red}9947&\color{red} 92391\\%& 544786& 2392663& 8526042& 25972362&   70015437\\%& 171120642\\%& 385916333\\
3 &-28& -29& -2& 0& \color{blue}58& \color{red}5365& \color{red}101850\\%& 894198& 5096487& 21888529&   76804254& 231399682\\%& 618759263\\%& 1503039657\\
4& -300& -707& -464& -58& 0& \color{blue}928& \color{red}49775\\%& 843165& 7032993& 38869314&   163555964& 565747296\\%& 1686960932\\%& 4476323164\\
5 &-1925& -6999& -9947& -5365& -928& 0& \color{blue}10646\\%& 391587& 5965996&   47459818& 254962664& 1052692147\\%& 3592900602\\%& 10609171946\\
6 &-8918& -42889& -92391& -101850& -49775& -10646& 0\\%& 97429& 2702949&   37329322& 284088584& 1486376037\\%& 6029111166\\%& 20319991572\\
7 &-32928& -193102& -544786& -894198& -843165& -391587& -97429\\%& 0&   753333& 16753135& 211306917& 1542462875\\%& 7872407942\\%& 31403671352\\
8 &-102816& -699762& -2392663& -5096487& -7032993& -5965996& -2702949\\%& -753333& 0& 5100348& 94814881& 1099832018\\%& 7718213190\\%& 38478234720\\
\end{tabular}
\caption{Series coefficients $ b^{SO(8)}_{k,l} $ for the elliptic genus of one $ SO(8) $  instanton.}\label{tb:so8red}
\end{center}
\end{table}

 At first glance, however, for other $H^{(1)}_G$ theories this relation seems to fail: the states under the diagonal in Tables \ref{tb:SO8} and \ref{tb:E6}-\ref{tb:E8} contribute to the elliptic genus with an opposite sign compared to the ones above the diagonal, and therefore the elliptic genus vanishes when we set $ v^2\to q^{1/2} $. However, a closer look at the expansion coefficients hints at a possible relation. For example, if we isolate the coefficients $ b^{SO(8)}_{k,l} $ with $ k > l+1 $ in the coefficient table for $ G= SO(8) $ (shown in red in Table \ref{tb:so8red}), and sum along anti-diagonals (that is, set $ v =  q^{1/4} $ in the sum $v^{-2}\sum_{l\geq 0}\sum_{k>l+1}b^{SO(8)}_{k,l} v^{2k}(q/v^2)^{2l}$), we find the following expression:
\begin{equation} 1+ 28\, q^{1/2} + 329\, q +2632 \, q^{3/2}+1638\mathbf{\color{red}1\color{black}}\, q^{2}+85764\, q^{5/2}+393\mathbf{\color{red}674}\, q^3+\mathcal{O}(q^{7/2}),\end{equation}
which disagrees from the Schur index of $ H^{(1)}_{SO(8)} $ \cite{Beem:2013sza,Cordova:2016uwk} (with $ q\to q^{1/2}) $
\begin{equation}1 + 28\, q^{1/2} + 329\, q + 2632\, q^{3/2} + 1638\mathbf{\color{red}0}\, q^2 + 85764\, q^{5/2} + 393\mathbf{\color{red}589}\, q^3+\mathcal{O}(q^{7/2})\end{equation}
by a small subleading correction
\begin{equation} \mathbf{\color{red}1}\cdot q^{2}+\mathbf{\color{red}85}\, q^3 +\mathcal{O}(q^{7/2}).\end{equation}
One also notices the existence of another sequence of coefficients, marked in blue in Table \ref{tb:so8red}, which is given by:
\begin{equation} 2\cdot 1,\qquad 2\cdot 29,\qquad 2\cdot 464,\qquad \dots\end{equation}
which is essentially a repetition of the coefficients  $b^{SO(8)}_{1,0}, b^{SO(8)}_{2,1}, b^{SO(8)}_{3,2},\dots $, multiplied by a factor of 2.\newline

\noindent A similar pattern holds for the case $ G = E_6 $: if we isolate the terms $ b^{E_6}_{k,l} $ with $ k > l+3 $ in Table \ref{tb:E6} and sum over anti-diagonals, we obtain:
\begin{align}q^{-2}&\sum_{l\geq 0}\sum_{k>l+1}b^{SO(8)}_{k,l} q^{(k+l)/2} =\nonumber\\ &1 + 78\, q^{1/2}+ 2509\, q+49270\, q^{3/2}+69842\mathbf{\color{red}6}\, q^2+7815106\, q^{5/2}+72903\mathbf{\color{red}429}\, q^3+\mathcal{O}(q^{7/2}), \end{align}
which is again very close to the Schur index of the 4d $ H^{(1)}_{E_6} = T_3 $ SCFT:\footnote{ We are grateful to Wenbin Yan for providing us with code to compute the Schur index of the $ T_3 $ theory to high orders in $ q $.}
\begin{equation}\mathcal{I}_{H^{(1)}_{E_6}}(q^{1/2}) = 1 + 78\, q^{1/2}+ 2509\, q+49270\, q^{3/2}+69842\mathbf{\color{red}5}\, q^2+7815106\, q^{5/2}+72903\mathbf{\color{red}350}\, q^3+\mathcal{O}(q^{7/2}). \end{equation}
We recognize the difference between the two series expansions,
\begin{equation} \mathbf{\color{red}1}\cdot q^2+\mathbf{\color{red}79}\, q^{3}+\mathcal{O}(q^{7/2})\end{equation}
as consisting of the diagonal coefficients $ b_{k+3,k} $. As in the $ G = SO(8) $ case, here we also notice that there are additional sequences of coefficients
\begin{equation} b^{E_6}_{5+n,5+n} = 2\cdot b^{E_6}_{4+n,n} \end{equation}
and 
\begin{equation} b^{E_6}_{5+n,4+n} = -1\cdot b^{E_6}_{4+n,n},\end{equation}
as well as analogous sequences in the bottom left half of the table.\newline

\begin{table}
\begin{center}{
\begin{tabular}{c| llllll}\,\,\,\,\,\,\,\,l\,\,\,\,\,\,\,k&10&11&12&13&14&15\\
\hline
0& 1& 248& 27000& 1763125& 79143000&   2642777280\\
1&  0& 249& 57877& 5943753& 368338125&   15776893240\\
2&  0& 0& 31374& 6815877& 659761497&   38811914750\\
3& 0& -1& 0& 2666375& 539686750&   49211333622\\
4& 0& 0& -249& 0& 171756125& 32299833875\\
5& 0& 0& 0& -31374& 0& 8931266291\\
6& 0& 0& 0& -248& -2666375& 0\\
7& 0& 0& 0& 0& -57877& -171756125\\
8& 0& 0& 0& 0& 0& -6815877\\
9&  0& 0& 0& 0& 
  0& -27000
\end{tabular}
\caption{Series expansion coefficients $ b^{E_8}_{k,l} $ for one $ E_8 $ instanton.}
\label{tb:E8rehashed}}
\end{center}
\end{table}
In the $ E_7 $ and $ E_8 $ cases (Tables \ref{tb:E7} and \ref{tb:E8}) we see a similar pattern of repeating sequences; for example, zooming into a small region in the table of $ b^{E_8}_{k,l} $ coefficients (Table \ref{tb:E8rehashed}), we see that the additional sequences of coefficients (such as the one starting with $ b_{11,3} = -1 $ in this example) consist of additional copies of the same sequences of coefficients as in the top sequence,
\begin{align}
&\{1, 249, 31374, 2666375, 171756125\dots\},\\
&\{248, 57877, 6815877\dots\},\\
&\{27000,\dots\}.
\end{align}

By inspection, we find from the data at hand that all the properties discussed above are simultaneously satisfied if we make the following conjecture:\newline\\
\noindent \emph{The elliptic genus of the theory $ h^{(1)}_G, $ for $ G = SU(3),SO(8),F_4,E_6,E_7,E_8 $ can be written as}
\begin{align}\mathbb{E}_{h_{G}^{(1)}}(\epsilon_+, \tau)&=v^{\frac{h^\vee_G}{3}-1}\sum_{n\geq 0}q^{2n}\bigg[\left(\frac{v}{q^{1/4}}\right)^{\frac{2h^\vee_G}{3}}L_G(q^n\, v; q)-(-1)^{h^\vee_G} \left(\frac{q^{1/4}}{v}\right)^{\frac{2h^\vee_G}{3}}L_G(q^{n+1/2}/v; q)\nonumber\\
&+(1+(-1)^{h^\vee_G})q^{\frac{1}{2}\left(\frac{h^\vee_G}{3}+1\right)}\left(\left(\frac{v}{q^{1/4}}\right)^2L_G(q^{n+1/2}\,v,q)-\left(\frac{q^{1/4}}{v}\right)^2L_G(q^{n+1}/v,q)\right)\nonumber\\
&-q^2\left(-(-1)^{h^\vee_G}\left(\frac{v}{q^{1/4}}\right)^{4-2\frac{h^\vee_G}{3}}L_G(q^{n+1}\, v,q)+ \left(\frac{q^{1/4}}{v}\right)^{4-2\frac{h^\vee_G}{3}}L_G(q^{n+3/2}/v,q)\right)\bigg] \label{eq:conj} \end{align}
\emph{where}
\begin{equation} L_G(v,q) = \sum_{k,l\geq 0} h^G_{k,l}v^{2k}q^l\end{equation}
\emph{is a series involving only positive powers of $ v,q $.}\\

The coefficients $ h^G_{k,l} $ are uniquely determined by requiring that it satisfies \eqref{eq:conj}, where $ \mathbb{E}_{h_{G}{(1)}}(\epsilon_+,\tau) $ is the elliptic genus determined by modularity in Section \ref{sec:hilb}. We find that the function $ L_G(v,q) $ thus obtained satisfies the following additional properties:
\begin{enumerate}
\item The coefficients $ h^G_{k,l} $ are positive integers, which can be expressed as linear combinations of dimensions of irreducible representations of $ G $ with positive coefficients.
\item $ L_{G}(v,0) $ is the Hilbert series of the reduced moduli space of one $ G $-instanton (that is, the Hall-Littlewood index of the $ H^{(1)}_{G} $ theory).
\item $h^G_{0,1} = \text{dim}(G) + 1.$
\end{enumerate}

Remarkably, for the cases $ G = SU(3), SO(8), E_6 $ for which a series expansion of the Schur index is known, we also find that $ L_G(q^{1/4},q) $ coincides with $ \mathcal{I}_{H^{(1)}_G}(q^{1/2}) $, where $ \mathcal{I}_{H^{(1)}_G}(q) $ is the Schur index of the 4d SCFT $ H^{(1)}_G $! We discuss the various cases in turn.\newline

\noindent$\mathbf{G = SU(3).}$ In this case, setting $ h^\vee_{SU(3)} = 3 $ in Equation (\ref{eq:conj}) one finds that the right hand side collapses to just two terms, and one has the relation
\begin{equation} \mathbb{E}_{h_{SU(3)}^{(1)}}(\epsilon_+,\tau) = q^{-1/2}v^2 L_{SU(3)}(v,q) + q^{1/2} v^{-2}L_{SU(3)}(q^{1/2}/v,q).\label{eq:Hsu3}\end{equation}
Using
\begin{equation} L_{SU(3)}(v,q) = \sum_{k,l\geq 0}h^{SU(3)}_{k,l}v^{2k}q^l,\end{equation}
one sees that 
\begin{equation}h^{SU(3)}_{k,l} = b^{SU(3)}_{k+l,l} \end{equation}
are just the coefficients appearing in the upper right half of Table \ref{tb:SU3}. The two terms in Equation \ref{eq:Hsu3} correspond respectively to the upper right and  bottom left halves of the table, and we recover the results of Section \ref{sec:caseSU3}. In particular, Equation \eqref{eq:reln}, which we have verified to hold for the first $ 15 $ coefficients in the $ q $-expansion, is equivalent to the statement
\begin{equation} L_{SU(3)}(q^{1/4},q) = \mathcal{I}_{H_{SU(3)}^{(1)}}(q^{1/2}).\end{equation}
We note that $ H(v,q) $ is has an extremely simple form:
\begin{align} L_{SU(3)}(v,q) = (q,q)_{\infty}^{-8}&\bigg((1+8v^2+27 v^4+\mathcal{O}(v^6))+q+(1 + 8 v^2) q^2 + (1 + 27 v^4) q^3\nonumber\\& + (1 + 8 v^2 + 64 v^6) q^4 + (1 + 125 v^8) q^5+\mathcal{O}(q^6)\bigg),\end{align}
where $ (q,q)_\infty = q^{-1/24}\eta(q) $ is the $ q $-Pochhammer symbol. 
 This infinite series is an expansion of the following sum:
\begin{equation}L_{SU(3)}(v,q) = (q,q)_{\infty}^{-8}\left(\sum_{n\geq 1}\frac{n^3v^{2n-2}}{1-q^n}\right),\label{eq:summedup}\end{equation}
which also gives the following formula for the Schur index of the $ (A_1,D_4) $ Argyres-Douglas theory:
\begin{equation} \mathcal{I}_{(A_1,D_4)}(q) = (q^2,q^2)_\infty^{-8}\left(\sum_{n\geq 1}\frac{n^3q^{n-1}}{1-q^{2n}}\right).\end{equation}\\
Note that the $ q^0 $ term in Equation \eqref{eq:summedup} coincides with the Hilbert series of one SU(3) instanton, since 
\begin{equation} n^3 = \dim(n\cdot \text{Adj}_{SU(3)}).\end{equation}
We have also been able to resum Equation \eqref{eq:summedup} into the following closed form:
\begin{equation}L_{SU(3)}(v,q) = -(q,q)_{\infty}^{-8}\, q\frac{\partial}{\partial q}\log\left[\,\prod_{j=0}^\infty \left(q^{-v^{2j}j^3}(1-v^{2j-2}q^j)^{j^2}\right)\right].\end{equation}
It would be interesting to find a field theoretic interpretation for this formula.\newline\\

\noindent$\mathbf{G = SO(8).}$
We use Equation \eqref{eq:conj} with
\begin{equation} h^\vee_G = 6,\end{equation}
to solve for $ L_{SO(8)}(v,q) $, and find:
\begin{align} L_{SO(8)}&=  (1\! +\! 28 v^2\! +\! 300 v^4\! +\! 1925 v^6\! +\! 8918 v^8\! +\! 32928 v^{10}\! +\! 
   102816 v^{12}\! +\! 282150 v^{14}\!  +\! \mathcal{O}(v^{18}))+\nonumber\\ %1591876 v^18 + 
%   3383380 v^20 + 6782139 v^22 + 12931100 v^24 + 23609600 v^26 + 
%   41505024 v^28 + 70570332 v^30 + 116486397 v^32 + 187250700 v^34 + 
%   293916700 v^36 + 451511137 v^38 + 680159634 v^40 + 
%   1006454240 v^42 + 
%   1465100000 v^44) + 
&(29\! +\! 707 v^2\! +\! 6999 v^4\! +\! 42889 v^6\! +\! 
    193102 v^8\! +\! 699762 v^{10}\! +\! 2156994 v^{12} +\mathcal{O}(v^{16}))q+\nonumber\\ 
%    14426643 v^16 + 32695949 v^18 + 69214145 v^20 + 138303711 v^22 + 
%    263023852 v^24 + 479230612 v^26 + 841026516 v^28 + 
%    1427930028 v^30 + 2354143833 v^32 + 3780353031 v^34 + 
%    5928539771 v^36 + 9100359653 v^38 + 13699687386 v^40) q + 
&(463 + 
    9947 v^2 + 92391 v^4 + 544786 v^6 + 2392663 v^8 + 8526042 v^{10} + \mathcal{O}(v^{14})q^2+\nonumber\\
    %70015437 v^14 + 171120642 v^16 + 385916333 v^18 + 
    %813809945 v^20 + 1621184292 v^22 + 3075525313 v^24 + 
    %5592276868 v^26 + 9797701188 v^28 + 16611545223 v^30 + 
    %27353873412 v^32 + 43881026319 v^34 + 
    %68756302139 v^36) q^2 + 
 &   (5280 + 101850 v^2 + 894198 v^4 + 
    5096487 v^6 + 21888529 v^8  +\mathcal{O}(v^{12}))q^3+\nonumber\\ 
    %231399682 v^12 + 
%    618759263 v^14 + 1503039657 v^16 + 3373719626 v^18 + 
   % 7088023590 v^20 + 14078114559 v^22 + 26643109593 v^24 + 
%    48349757662 v^26 + 84569735778 v^28 + 143185964551 v^30 + 
   % 235505118881 v^32) q^3 +
&    (47897 + 842537 v^2 + 7032993 v^4 + 
    38869314 v^6  +\mathcal{O}(v^{10}))q^4+\nonumber\\
    % 565747296 v^{10} +\mathbb{O 1686960932 v^12 + 
%    4476323164 v^14 + 10810009350 v^16 + 24153984451 v^18 + 
%    50564191035 v^20 + 100140173334 v^22 + 189072266280 v^24 + 
%    342448917866 v^26 + 598019451678 v^28) q^4 
&+ \mathcal{O}(q^5).
%(366384 + 
%    5951291 v^2 + 47455968 v^4 + 254962664 v^6 + 1052692147 v^8 + 
%    3592900602 v^10 + 10609171946 v^12 + 27945594649 v^14 + 
%    67108405140 v^16 + 149290684613 v^18 + 311438534545 v^20 + 
%    615058864542 v^22 + 1158614498465 v^24) q^5 + (2450888 + 
%    37134593 v^2 + 284002506 v^4 + 1486358201 v^6 + 6029111166 v^8 + 
%    20319991572 v^10 + 59446988560 v^12 + 155495404502 v^14 + 
%    371392453929 v^16 + 822706607523 v^18 + 
%    1710462812995 v^20) q^6 + (14696012 + 209416671 v^2 + 
%    1541366304 v^4 + 7872021738 v^6 + 31403605496 v^8 + 
%    104584738275 v^10 + 303273841325 v^12 + 787960287976 v^14 + 
%    1872218220684 v^16) q^7 + (80404637 + 1084923467 v^2 + 
%    7707927825 v^4 + 38473447469 v^6 + 151083091789 v^8 + 
%    497484015483 v^10 + 1430458658949 v^12) q^8 + (406775833 + 
%    5226867454 v^2 + 35938269435 v^4 + 175553715735 v^6 + 
%    679168383970 v^8) q^9 + (1922660146 + 23639303064 v^2 + 
%    157662585462 v^4) q^10 + 8560163018 q^11
\label{eq:so8rescaled}
\end{align}

We have not been able to resum this series as in Equation \eqref{eq:summedup} for the $ G = SU(3) $ case.  In the following Table \ref{tb:SO8b} we display how the various copies of $ L_{SO(8)}(v,q) $ are intertwined to give the coefficients $ b^{SO(8)}_{k,l} $ of the elliptic genus of Table \ref{tb:SO8}.\newline\\
\noindent If we now take the limit $ v\to q^{1/4} $, we obtain
\begin{align} L_{SO(8)}(q^{1/4},q) &= 1\! +\! 28 q^{1/2}\! + 329 q\! +\! 2632 q^{3/2}\! +\! 16380 q^2\! +\! 85764 q^{5/2}\! +\! 
 393589 q^3\! +\! 1628548 q^{7/2} \nonumber\\&+\! 6190527 q^4\! +\! 21921900 q^{9/2}\! +\! 
 73070291 q^5\! + 231118384 q^{11/2} + 698128389 q^6 \nonumber\\&+ 
 2024433460 q^{13/2} + 5659730075 q^7 + 15309703500 q^{15/2} + 
 40191125219 q^8\nonumber\\
&+\mathcal{O}(q^{17/2}), \end{align}
in perfect agreement with the expression for the vacuum character of the $ \widehat{so(8)}_{-2} $ algebra given in Appendix C of \cite{Cordova:2016uwk}, which captures the Schur index of the $ \mathcal{H}^{(1)}_{SO(8)} $ theory with $ q \to q^{1/2} $.\newline\\

\noindent$\mathbf{G = E_6.}$ Proceeding as above for $ G = E_6 $, using $ h^\vee_{E_6} = 12$ we find:
\begin{align} L_{E_6}(v,q)& =(1\! +\! 78 v^2\! +\! 2430 v^4\! +\! 43758 v^6\! +\! 537966 v^8\! +\! 4969107 v^{10} \!+\!
   36685506 v^{12}\!+\!\mathcal{O}(v^{14}))\nonumber\\
   &  +(79 + 5512 v^2 + 157221 v^4 + 2644707 v^6 + 
    30843384 v^8 + 273370383 v^{10}\!+\!\mathcal{O}(v^{12}))\, q\nonumber\\
    &+ (3238 + 201292 v^2 + 5283549 v^4 + 
    83526287 v^6 + 928768412 v^8\!+\!\mathcal{O}(v^{10}))\, q^2 \nonumber\\&
    + (90911 + 5048576 v^2 + 122611239 v^4 + 
    1830734165 v^6\!+\!\mathcal{O}(v^{8}))\, q^3 \nonumber \\ 
    &+ (1956516 + 97616506 v^2 + 
    2205133146 v^4\!+\!\mathcal{O}(v^{6}))\, q^4+\mathcal{O}(q^5).\end{align}
We display the contribution of the various copies of $ L_{E_6}(v,q) $ to the elliptic genus of one $ E_6 $ string in Table \ref{tb:E6b} of the Appendix.\newline\\
\noindent In the limit $ v \to q^{1/4} $ we find:
\begin{align}L_{E_6}(q^{1/4},q)&= 1 + 78 q^{1/2} + 2509 q + 49270 q^{3/2} + 698425 q^2 + 
 7815106 q^{5/2} + 72903350 q^3 \nonumber\\
 &+ 587906696 q^{7/2} + 
 4204567965 q^4 + 27174694560 q^{9/2} + 161016744070 q^5 \nonumber\\
 &+ 
 884547201850 q^{11/2} + 4545922103619 q^6 + 
 22017119036040 q^{13/2} \nonumber\\
 &+ 101105788757675 q^7 + 
 442470577988634 q^{15/2} + 1853392626320950 q^8\nonumber\\
 &+\mathcal{O}(q^{17/2}),\end{align}
in perfect agreement with the first 17 terms in the expansion of the Schur index of the $ T_3 $ theory. \newline\\

\noindent$\mathbf{G = E_7.}$ We set $ h^\vee_{E_7} = 18 $ in Equation \eqref{eq:conj} and find:

\begin{align} L_{E_7}(v,q) &= (1 + 133 v^2 + 7371 v^4 + 238602 v^6+ 5248750 v^8 + 85709988 v^{10} + 
\mathcal{O}(v^{12})\nonumber\\ &+ (134 + 16283 v^2 + 835562 v^4 + 
    25353429 v^6 + 528271250 v^8+\mathcal{O}(v^{10})) q\nonumber\\ & + (9178 + 1014581 v^2 + 
    48250384 v^4 + 1375996758 v^6 +\mathcal{O}(v^8))q^2\nonumber\\ &+ (426533 + 42814809 v^2 + 
    1890508984 v^4 +\mathcal{O}(v^6))q^3\nonumber\\&+ (15077814 + 1374731795 v^2 +\mathcal{O}(v^4)) q^4\nonumber\\& + \mathcal{O}(v^5).\end{align}
We display the contribution of the various copies of $ L_{E_7}(v,q) $ to the elliptic genus of one $ E_7 $ string in Table \ref{tb:E7b} of the Appendix.\newline

\noindent In the limit $ v \to q^{1/4} $, this gives:
\begin{align}L_{E_7}(q^{1/4},q)&= 1 + 133 q^{1/2} + 7505 q + 254885 q^{3/2} + 6093490 q^2 + 
 112077998 q^{5/2} \nonumber\\& +1678245091 q^3 + 21264679635 q^{7/2} + 
 234433785700 q^4 + 2296105563465 q^{9/2} \nonumber\\&+ 20303111086038 q^5 + 
 164158274895703 q^{11/2} + 1226192258964745 q^6 \nonumber\\&+ 
 8533333787379775 q^{13/2} + 55718714973652300 q^7 + 
 343388965671840483 q^{15/2} \nonumber\\&+ 2007596030844978734 q^8+\mathcal{O}(q^{17/2}). \label{eq:schurE7}\end{align}
The Schur index of the $ H^{(1)}_{E_7} $ theory can be computed by the techniques of \cite{Gadde:2010en, Gaiotto:2012uq}. It is natural to conjecture that $ L_{E_7}(q^{1/4},q) $ coincides with the Schur index $ \mathcal{I}_{H^{(1)}_{E_7}}(q^{1/2}) $; we have verified this up to $ \mathcal{O}(q^{7/2}) $.\footnote{ We thank Wenbin Yan for providing us with code for computing the Schur index of the $ H^{(1)}_{E_7} $ theory.} \newline\\

\noindent$\mathbf{G = E_8.}$ We set $ h^\vee_{E_8} = 30 $ in Equation \eqref{eq:conj} and find:

\begin{align} L_{E_8}(v,q) &= (1 + 248 v^2 + 27000 v^4 + 1763125 v^6 + 79143000 v^8 + 
   2642777280 v^{10}+\mathcal{O}(v^{12}))\nonumber\\&+
   (249 + 57877 v^2 + 
    5943753 v^4 + 368338125 v^6 + 15776893240 v^8 + 
    +\mathcal{O}(v^{10}))q\nonumber\\& + (31373 + 6815877 v^2 + 
    659761497 v^4 + 38811914750 v^6 +\mathcal{O}(v^{8})) q^2\nonumber\\&+ (2666126 + 
    539686750 v^2 + 49211333622 v^4 +\mathcal{O}(v^{6}) ) q^3 \nonumber\\&+ (171724751 + 
    32299833627 v^2+\mathcal{O}(v^4)) q^4+\mathcal{O}(q^{5}).\end{align}
We display the contribution of the various copies of $ L_{E_8}(v,q) $ to the elliptic genus of one $ E_8 $ string in Table \ref{tb:E8b} of the Appendix.\newline

\noindent In the limit $ v \to q^{1/4} $, we find:
\begin{align}L_{E_8}(q^{1/4},q) &=1 + 248 q^{1/2} + 27249 q + 1821002 q^{3/2} + 85118126 q^2 + 
 3017931282 q^{5/2} \nonumber\\&+ 85616292063 q^3 + 2018221136220 q^{7/2} + 
 40655908880933 q^4 \nonumber\\&+ 715118758926278 q^{9/2} + 
 11171613223900451 q^5 + 157140768554366660 q^{11/2} \nonumber\\&+ 
 2012705625856030235 q^6 + 23694966834840175472 q^{13/2}\nonumber\\& + 
 258431445654249301583 q^7 + 2628885836402784435498 q^{15/2} \nonumber\\&+ 
 25087207661618093562092 q^8 + \mathcal{O}(q^{17/2})\label{eq:schurE8}\end{align}
We conjecture that this expression agrees with the Schur index $ \mathcal{I}_{H^{(1)}_{E_8}}(q^{1/2}) $; we have checked up to $ \mathcal{O}(q^{5/2}) $ that the two quantities agree.\footnote{ We thank Wenbin Yan for providing us with code to compute the Schur index of the $ H^{(1)}_{E_8} $ theory.}\newline\\

\noindent$\mathbf{G = F_4.}$ We set $ h^\vee_{F_4} = 9 $ in Equation \eqref{eq:conj} and find:

\begin{align} L_{F_4}(v,q) &=(1 + 52 v^2 + 1053 v^4 + 12376 v^6 + 100776 v^8 + 627912 v^{10} + 
   3187041 v^{12} +\mathcal{O}(v^{14})\nonumber\\&+ (53 + 2432 v^2 + 44980 v^4 + 495872 v^6 + 
    3856722 v^8 + 23235328 v^{10} +\mathcal{O}(v^{12}))q\nonumber\\&+ (1483 + 59996 v^2 + 1023464 v^4 + 
    10670660 v^6 + 79721160 v^8 +\mathcal{O}(v^{10}))q^2+\nonumber\\&+ (28771 + 1034880 v^2 + 16410602 v^4 + 
    162744192 v^6 +\mathcal{O}(v^{8}))q^3+\nonumber\\&+ (432526 + 13979228 v^2 + 
    207409930 v^4 +\mathcal{O}(v^{6}))q^4+\nonumber\\&+\mathcal{O}(q^5).\end{align}

We display the contribution of the various copies of $ L_{F_4}(v,q) $ to the elliptic genus of one $ F_4 $ string in Table \ref{tb:F4b} of the Appendix.\newline

\noindent In the limit $ v \to q^{1/4} $ we find:
\begin{align} L_{F_4}(q^{1/4},q) &= 1 + 52 q^{1/2} + 1106 q + 14808 q^{3/2} + 147239 q^2 + 
 1183780 q^{5/2} + 8095998 q^3 \nonumber\\&+ 48688888 q^{7/2} + 263508351 q^4 + 
 1305275544 q^{9/2} + 5993906570 q^5\nonumber\\& + 25771913376 q^{11/2} + 
 104583612240 q^6 + 403149160444 q^{13/2} \nonumber\\&+ 1484121980708 q^7 + 
 5241010219736 q^{15/2} + 17821566681691 q^8+\mathcal{O}(q^{17/2}).\label{eq:F4inlimit}
\end{align}

It is interesting to remark that Schur indices can be identified with vacuum characters of chiral algebras \cite{Beem:2013sza}. The properties of the functions $L_{G}(v,q)$ hint at a similar relation among (non-supersymmetric) chiral algebras and 2d $(0,4)$ BPS strings of 6d $(1,0)$ theories. Understanding the details of such relation goes beyond the scope of the present work and we leave it to future work \cite{DZL2}.

\hyphenation{Ne-me-schan-sky}

\section*{Acknowledgments} 
We thank C. Cordova, A. Gadde, A. Hanany, J.J. Heckman, S. Kim, K. Lee, E. Looijenga, N. Mekareeya, D. R. Morrison, L. Rastelli, S. Shakirov, S.-H. Shao, J. Song, W. Yan, A. Zaffaroni and especially T. Dumitrescu, A. Klemm and C. Vafa for many enlightening discussions. We gratefully acknowledge support from the Simons Center for Geometry and Physics, Stony Brook University at which part of the research for this paper was performed. We also thank the Institute Henri Poincar\'e and the ENS (Paris) for hospitality at different stages of this work. MdZ also thanks the organizers of the 2016 Amsterdam String Workshop for hospitality. This project has received funding from the European Union's Horizon 2020 research and innovation programme under the Marie Sklodowska-Curie grant agreement No 708045. The work of MdZ is supported by NSF grant PHY-1067976.

\appendix

\section{Explicit expressions for the elliptic genera}
\label{sec:appdata}
In this Appendix we collect the results of our computations for the elliptic genera of one $ G $-instanton, for $ G = SU(3), SO(8), F_4,E_6, E_7, E_8 $, as fixed by the modularity constraints discussed in Section \ref{sec:hilb}. In the first part of the Appendix we provide the explicit expressions for the numerator of
\begin{align} \widetilde{\mathbb{E}}_{h_{G}^{(1)}}(2\epsilon_+,m_\alpha,\tau) = \frac{\mathcal{N}_{G,1}(2\epsilon_+,m_\alpha,\tau)}{\eta(\tau)^{4(h^\vee_G-1)}\displaystyle{\prod_{\alpha\in \Delta_+}\varphi_{-1,1/2}( 2\epsilon_+ + m_{\alpha},\tau)\varphi_{-1,1/2}( 2\epsilon_+ - m_{\alpha},\tau)}},\label{eq:ansatz2}\end{align}
 in the limit $ m_{\alpha}\to 0 $. In the second part of the Appendix we provide extensive tables of series coefficients of the elliptic genus \eqref{eq:ansatz2}, expanded in powers of $ v^2 = e^{4\pi i  \epsilon_+} $ and $ q v^{-2} = e^{2\pi i (\tau-2\epsilon_+)} $.
 
 \subsection{Explicit form of the numerator terms}
  We write the expressions for the numerators of the Ansatz in terms of the Jacobi forms $ \phi_{-2,1}(2\epsilon_+,\tau),\phi_{0,1}(2\epsilon_+,\tau), \phi_{0,3/2}(2\epsilon_+,\tau) $ and of the Eisenstein series $ E_4(\tau), E_6(\tau)$. 

For conciseness, in what follows we drop the arguments of these functions and also write $\mathcal{N}_{G,1}(2\epsilon_+,0,\tau) = \mathcal{N}_{G,1}$. We find the following results:

\begin{align} \mathcal{N}_{SU(3),1} = \frac{1}{24}\phi_{-2,1}(E_4\phi_{-2,1}^2-\phi_{0,1}^2).\hspace{3.4in}\end{align}

\begin{equation}\mathcal{N}_{SO(8),1}=\frac{1}{144} \phi_{-2,1}^7\phi_{0,3/2} (2 E_6 \phi_{-2,1}^3 - 9 E_4 \phi_{-2,1}^2 \phi_{0,1} + 7 \phi_{0,1}^3).\hspace{2.3in}\end{equation}
\begin{flushleft}\begin{align}\mathcal{N}_{F_4,1}=\frac{1}{746496}\phi_{-2,1}^{16} \bigg(&\phi_{-2,1}^6 \phi_{0,1} \left(56 E_6^2-81 E_4^3\right)+45 E_4^2 E_6 \phi_{-2,1}^7+486 E_4^2 \phi_{-2,1}^4 \phi_{0,1}^3\nonumber\\
-&366 E_4 E_6 \phi_{-2,1}^5 \phi_{0,1}^2-453 E_4 \phi_{-2,1}^2 \phi_{0,1}^5+209 E_6 \phi_{-2,1}^3 \phi_{0,1}^4+104 \phi_{0,1}^7\bigg).\hspace{0.3in}\end{align}
\begin{align}\mathcal{N}_{E_6,1}=\frac{1}{23887872}\phi_{-2,1}^{25} \phi_{0,3/2} &\bigg(9 \phi_{-2,1}^8 \left(23 E_4^4-64 E_4 E_6^2\right)+4 \phi_{-2,1}^6 \phi_{0,1}^2 \left(512 E_6^2-1845 E_4^3\right)\nonumber\\
&\!\!\!\!+4656 E_4^2 E_6 \phi_{-2,1}^7 \phi_{0,1}+23010 E_4^2 \phi_{-2,1}^4 \phi_{0,1}^4-14880 E_4 E_6 \phi_{-2,1}^5 \phi_{0,1}^3\nonumber\\
&-18564 E_4 \phi_{-2,1}^2 \phi_{0,1}^6+7280 E_6 \phi_{-2,1}^3 \phi_{0,1}^5+4199 \phi_{0,1}^8\bigg).\end{align}
\begin{align}&\mathcal{N}_{E_7,1}=\frac{1}{2972033482752} \phi_{-2,1}^{46}  \phi_{0,3/2}\bigg(12 (6399 E_4^5 E_6 - 
      10528 E_4^2 E_6^3) \phi_{-2,1}^{13} \nonumber\\
      &+ (1472256 E_4^3 E_6^2\!-\!151875 E_4^6 \!-\! 
      60416 E_6^4) \phi_{-2,1}^{12} \phi_{0,1} 
      - 180 E_4 E_6 (26739 E_4^3 - 8704 E_6^2) \phi_{-2,1}^{11} \phi_{0,1}^2 \nonumber\\
      &+ 
   18 E_4^2 (258993 E_4^3 - 627040 E_6^2) \phi_{-2,1}^{10} \phi_{0,1}^3 + 
   280 E_6 (106623 E_4^3 - 5680 E_6^2) \phi_{-2,1}^9 \phi_{0,1}^4 \nonumber\\
   &- 
   567 E_4 (45667 E_4^3 - 29056 E_6^2) \phi_{-2,1}^8 \phi_{0,1}^5 - 51471000 E_4^2 E_6 \phi_{-2,1}^7 \phi_{0,1}^6 \nonumber\\
   &+ 
   228 (217503 E_4^3 - 25648 E_6^2) \phi_{-2,1}^6 \phi_{0,1}^7 + 31668516 E_4 E_6 \phi_{-2,1}^5 \phi_{0,1}^8 \nonumber\\
   &- 
   40739325 E_4^2 \phi_{-2,1}^4 \phi_{0,1}^9 - 6249100 E_6 \phi_{-2,1}^3 \phi_{0,1}^{10} + 14827410 E_4 \phi_{-2,1}^2 \phi_{0,1}^{11} - 
   1964315 \phi_{0,1}^{13}\bigg).\end{align}
\begin{align}&\mathcal{N}_{E_8,1} =  \frac{\phi_{-2,1}^{91} \phi_{0,3/2}}{92010239818739402932224} \bigg(2877420 E_4 E_6 \phi_{-2,1}^{11} \phi_{0,1}^{12} \left(333172971 E_4^3-32233088 E_6^2\right)+\nonumber\\
&29638480 E_6 \phi_{-2,1}^9 \phi_{0,1}^{14}\!\left(481040 E_6^2\!-\!23057271 E_4^3\right)\!+\!820244934 E_4 \phi_{-2,1}^8 \phi_{0,1}^{15} \left(539755 E_4^3\!-\!134912 E_6^2\right)\nonumber\\
&-115263177 \phi_{-2,1}^6 \phi_{0,1}^{17} \left(1982439 E_4^3-108880 E_6^2\right)+278766529364394 E_4^2 E_6 \phi_{-2,1}^7 \phi_{0,1}^{16}\nonumber\\
&+71015153903967 E_4^2 \phi_{-2,1}^4 \phi_{0,1}^{19}-33264 E_4 E_6 \phi_{-2,1}^{17} \phi_{0,1}^6 (2345906637 E_4^6-3740993360 E_4^3 E_6^2\nonumber\\
&+159614976 E_6^4)\!+\!1716 E_6 \phi_{-2,1}^{15} \phi_{0,1}^8 \left(202247657541 E_4^6\!-\!142801148160 E_4^3 E_6^2\!+\!1237560320 E_6^4\right)\nonumber\\
&-858 E_4 \phi_{-2,1}^{14} \phi_{0,1}^9 \left(193760793603 E_4^6-715814423280 E_4^3 E_6^2+41243970560 E_6^4\right)\nonumber\\
&+806 \phi_{-2,1}^{12} \phi_{0,1}^{11} \left(481766368221 E_4^6-828424091520 E_4^3 E_6^2+10039040000 E_6^4\right)\nonumber\\
&-531960 E_4^2 E_6 \phi_{-2,1}^{13} \phi_{0,1}^{10} \left(1457598645 E_4^3-418811552 E_6^2\right)\nonumber\\
&-12284370 E_4^2 \phi_{-2,1}^{10} \phi_{0,1}^{13} \left(43165017 E_4^3-31257376 E_6^2\right)+6 E_4 E_6 \phi_{-2,1}^{23} (73362915 E_4^9\nonumber\\
&-1968261120 E_4^6 E_6^2+2153134080 E_4^3 E_6^4-80478208 E_6^6)\nonumber\\
&+4 E_6 \phi_{-2,1}^{21} \phi_{0,1}^2 (-47714905305 E_4^9+408586731840 E_4^6 E_6^2\nonumber\\
&-158043820032 E_4^3 E_6^4+1049559040 E_6^6)+3 E_4 \phi_{-2,1}^{20} \phi_{0,1}^3 (46391070465 E_4^9\nonumber\\
&-1684505859840 E_4^6 E_6^2+1886854717440 E_4^3 E_6^4-69807374336 E_6^6)\nonumber\\
&-7 \phi_{-2,1}^{18} \phi_{0,1}^5 \left(568895485455 E_4^9\!-\!9050573631168 E_4^6 E_6^2\!+\!4143604654080 E_4^3 E_6^4\!-\!29575086080 E_6^6\right)\nonumber\\
&+18 E_4^2 E_6 \phi_{-2,1}^{19} \phi_{0,1}^4 \left(405308228085 E_4^6-1442655164160 E_4^3 E_6^2+199133118464 E_6^4\right)\nonumber\\
&+99 E_4^2 \phi_{-2,1}^{16} \phi_{0,1}^7 \left(384852307779 E_4^6-2932139934720 E_4^3 E_6^2+519389409280 E_6^4\right)\nonumber\\
&+3 E_4^2 \phi_{-2,1}^{22} \phi_{0,1} \left(-258037569 E_4^9\!+\!28032966000 E_4^6 E_6^2\!-\!80889477120 E_4^3 E_6^4\!+\!10505617408 E_6^6\right)\nonumber\\
&-60306155259108 E_4 E_6 \phi_{-2,1}^5 \phi_{0,1}^{18}\!-\!12164845368165 E_4 \phi_{-2,1}^2 \phi_{0,1}^{21}\!+\!5355592300450 E_6 \phi_{-2,1}^3 \phi_{0,1}^{20}\nonumber\\
&+881510533925 \phi_{0,1}^{23}\bigg).\end{align}\end{flushleft}

\subsection{Tables of coefficients}
In Tables \ref{tb:SU3} - \ref{tb:E8} we display several numerical coefficients of the series expansion of the elliptic genera of the theories $ h^{(1)}_{G} $ for  $ G = SU(3),SO(8),E_6,E_7,E_8 $. For $ G = SU(3), F_4 $, the elliptic genera display the following symmetry:
\begin{equation}\mathbb{E}_{h_{G}^{(1)}}(\tau-2\epsilon_+,\tau) =q^{\frac{1}{2}(\frac{h^\vee_G}{3}-1)}v^{2(1-\frac{h^\vee_G}{3})}\mathbb{E}_{h_{G}^{(1)}}(2\epsilon_+,\tau);\end{equation}
while on the other for $ G = SO(8),E_6,E_7,E_8 $ one has:
\begin{equation}\mathbb{E}_{h_{G}^{(1)}}(\tau-2\epsilon_+,\tau) = -q^{\frac{1}{2}(\frac{h^\vee_G}{3}-1)}v^{2(1-\frac{h^\vee_G}{3})}\mathbb{E}_{h_{G}^{(1)}}(2\epsilon_+,\tau).\end{equation}
Also, the leading order term in the $ q $-expansion of $ \mathbb{E}_{h_{G}^{(1)}}(2\epsilon_+,\tau) $ is proportional to $ q^{-4\frac{h^\vee_G-1}{6}} $.\newline

We therefore find it convenient to rescale the elliptic genus and rewrite it in terms of the variables $ p=v^2,\tilde{p}=q\, v^{-2} $ as follows:
\begin{equation} \mathbb{E}_{h_{G}^{(1)}}(2\epsilon_+,\tau) \rightarrow \mathcal{E}_{G}(p,\tilde p) = q^{\frac{h^\vee_G-1}{6}}v^{1-\frac{h^\vee_G}{3}}\mathbb{E}_{h_{G}^{(1)}}(2\epsilon_+,\tau). \end{equation}
The rescaled elliptic genus then has the following expansion:
\begin{equation} \mathcal{E}_G(p,\tilde p) = \sum_{k,l\geq 0} b^G_{k,l}\, p^k{\tilde p}^l,\end{equation}
where 
\begin{equation} b^G_{k,l}= b^G_{l,k}\text{ for $G=SU(3),F_4$,}\qquad\text{ and }\qquad b^G_{k,l}=-b^G_{l,k}\text{ for $G=SO(8),E_6,E_7,E_8$.}\end{equation}
For instance,
\begin{equation} \mathcal{E}_{SU(3)}(p,\tilde p) = (p + 8 p^2 +\mathcal{O}(p^3))+\tilde p(1+9p^2+\mathcal{O}(p^3))+\tilde p^2(8+9 p +\mathcal{O}(p^3))+\mathcal{O}(\tilde p^3).\end{equation}
In Tables \ref{tb:SU3} - \ref{tb:E8} we display the expansion coefficients $ b^G_{k,l} $ for all $ G $.\newline

\noindent Finally, in Tables \ref{tb:SO8b}-\ref{tb:E8b} we display portions of Tables \ref{tb:SO8}--\ref{tb:E8}, now including information about how the series coefficients $ b^G_{k,l} $ arise as sum of contributions from the different terms in Equation \eqref{eq:conj}, which we repeat here for convenience:
\begin{align}\mathbb{E}_{h_{G}^{(1)}}(\epsilon_+, \tau)&=v^{\frac{h^\vee_G}{3}-1}\sum_{n\geq 0}q^{2n}\bigg[\left(\frac{v}{q^{1/4}}\right)^{\frac{2h^\vee_G}{3}}L_G(q^n\, v; q)-(-1)^{h^\vee_G} \left(\frac{q^{1/4}}{v}\right)^{\frac{2h^\vee_G}{3}}H_G(q^{n+1/2}/v; q)\nonumber\\
&+(1+(-1)^{h^\vee_G})q^{\frac{1}{2}\left(\frac{h^\vee_G}{3}+1\right)}\left(\left(\frac{v}{q^{1/4}}\right)^2L_G(q^{n+1/2}\,v,q)-\left(\frac{q^{1/4}}{v}\right)^2L_G(q^{n+1}/v,q)\right)\nonumber\\
&-q^2\left(-(-1)^{h^\vee_G}\left(\frac{v}{q^{1/4}}\right)^{4-2\frac{h^\vee_G}{3}}L_G(q^{n+1}\, v,q)+ \left(\frac{q^{1/4}}{v}\right)^{4-2\frac{h^\vee_G}{3}}L_G(q^{n+3/2}/v,q)\right)\bigg]. \label{eq:conj2} \end{align}

Each entry in Tables \ref{tb:SO8b}-\ref{tb:E8b} is schematically written as a sum of integers with subscripts; the two subscripts $ m,n $ indicate that the integer arises from the $ m $-th occurrence of the function $ L_G(v,q) $ in the $ n $-th term of the sum in Equation \eqref{eq:conj2}. Thus, for example, the $ k =5, l=6 $ entry in Table \ref{tb:SO8b} for $ G = SO(8) $,

\begin{equation} 2\cdot5280_{3, 0} + 2\cdot29_{3, 1} + 28_{5, 0}\end{equation}

indicates that $ b^{SO(8)}_{5,6} = 10646 $ arises as the sum of three terms: $ 2\cdot 5280_{3,0} $ comes from the $ L_{SO(8)}(q^{n+1/2}v,q) $ term in Equation \eqref{eq:conj2}, with $ n = 0 $; $ 2\cdot 29_{3,1} $ also comes from the same term, with $ n = 1 $; and $ 28_{5,0} $ comes from the $ L_{SO(8)}(q^{n+1}v,q) $ term, with $ v = 5 $.

\begin{sidewaystable}[p!]
\caption{Expansion coefficients $ b^{SU(3)}_{k, l} $ for one $SU(3)$ instanton.}
{\label{tb:SU3}\begin{center}\begin{tiny}
\begin{tabular}{c| lllllllllllllllll }\,\,\,\,\,\,\,\,l\,\,\,\,\,\,\,k&0&1&2&3&4&5&6&7&8&9&10&11&12&13&14&15\\
\hline
0&0& 1& 8& 27& 64& 125& 216& 343& 512& 729& 1000& 1331& 1728& 2197& 
  2744& 3375\\
1&1& 0& 9& 64& 216& 512& 1000& 1728& 2744& 4096& 5832& 8000& 10648& 
  13824& 17576& 21952\\
2&8& 9& 0& 53& 360& 1188& 2816& 5500& 9504& 15092& 22528& 32076& 
  44000& 58564& 76032& 96668\\
3&27& 64& 53& 0& 245& 1600& 5211& 12288& 24000& 41472& 65856& 98304& 
  139968& 192000& 255552& 331776\\
4&64& 216& 360& 245& 0& 971& 6168& 19818& 46528& 90750& 156816& 
  249018& 371712& 529254& 726000& 966306\\
5&125& 512& 1188& 1600& 971& 0& 3435& 21312& 67716& 158208& 308125& 
  532224& 845152& 1261568& 1796256& 2464000\\
6&216& 1000& 2816& 5211& 6168& 3435& 0& 11139& 67800& 213219& 495872& 
  964000& 1664280& 2642472& 3944448& 5616216\\
7&343& 1728& 5500& 12288& 19818& 21312& 11139& 0& 33667& 201536& 
  628074& 1454080& 2821500& 4867776& 7727447& 11534336\\
8&512& 2744& 9504& 24000& 46528& 67716& 67800& 33667& 0& 96004& 
  566496& 1750815& 4036096& 7816125& 13474296& 21384335\\
9&729& 4096& 15092& 41472& 90750& 158208& 213219& 201536& 96004& 0& 
  260564& 1518016& 4656339& 10690048& 20660750& 35586432\\
10&1000& 5832& 22528& 65856& 156816& 308125& 495872& 628074& 566496& 
  260564& 0& 677704& 3903616& 11890854& 27192832& 52450625\\
11&1331& 8000& 32076& 98304& 249018& 532224& 964000& 1454080& 1750815& 
  1518016& 677704& 0& 1698120& 9681344& 29302047& 66760704
\end{tabular}\end{tiny}
\end{center}}
\caption{Expansion coefficients $ b^{SO(8)}_{k, l} $ for one $SO(8)$ instanton.}
{\label{tb:SO8}\begin{center}\begin{tiny}
\begin{tabular}{c| lllllllllllll }\,\,\,\,\,\,\,\,l\,\,\,\,\,\,\,k&0&1&2&3&4&5&6&7&8&9&10&11\\
\hline
0&0& 0& 1& 28& 300& 1925& 8918& 32928& 102816& 282150& 698775& 
  1591876\\%& 3383380\\%& 6782139\\
1& 0& 0& 0& 29& 707& 6999& 42889& 193102& 699762& 2156994& 5864958& 
  14426643\\%& 32695949\\%& 69214145\\
2 &-1& 0& 0& 2& 464& 9947& 92391& 544786& 2392663& 8526042& 25972362& 
  70015437\\%& 171120642\\%& 385916333\\
3 &-28& -29& -2& 0& 58& 5365& 101850& 894198& 5096487& 21888529& 
  76804254& 231399682\\%& 618759263\\%& 1503039657\\
4& -300& -707& -464& -58& 0& 928& 49775& 843165& 7032993& 38869314& 
  163555964& 565747296\\%& 1686960932\\%& 4476323164\\
5 &-1925& -6999& -9947& -5365& -928& 0& 10646& 391587& 5965996& 
  47459818& 254962664& 1052692147\\%& 3592900602\\%& 10609171946\\
6 &-8918& -42889& -92391& -101850& -49775& -10646& 0& 97429& 2702949& 
  37329322& 284088584& 1486376037\\%& 6029111166\\%& 20319991572\\
7 &-32928& -193102& -544786& -894198& -843165& -391587& -97429& 0& 
  753333& 16753135& 211306917& 1542462875\\%& 7872407942\\%& 31403671352\\
8 &-102816& -699762& -2392663& -5096487& -7032993& -5965996& -2702949& 
-753333& 0& 5100348& 94814881& 1099832018\\%& 7718213190\\%& 38478234720\\
9&-282150& -2156994& -8526042& -21888529& -38869314& -47459818& 
-37329322& -16753135& -5100348& 0& 30977975& 496225009\\%& 5327733313\\%&36016902261\\
10& -698775& -5864958& -25972362& -76804254& -163555964& -254962664& 
-284088584& -211306917& -94814881& -30977975& 0& 171759770\\%& 2425495272\\%& 24244496105\\
11 &-1591876& -14426643& -70015437& -231399682& -565747296&
-1052692147& -1486376037& -1542462875& -1099832018& -496225009&
-171759770& 0\\%& 880831616\\%& 11160571613
\end{tabular}\end{tiny}
\end{center}}

\caption{Expansion coefficients $ b^{F_4}_{k, l} $ for one $F_4$ instanton.}
\begin{center}{\begin{tiny}
\begin{tabular}{c| llllllllllllll |}
\,\,\,\,\,\,\,\,l\,\,\,\,\,\,\,k&0&1&2&3&4&5&6&7&8&9&10&11\\
\hline
0&0& 0& 0& 1& 52& 1053& 12376& 100776& 627912& 3187041& 13748020& 
  51949755\\
  1& 0& 0& 0& 0& 53& 2432& 44980& 495872& 3856722& 23235328& 114994308& 
  486551936\\
2& 0& 0& 0& 0& 0& 1484& 59996& 1023464& 10670660& 79721160& 466152308& 
  2255183112\\
3& 1& 0& 0& 0& -1& 0& 28824& 1034880& 16410602& 162744192& 1172590352& 
  6672705920\\
4& 52& 53& 0& -1& 0& -53& 0& 434010& 13979280& 207409930& 1965196116& 
  13697411140\\
5& 1053& 2432& 1484& 0& -53& 0& -1484& 0& 5385595& 157185280& 
  2194218416& 19935570304\\
6& 12376& 44980& 59996& 28824& 0& -1484& -104& -28824& 0& 57257172& 
  1528151788& 20167290000\\
7& 100776& 495872& 1023464& 1034880& 434010& 0& -28824& -4864& -434010&
   0& 535734136& 13181359488\\
8& 627912& 3856722& 10670660& 16410602& 13979280& 5385595& 
  0& -434010& -119992& -5386648& 0& 4498964466\\
9& 3187041& 23235328& 79721160& 162744192& 207409930& 157185280& 
  57257172& 0& -5386648& -2069760& -57302152& 0\\
10& 13748020& 114994308& 466152308& 1172590352& 1965196116& 2194218416& 
  1528151788& 535734136& 0& -57302152& -27958560& -536757600\\
11& 51949755&\! 486551936&\! 2255183112&\! 6672705920&\! 13697411140&\!
  19935570304&\! 20167290000&\! 13181359488&\! 4498964466&\! 
  0&\! -536757600&\! -314370560
\end{tabular}\end{tiny}\label{tb:F4}}
\end{center}
\end{sidewaystable}
\begin{sidewaystable}[p!]
\caption{Expansion coefficients $ b^{E_6}_{k, l} $ for one $E_6$ instanton.}
\begin{center}{\begin{tiny}\label{tb:E6}
\begin{tabular}{c|lllllllllllllllllll |}
\,\,\,\,\,\,\,\,l\,\,\,\,\,\,\,k&0&1&2&3&4&5&6&7&8&9&10&11\\
\hline
0&0& 0& 0& 0& 1& 78& 2430& 43758& 537966& 4969107& 36685506& 
  225961450\\
1& 0& 0& 0& 0& 0& 79& 5512& 157221& 2644707& 30843384& 273370383& 
  1953225274\\
2& 0& 0& 0& 0& 0& 0& 3239& 201292& 5283549& 83526287& 928768412& 
  7930066131\\
3& 0& 0& 0& 0& 0& -1& 0& 90990& 5048576& 122611239& 1830734165& 
  19486798202\\
4& -1& 0& 0& 0& 0& 2& -79& 0& 1959755& 97616584& 2205133146& 
  31229737630\\
5& -78& -79& 0& 1& -2& 0& 158& -3083& 0& 34418248& 1549515786& 
  32718288609\\
6& -2430& -5512& -3239& 0& 79& -158& 0& 6400& -79966& 4860& 512609433& 
  21005763831\\
7& -43758& -157221& -201292& -90990& 0& 3083& -6400& 0& 
  176468& -1557171& 314442& 6653537113\\
8& -537966& -2644707& -5283549& -5048576& -1959755& 0& 79966& -176468& 
  0& 3718218& -24321096& 10567098\\
9& -4969107& -30843384& -83526287& -122611239& -97616584& -34418248& 
-4860& 1557171& -3718218& 0& 63787920& -317376265\\
10& -36685506& -273370383& -928768412& -1830734165& -2205133146& 
-1549515786& -512609433& -314442& 24321096& -63787920& 0& 927602282\\
11& -225961450& -1953225274& -7930066131& -19486798202& -31229737630& 
-32718288609& -21005763831& -6653537113& -10567098& 
  317376265& -927602282& 0
\end{tabular}\end{tiny}}
\end{center}
\caption{Expansion coefficients $ b^{E_7}_{k, l} $ for one $E_7$ instanton.}
\begin{center}{\begin{tiny}\label{tb:E7}
\begin{tabular}{c| lllllllllllllll |}\,\,\,\,\,\,\,\,l\,\,\,\,\,\,\,k&0&1&2&3&4&5&6&7&8&9&10&11&12\\
\hline
0&0& 0& 0& 0& 0& 0& 1& 133& 7371& 238602& 5248750& 85709988&1101296924\\
1&0& 0& 0& 0& 0& 0& 0& 134& 16283& 835562& 25353429& 528271250&8241562056\\
2&0& 0& 0& 0& 0& 0& 0& 0& 9179& 1014581& 48250384& 1375996758&27238257500\\
3&0& 0& 0& 0& 0& 0& 0& -1& 0& 426667& 42814809& 1890508984&50809770626\\
4&0& 0& 0& 0& 0& 0& 0& 0& -134& 0& 15086993& 1374731928&56496374025\\
5&0& 0& 0& 0& 0& 0& 0& 0& 0& -9179& 0& 431738223&35790797572\\
6&-1& 0& 0& 0& 0& 0& 0& 2& 0& -133& -426667& 0&10396215666\\
7&-133& -134& 0& 1& 0& 0& -2& 0& 268& 266& -16283& -15086993&0\\
8&-7371& -16283& -9179& 0& 134& 0& 0& -268& 0& 18358& 32566& -999839&-431738223\\
9&-238602& -835562& -1014581& -426667& 0& 9179& 133& -266& -18358& 0&   853334& 2021791&-41143685\\
10&-5248750& -25353429& -48250384& -42814809& -15086993& 0& 426667& 16283& -32566& -853334& 0& 30173986&84794056\\
11&-85709988& -528271250& -1375996758& -1890508984& -1374731928& 
-431738223& 0& 15086993& 999839& -2021791& -30173986& 0&863476446
\end{tabular}\end{tiny}}
\end{center}
\caption{Expansion coefficients $ b^{E_8}_{k, l} $ for one $E_8$ instanton.}
\begin{center}{\begin{tiny}
\begin{tabular}{c| lllllllllllllllllll}\,\,\,\,\,\,\,\,l\,\,\,\,\,\,\,k&0&1&2&3&4&5&6&7&8&9&10&11&12&13&14&15&16&17&18\\
\hline
0&0& 0& 0& 0& 0& 0& 0& 0& 0& 0& 1& 248& 27000& 1763125& 79143000& 
  2642777280& 69176971200& 1473701482500&26284473168750\\
1& 0& 0& 0& 0& 0& 0& 0& 0& 0& 0& 0& 249& 57877& 5943753& 368338125& 
  15776893240& 505168052220& 12734438533800&262256288252820\\
2& 0& 0& 0& 0& 0& 0& 0& 0& 0& 0& 0& 0& 31374& 6815877& 659761497& 
  38811914750& 1587614120010& 48797287538220&1186023441073800\\
3& 0& 0& 0& 0& 0& 0& 0& 0& 0& 0& 0& -1& 0& 2666375& 539686750& 
  49211333622& 2749381821611& 107502041857010&3175264298363625\\
4& 0& 0& 0& 0& 0& 0& 0& 0& 0& 0& 0& 0& -249& 0& 171756125& 32299833875&
   2773706709375& 147262874984139&5509358828372000\\
5& 0& 0& 0& 0& 0& 0& 0& 0& 0& 0& 0& 0& 0& -31374& 0& 8931266291& 
  1557460109793& 125962776469125&6360231933621889\\
6& 0& 0& 0& 0& 0& 0& 0& 0& 0& 0& 0& 0& 0& -248& -2666375& 0& 
  389911527335& 62985386345455&4799599023251403\\
7& 0& 0& 0& 0& 0& 0& 0& 0& 0& 0& 0& 0& 0& 0& -57877& -171756125& 0& 
  14678263182211&2196144945999203\\
8& 0& 0& 0& 0& 0& 0& 0& 0& 0& 0& 0& 0& 0& 0& 0& -6815877& -8931266291& 
  0&485801431557625\\
9& 0& 0& 0& 0& 0& 0& 0& 0& 0& 0& 0& 0& 0& 0& 
  0& -27000& -539686750& -389911527335&0\\
10& -1& 0& 0& 0& 0& 0& 0& 0& 0& 0& 0& 2& 0& 0& 0& 
  0& -5943753& -32299833875&-14678263182211\\
11& -248& -249& 0& 1& 0& 0& 0& 0& 0& 0& -2& 0& 498& 496& 0& 0& 
  0& -65976149&-1557460109793
\end{tabular}\end{tiny}
\label{tb:E8}}
\end{center}
\end{sidewaystable}

\begin{sidewaystable}[p!]
\caption{Expansion coefficients $ b^{SO(8)}_{k, l} $ for one $SO(8)$ instanton (detailed version).}
{\label{tb:SO8b}\begin{center}\begin{tiny}
\begin{tabular}{c| lllllllllllll }\,\,\,\,\,\,\,\,l\,\,\,\,\,\,\,k&0&1&2&3&4&5&6\\
\hline
0& 0& 0&1$_{1, 0}$& 28$_{1, 0}$& 300$_{1, 0}$& 1925$_{1, 0}$& 
8918$_{1, 0}$\\                   1& 0& 0& 0& 29$_{1, 0}$& 707$_{1, 
0}$& 6999$_{1, 0}$& 42889$_{1, 0}$\\                   2& -1$_{2, 
0}$& 0& 0& 2$\cdot$1$_{3, 0}$& 463$_{1, 0}$ +1$_{1, 1}$& 9947$_{1, 0}$& 
92391$_{1, 0}$\\                   3& -28$_{2, 0}$& -29$_{2, 0}$& 
-2$\cdot$1$_{4, 0}$&1$_{5, 0}$ -1$_{6, 0}$& 2$\cdot$29$_{3, 0}$& 5280$_{1, 0}$ + 
29$_{1, 1}$ + 2$\cdot$28$_{3, 0}$& 101850$_{1, 0}$\\                   4& 
-300$_{2, 0}$& -707$_{2, 0}$& -463$_{2, 0}$ -1$_{2, 1}$& -2$\cdot$29$_{4, 
0}$& 29$_{5, 0}$ - 29$_{6, 0}$& 2$\cdot$463$_{3, 0}$ + 2$\cdot$1$_{3, 1}$& 47897$_{1, 
0}$ + 463$_{1, 1}$ +1$_{1, 2}$ + 2$\cdot$707$_{3, 0}$\\                   5& 
-1925$_{2, 0}$& -6999$_{2, 0}$& -9947$_{2, 0}$& -5280$_{2, 0}$ \!- \!
29$_{2, 1}$ \!-\! 2$\cdot$28$_{4, 0}$&\! -\!2$\cdot$463$_{4, 0}$ \!-\! 2$\cdot$1$_{4, 1}$& 463$_{5, 0}$ 
+1$_{5, 1}$ - 463$_{6, 0}$ -1$_{6, 1}$& 2$\cdot$5280$_{3, 0}$ + 2$\cdot$29$_{3, 1}$ 
+ 28$_{5, 0}$\\                   6& -8918$_{2, 0}$& -42889$_{2, 0}$& 
-92391$_{2, 0}$& -101850$_{2, 0}$& -47897$_{2, 0}$\! -\! 463$_{2, 1}$ 
\!-\!1$_{2, 2}$\!-\! 2$\cdot$707$_{4, 0}$& \!-\!2$\cdot$5280$_{4, 0}$ \!-\! 2$\cdot$29$_{4, 1}$ \!-\! 28$_{6, 
0}$& 5280$_{5, 0}$ + 29$_{5, 1}$ - 5280$_{6, 0}$ - 29$_{6, 1}$\\      
             7& -32928$_{2, 0}$& -193102$_{2, 0}$& -544786$_{2, 0}$& 
-894198$_{2, 0}$& -842537$_{2, 0}$ - 28$_{2, 1}$ - 2$\cdot$300$_{4, 0}$& 
-366384$_{2, 0}$ \!-\! 5280$_{2, 1}$ \!-\! 29$_{2, 2}$ \!-\! 2$\cdot$9947$_{4, 0}$& 
-2$\cdot$47897$_{4, 0}$ - 2$\cdot$463$_{4, 1}$ - 2$\cdot$1$_{4, 2}$ - 707$_{6, 0}$\\          
         8& -102816$_{2, 0}$& -699762$_{2, 0}$& -2392663$_{2, 0}$& 
-5096487$_{2, 0}$& -7032993$_{2, 0}$& -5951291$_{2, 0}$ - 707$_{2, 
1}$ - 2$\cdot$6999$_{4, 0}$& -2450888$_{2, 0}$ \!-\! 47897$_{2, 1}$ \!-\! 463$_{2, 
2}$ \!-\!1$_{2, 3}$ \!-\! 2$\cdot$101850$_{4, 0}$\\                   9& -282150$_{2, 
0}$& -2156994$_{2, 0}$& -8526042$_{2, 0}$& -21888529$_{2, 0}$& 
-38869314$_{2, 0}$& -47455968$_{2, 0}$ - 2$\cdot$1925$_{4, 0}$& 
-37134593$_{2, 0}$ - 9947$_{2, 1}$ - 2$\cdot$92391$_{4, 0}$
\end{tabular}\end{tiny}
\end{center}}\vspace{1in}
\caption{Expansion coefficients $ b^{F_4}_{k, l} $ for one $F_4$ instanton  (detailed version).}
{\label{tb:F4b}\begin{center}\begin{tiny}
\begin{tabular}{c| lllllllllllll }\,\,\,\,\,\,\,\,l\,\,\,\,\,\,\,k&0&1&2&3&4&5&6\\
\hline
0& 0& 0& 0&$1_{1, 0}$& 52$_{1, 0}$& 1053$_{1, 0}$& 12376$_{1, 0}$\\
1& 0& 0& 0& 0& 53$_{1, 0}$& 2432$_{1, 0}$& 
44980$_{1, 0}$\\
2& 0& 0& 0& 0& 0& 1483$_{1, 0}$ 
+1$_{1, 1}$& 59996$_{1, 0}$\\
3&$1_{2, 0}$& 0& 0& 0& 
-1$_{6, 0}$& 0& 28771$_{1, 0}$ + 53$_{1, 1}$\\
4& 
52$_{2, 0}$& 53$_{2, 0}$& 0& -1$_{5, 0}$& 0& -53$_{6, 0}$& 0\\
5& 1053$_{2, 0}$& 
2432$_{2, 0}$& 1483$_{2, 0}$ +1$_{2, 1}$& 0& -53$_{5, 0}$& 0& 
-1483$_{6, 0}$ -1$_{6, 1}$\\
6& 12376$_{2, 0}$& 44980$_{2, 0}$& 
59996$_{2, 0}$& 28771$_{2, 0}$ + 53$_{2, 1}$& 0& -1483$_{5, 0}$ 
-1$_{5, 1}$\\
7& 100776$_{2, 0}$& 495872$_{2, 0}$& 1023464$_{2, 0}$& 
1034880$_{2, 0}$& 432526$_{2, 0}$ + 1483$_{2, 1}$ +1$_{2, 2}$& 0& 
-28771$_{5, 0}$ - 53$_{5, 1}$\\
8& 
627912$_{2, 0}$& 3856722$_{2, 0}$& 10670660$_{2, 0}$& 16410602$_{2, 
0}$& 13979228$_{2, 0}$ + 52$_{2, 1}$& 5356771$_{2, 0}$ + 28771$_{2, 
1}$ + 53$_{2, 2}$& 0\\
9& 3187041$_{2, 0}$& 23235328$_{2, 0}$& 79721160$_{2, 0}$& 
162744192$_{2, 0}$& 207409930$_{2, 0}$& 157182848$_{2, 0}$ + 
2432$_{2, 1}$& 56823162$_{2, 0}$ + 432526$_{2, 1}$ + 1483$_{2, 2}$ 
+1$_{2, 3}$
\end{tabular}\end{tiny}
\end{center}}

\end{sidewaystable}

\begin{sidewaystable}[p!]
\caption{Expansion coefficients $ b^{E_6}_{k, l} $ for one $E_6$ instanton  (detailed version).}
{\label{tb:E6b}\begin{center}\begin{tiny}
\begin{tabular}{c| lllllllllllll }\,\,\,\,\,\,\,\,l\,\,\,\,\,\,\,k&0&1&2&3&4&5&6&7\\
\hline 
0& 0& 0& 0& 0&$1_{1, 0}$& 78$_{1, 0}$& 2430$_{1, 0}$& 43758$_{1, 0}$\\%& 537966$_{1, 0}$\\
                   1& 0& 0& 0& 0& 0& 79$_{1, 0}$& 
5512$_{1, 0}$& 157221$_{1, 0}$\\%& 2644707$_{1, 0}$\\                   
2& 0& 0& 0& 0& 0& 0& 3238$_{1, 0}$ +1$_{1, 1}$& 201292$_{1, 0}$\\%& 5283549$_{1, 0}$\\
                   3& 0& 0& 0& 0& 0& -1$_{6, 0}$& 
0& 90911$_{1, 0}$ + 79$_{1, 1}$\\%& 5048576$_{1, 0}$\\                   
4& -1$_{2, 0}$& 0& 0& 0& 0& 2$\cdot$1$_{3, 0}$& -79$_{6, 0}$& 0\\%& 1956516$_{1, 0}$ + 3238$_{1, 1}$ +1$_{1, 2}$\\                   
5& -78$_{2, 0}$& 
-79$_{2, 0}$& 0&$1_{5, 0}$& -2$\cdot$1$_{4, 0}$& 0& 2$\cdot$79$_{3, 0}$& 2$\cdot$78$_{3, 0}$ 
- 3238$_{6, 0}$ -1$_{6, 1}$\\%& 0\\ 
                  6& -2430$_{2, 0}$& 
-5512$_{2, 0}$& -3238$_{2, 0}$ -1$_{2, 1}$& 0& 79$_{5, 0}$& -2$\cdot$79$_{4, 
0}$& 0& 2$\cdot$3238$_{3, 0}$ + 2$\cdot$1$_{3, 1}$ - 78$_{6, 0}$\\%& 11024$_{3, 0}$ - 90911$_{6, 0}$ - 79$_{6, 1}$\\ 
                 7& -43758$_{2, 0}$& 
-157221$_{2, 0}$& -201292$_{2, 0}$& -90911$_{2, 0}$ \!-\! 79$_{2, 1}$& 0& 
-2$\cdot$78$_{4, 0}$ + 3238$_{5, 0}$ +1$_{5, 1}$& -2$\cdot$3238$_{4, 0}$ - 2$\cdot$1$_{4, 
1}$ + 78$_{5, 0}$& 0\\%& 181822$_{3, 0}$ + 158$_{3, 1}$ - 5512$_{6, 0}$\\
                   8& -537966$_{2, 0}$& -2644707$_{2, 0}$& 
-5283549$_{2, 0}$& -5048576$_{2, 0}$& -1956516$_{2, 0}$ \!-\! 3238$_{2, 
1}$ \!-\!1$_{2, 2}$& 0& -2$\cdot$5512$_{4, 0}$ \!+\! 90911$_{5, 0}$ \!+\! 79$_{5, 1}$& -2$\cdot$90911$_{4, 0}$ - 2$\cdot$79$_{4, 1}$ + 5512$_{5, 0}$\\%& 0\\                  
% 9& -4969107$_{2, 0}$& -30843384$_{2, 0}$& -83526287$_{2, 0}$& 
%-122611239$_{2, 0}$& -97616506$_{2, 0}$ - 78$_{2, 1}$& -34327258$_{2, 
%0}$ \!-\! 90911$_{2, 1}$ \!-\! 79$_{2, 2}$& -4860$_{4, 0}$& -402584$_{4, 0}$ 
%\!+\! 1956516$_{5, 0}$ \!+\! 3238$_{5, 1}$ \!+\! 1$_{5, 2}$\\%& \!-\! 3913032$_{4, 0}$ \!-\! 6476$_{4, 1}$ \!-\! 2$_{4, 2}$ \!+\!201292$_{5, 0}$
\end{tabular}\end{tiny}
\end{center}}
\caption{Expansion coefficients $ b^{E_7}_{k, l} $ for one $E_7$ instanton  (detailed version).}
{\label{tb:E7b}\begin{center}\begin{tiny}
\begin{tabular}{c| lllllllllllll }\,\,\,\,\,\,\,\,l\,\,\,\,\,\,\,k&6&7&8&9&10&11&12\\
\hline
0&1$_{1,0}$& 133$_{1, 0}$& 7371$_{1, 0}$& 238602$_{1, 0}$& 5248750$_{1, 
0}$& 85709988$_{1, 0}$& 1101296924$_{1, 0}$\\%& 11604306012$_{1, 0}$\\   
1&      0& 134$_{1, 0}$& 16283$_{1, 0}$& 835562$_{1, 0}$& 25353429$_{1, 
0}$& 528271250$_{1, 0}$& 8241562056$_{1, 0}$\\%& 101887800238$_{1, 0}$\\ 
2&        0& 0& 9178$_{1, 0}$ +1$_{1, 1}$& 1014581$_{1, 0}$& 
48250384$_{1, 0}$& 1375996758$_{1, 0}$& 27238257500$_{1, 0}$\\%& 407138342352$_{1, 0}$\\ 
3&        0& -1$_{6, 0}$& 0& 426533$_{1, 0}$ + 
134$_{1, 1}$& 42814809$_{1, 0}$& 1890508984$_{1, 0}$& 
50809770626$_{1, 0}$\\%& 958188744375$_{1, 0}$\\ 
4&        0& 0& -134$_{6, 
0}$& 0& 15077814$_{1, 0}$ + 9178$_{1, 1}$ +1$_{1, 2}$& 1374731795$_{1, 
0}$ + 133$_{1, 1}$& 56496374025$_{1, 0}$\\%& 1434911790496$_{1, 0}$\\    
 5&    0& 0& 0& -9178$_{6, 0}$ -1$_{6, 1}$& 0& 431311556$_{1, 0}$ + 
426533$_{1, 1}$ + 134$_{1, 2}$& 35790781289$_{1, 0}$ + 16283$_{1, 
1}$\\%& 1372710611055$_{1, 0}$\\
6&         0& 2$\cdot$1$_{3, 0}$& 0& -133$_{6, 
0}$& -426533$_{6, 0}$ - 134$_{6, 1}$& 0& 10381128673$_{1, 0}$ + 
15077814$_{1, 1}$ + 9178$_{1, 2}$ +1$_{1, 3}$\\%& 786475367467$_{1, 0}$ + 1014581$_{1, 1}$\\ 
7&        -2$\cdot$1$_{4, 0}$& 0& 2$\cdot$134$_{3, 0}$& 2$\cdot$133$_{3, 
0}$& -16283$_{6, 0}$& -15077814$_{6, 0}$ - 9178$_{6, 1}$ -1$_{6, 2}$& 
0\\%& 215981422537$_{1, 0}$ + 431311556$_{1, 1}$ + 426533$_{1, 2}$ + 134$_{1, 3}$\\
8&         0& -2$\cdot$134$_{4, 0}$& 0& 2$\cdot$9178$_{3, 0}$ + 2$\cdot$1$_{3, 
1}$& 2$\cdot$16283$_{3, 0}$& 2$\cdot$7371$_{3, 0}$ - 1014581$_{6, 0}$& 
-431311556$_{6, 0}$ - 426533$_{6, 1}$ - 134$_{6, 2}$\\%& 0\\         
9& 133$_{5, 0}$& -2$\cdot$133$_{4, 0}$& -2$\cdot$9178$_{4, 0}$ - 2$\cdot$1$_{4, 1}$& 0& 
2$\cdot$426533$_{3, 0}$ + 2$\cdot$134$_{3, 1}$& 2$\cdot$1014581$_{3, 0}$ - 7371$_{6, 0}$& 
2$\cdot$835562$_{3, 0}$ - 42814809$_{6, 0}$\\%& 477204$_{3, 0}$ - 10381128673$_{6, 0}$ - 15077814$_{6, 1}$ - 9178$_{6, 2}$ -$_{6, 3}$"
\end{tabular}\end{tiny}
\end{center}}
\caption{Expansion coefficients $ b^{E_8}_{k, l} $ for one $E_8$ instanton  (detailed version).}
{\label{tb:E8b}\begin{center}\begin{tiny}
\begin{tabular}{c| lllllllllllll }\,\,\,\,\,\,\,\,l\,\,\,\,\,\,\,k&10&11&12&13&14&15&16\\
\hline
0&$1_{1, 0}$& 248$_{1, 0}$& 27000$_{1, 0}$& 1763125$_{1, 0}$&  79143000$_{1, 0}$& 2642777280$_{1, 0}$& 69176971200$_{1, 0}$\\        
1& 0& 249$_{1, 0}$& 57877$_{1, 0}$& 5943753$_{1, 0}$& 368338125$_{1, 0}$& 15776893240$_{1, 0}$& 505168052220$_{1, 0}$\\     
2& 0& 0& 31373$_{1, 0}$ +1$_{1, 1}$& 6815877$_{1, 0}$& 659761497$_{1, 0}$& 38811914750$_{1, 0}$& 1587614120010$_{1, 0}$\\    
3& 0& -1$_{6, 0}$& 0& 2666126$_{1, 0}$ + 249$_{1, 1}$&  539686750$_{1, 0}$& 49211333622$_{1, 0}$& 2749381821611$_{1, 0}$\\    
4& 0& 0& -249$_{6, 0}$& 0& 171724751$_{1, 0}$ + 31373$_{1, 1}$ +1$_{1, 2}$& 32299833627$_{1, 0}$ + 248$_{1, 1}$& 2773706709375$_{1, 0}$\\
5& 0& 0& 0& -31373$_{6, 0}$ -1$_{6, 1}$& 0&  8928599916$_{1, 0}$ + 2666126$_{1, 1}$ + 249$_{1, 2}$& 
1557460051916$_{1, 0}$ + 57877$_{1, 1}$\\ 
6& 0& 0& 0& -248$_{6, 0}$& -2666126$_{6, 0}$ - 249$_{6, 1}$& 0& 389739771210$_{1, 0}$ + 171724751$_{1, 1}$ + 31373$_{1, 2}$ +1$_{1, 3}$\\
7& 0& 0& 0& 0& -57877$_{6, 0}$& -171724751$_{6, 0}$ - 31373$_{6, 1}$ 
-1$_{6, 2}$& 0\\
8& 0& 0& 0& 0& 0& -6815877$_{6, 0}$& -8928599916$_{6, 0}$ - 2666126$_{6, 1}$ - 249$_{6, 2}$\\
9&0& 0& 0& 0& 0& -27000$_{6, 0}$& -539686750$_{6, 0}$
\end{tabular}\end{tiny}
\end{center}}
\end{sidewaystable}

\newpage\bibliography{OEIS}

\end{document}